\documentclass[preprint,showpacs,preprintnumbers,amsmath,amssymb]{revtex4}

\usepackage{graphicx,psfrag}
\usepackage{dcolumn}
\usepackage{bm}

\def\barr{\begin{array}}
\def\earr{\end{array}}
\def\beq{\begin{equation}}
\def\eeq{\end{equation}}
\def\bea{\begin{eqnarray}}
\def\eea{\end{eqnarray}}
\def\bmath{\begin{displaymath}}
\def\emath{\end{displaymath}}
\def\bq{\begin{quote}}
\def\eq{\end{quote}}

\def\slash#1{\setbox0=\hbox{$#1$}#1\hskip-\wd0\hbox to\wd0{\hss\sl/\/\hss}}

\def\g5{\gamma_5}

\def\Li2{\mbox{Li$_2$}}

\def\lN1{\lambda_{N_1}}

\psfrag{log(mN)}{$\log{m_N}$}
\psfrag{BR(tau -> 3mu)}{BR($\tau \to 3 \mu$)}
\psfrag{BR(tau -> mu + gamma)}{BR($\tau \to \mu \gamma$)}
\psfrag{BR(tau -> 3e)}{BR($\tau \to 3 e$)}
\psfrag{BR(tau -> e + gamma)}{BR($\tau \to e \gamma$)}
\psfrag{BR(tau -> e +gamma)}{BR($\tau \to e \gamma$)}
\psfrag{BR(mu -> 3e)}{BR($\mu \to 3 e$)}
\psfrag{BR(mu -> e + gamma)}{BR($\mu \to e \gamma$)}
\psfrag{tanb}{$\tan{\beta}$}
\psfrag{tan b}{$\tan{\beta}$}
\psfrag{mod(theta1)}{$\vert \theta_1 \vert$}
\psfrag{mod(theta2)}{$\vert \theta_2 \vert$}
\psfrag{mod(theta3)}{$\vert \theta_3 \vert$}
\psfrag{mod(theta13)}{$\vert \theta_{13} \vert$}
\psfrag{M0}{$M_0$ (GeV)}
\psfrag{M1/2}{$M_{1/2}$ (GeV)}
\psfrag{M0=M1/2}{$M_0 = M_{1/2}$ (GeV)}
\psfrag{Deltas12}{$|\delta_{12}|$}
\psfrag{Deltas13}{$|\delta_{13}|$}
\psfrag{Deltas23}{$|\delta_{23}|$}
\psfrag{theta13}{$\theta_{13}$}
\psfrag{Ynu21}{$|Y_{\nu}^{12}|$}
\psfrag{Ynu22}{$|Y_{\nu}^{22}|$}
\psfrag{Ynu23}{$|Y_{\nu}^{32}|$}
\psfrag{Ynu31}{$|Y_{\nu}^{13}|$}
\psfrag{Ynu32}{$|Y_{\nu}^{23}|$}
\psfrag{Ynu33}{$|Y_{\nu}^{33}|$}

\begin{document}
\preprint{FTUAM-05-16}
\preprint{IFT-UAM/CSIC-05-45}
\title{Testing Supersymmetry with Lepton Flavor Violating\\ $\tau$ and $\mu$ decays}
\author{Ernesto Arganda}
\email{ernesto.arganda@uam.es}
\author{Mar\'{\i}a J. Herrero}%
\email{maria.herrero@uam.es}
\affiliation{%
Departamento de F\'{\i}sica Te\'orica, UAM/IFT, 28049, Madrid, Spain.
}%
\begin{abstract}
In this work the following lepton flavor violating $\tau$ and $\mu$ decays
are studied: $\tau^- \to \mu^- \mu^- \mu^+$,
$\tau^- \to e^- e^- e^+$, $\mu^- \to e^- e^- e^+$, 
$\tau^- \to \mu^- \gamma$, $\tau^- \to e^- \gamma$ and 
$\mu^- \to e^- \gamma$. We work in a supersymmetric 
scenario consisting of the minimal supersymmetric standard model 
particle content, extended by the addition of three heavy right handed
Majorana neutrinos and their supersymmetric partners, and where the 
generation of neutrino masses is done via the seesaw mechanism. Within 
this context, a significant lepton flavor mixing is generated in the
slepton sector due to the Yukawa neutrino 
couplings, which is transmited from
the high to the low energies via the renormalization group equations. 
This slepton mixing then generates via loops of supersymmetric particles
significant contributions to the rates of $l_j \to 3 l_i$ and the
correlated
$l_j \to l_i \gamma$ decays. We analize here in full detail these rates 
in terms of the relevant input parameters, which are the usual minimal
supergravity parameters and the seesaw parameters. For the 
$l_j \to 3 l_i$ decays, a full one-loop analytical computation 
of all the 
contributing supersymmetric loops is presented. This completes and corrects 
previous computations in the literature. In the numerical analysis 
compatibility with the most recent experimental upper bounds on all these 
$\tau$ and $\mu$
decays, with the neutrino data, and with the present lower bounds on the 
supersymmetric particle masses are required. Two typical scenarios with
degenerate and hierarchical heavy neutrinos are considered. 
We will show here that
the minimal supergravity and seesaw 
parameters do get important restrictions from these $\tau$ and $\mu$ 
decays in the
hierarchical neutrino case. 

\end{abstract}
\maketitle

\section{\label{sec:Intro} Introduction}

The present strong evidence for lepton flavor changing neutrino 
oscillations  in
neutrino data~\cite{neutrinodata} 
implies the existence of non-zero masses for the light neutrinos, 
and provides the first experimental clue for physics beyond the 
Standard Model (SM). These oscillations also give an important 
information on the neutrino mixing angles 
of the Maki-Nakagawa-Sakata matrix ($U_{MNS}$)~\cite{MNS}. 
The experimentally suggested smallness of the three neutrino masses 
can be explained in a very simple and elegant way by the seesaw 
mechanism of neutrino mass generation~\cite{seesaw}. 
This mechanism is usually implemented by the introduction 
of three heavy right-handed (RH) Majorana neutrinos whose masses, 
$m_{M_i}$, can be much higher than the SM particle masses. 
The smallness of the light neutrino masses, $m_{\nu_i}$, appears naturally due to the induced large suppression 
by the ratio of the two very distant mass scales that are involved 
in the $3 \times 3$ seesaw mass matrices, the Majorana matrix $m_M$ and the 
Dirac matrix $m_D$. For instance, in the one generation case, where the seesaw
model predicts $m_\nu \sim m_D^2/m_M$, light
neutrino masses in the 0.1 - 1 eV range can be generated with $m_D$ being
of the order of the electroweak scale, $v=174$ GeV, and  large $m_M$ of
the order of $10^{14}$ GeV. This huge separation between
$m_M$ and the electroweak scale has, however, a serious 
drawback since 
it leads to the well known
hierarchy problem of the SM, where a tree level Higgs boson mass 
of the order of $v$ is driven by the radiative corrections involving the Majorana
neutrinos 
to very unnatural high values related to the 
new scale $m_M$. 

The most elegant solution to this hierarchy problem 
is provided by 
the introduction of the symmetry relating fermions and bosons, called
supersymmmetry (SUSY). When the seesaw mechanism for the neutrino mass
generation is implemented in a SUSY context, the SUSY scalar partners 
of the
neutrinos, i.e. the sneutrinos, do also contribute to the 
radiative corrections
of the Higgs boson masses and cancel the
dangerous contributions from the Majorana neutrinos, solving in this way 
the hierarchy problem of the simplest non-SUSY version of the seesaw
mechanism.

The best evidence of supersymmetry would be obviously the discovery of 
the SUSY particles in the present or next generation colliders.
However, there are alternative ways to test
supersymmetry which are indirect and
complementary to the direct SUSY particle searches. These refer to the 
potential measurement of the SUSY particle contributions, via radiative
corrections, to rare processes which are being explored at present and 
whose rates are predicted to be highly suppressed in the SM. Among these 
processes, the Lepton Flavor Violating (LFV) $\tau$ and $\mu$ decays are
probably the most interesting ones for various reasons. 
On one hand, they get
vanishing rates
in the SM with massless neutrinos and highly suppressed rates
in the SM with massive netrinos. The smallness of these rates in the
non-SUSY version of the seesaw mechanism for neutrino mass generation 
is due to their suppression by inverse powers of the heavy scale 
$m_M$.      
On the other hand, although these decays have not
been seen so far in the present experiments, there are very 
restrictive upper
bounds on their possible rates which imply important restrictions on the 
new physics beyond the SM. These restrictions apply even more 
severely to the case
of softly broken SUSY theories with massive neutrinos and the seesaw
mechanism, since these give rise to higher rates~\cite{borzumati}, 
being suppressed  
by inverse powers of the SUSY breaking scale, 
$m_{SUSY}\leq 1$ TeV, instead of inverse powers of $m_M$.            
 
We will be devoted here in particular to the LFV  $\tau$ and $\mu$ 
decays of type
$l_j\to l_i \gamma$ and $l_j\to 3l_i$ where the present experimental 
upper bounds are the most restrictive ones~\cite{Aubert:2003pc,Bellgardt:1987du,Aubert:2005ye,Aubert:2005wa,mue}, 
specifically,
\begin{eqnarray}
BR(\tau^- \to \mu^- \mu^- \mu^+) &<& 1.9 \times
10^{-7} \nonumber \\
BR(\tau^- \to e^- e^- e^+) &<& 2.0 \times 10^{-7}
\nonumber \\
BR(\mu^- \to e^- e^- e^+) &<& 1.0 \times 10^{-12}
\nonumber \\
BR(\tau^- \to \mu^- \gamma) &<& 6.8 \times 10^{-8} \nonumber \\
BR(\tau^- \to e^- \gamma) &<& 1.1 \times 10^{-7} \nonumber  \\
BR(\mu^- \to e^- \gamma) &<& 1.2 \times 10^{-11} \nonumber
\label{cotas}
\end{eqnarray}
Our aim in this paper is to analize the branching ratios that can be
generated for all these processes in the context of the SUSY-seesaw
scenario with the minimal SUSY content, i.e the Minimal Supersymmetric Standard
Model (MSSM), enhanced by the
addition of three RH neutrinos and their corresponding
SUSY partners. These LFV processes are 
induced
by loops of SUSY particles which transmit the lepton flavor mixing from
the slepton mass matrices to the observable charged lepton sector. The
intergenerational mixing in the slepton sector, $(M_{\tilde l})_{ij},\, i\neq j$, 
 is induced in turn by 
the radiative corrections involving the neutrino Yukawa couplings $Y_{\nu}$ or,
simmilarly, by the running of the soft SUSY parameters in the slepton
sector, via the Renormalization Group Equations (RGEs), from the high
energy scale, $M_X>m_M$, where the heavy Majorana neutrinos are still active, 
down to the electroweak scale. We will assume here a Minimal Supergravity
scenario (mSUGRA) with universal soft parameters at $M_X$ and the
breaking of the electroweak symmetry being generated radiatively. This scenario
is also refered to in the literature as the constrained MSSM (CMSSM).    

The above LFV processes have previously been studied in the SUSY-seesaw context 
by several authors~\cite{Hisano:1995cp,todos,Casas:2001sr,Babu:2002et,Ellis:2002fe}, under some specific 
assumptions for both seesaw
parameters, $m_D$ (or $Y_{\nu}$, since they are related by $m_D=Y_{\nu}v\sin\beta$,
with $\tan\beta=v_2/v_1$ being the ratio between the two MSSM Higgs vacuum 
spectation values) and $m_M$, and for the mSUGRA parameters, $M_0$, $M_{1/2}$,
$A_0$, $\mbox{sign}(\mu)$ and $\tan\beta$. 

Our present study of these decay channels updates, completes and corrects the previous
anlayses in several respects. First we include, by the first time to our 
knowledge, the full set of SUSY one-loop contributions to the $l_j \to 3 l_i$
decays, namely, the photon, the Z boson, and the 
Higgs bosons penguin diagrams, and the box diagrams. 
The most complete computation so
far of
these $l_j \to 3 l_i$ decays was done in~\cite{Hisano:1995cp} where the 
contributions from the photon and
Z boson penguin diagrams and from the box diagrams were included, 
but they focused
on the particular choice of degenerate heavy Majorana neutrinos and they
presented 
numerical results just for $\mu \to 3e$ decays. 
We extend this
previous study by including in addition the Higgs penguin diagrams mediated 
by the three neutral
MSSM bosons, $H_0$, $h_0$ and $A_0$, and correct
their results for the Z penguin contributions. We also extend their study in that we present results
for the three decays, $\mu \to 3e$, $\tau \to 3\mu$ and $\tau \to 3e$ and
consider both possible scenarios, degenerate and hierarchical heavy neutrinos. 

The contributions from the Higgs penguin diagrams, 
in the
SUSY-seesaw model, were firstly analized in~\cite{Babu:2002et}. They worked
in the large $\tan \beta$ limit and used the mass insertion approximation to
account for the induced effect from the intergenerational slepton mixing 
in the SUSY contributing loops. 
There it was concluded that these Higgs-mediated 
contributions 
can be
very relevant in the large $\tan\beta$ region, because the radiatively induced
LFV Higgs-$\tau$-$\mu$ 
couplings grow as $\tan^2 \beta$ (and, in consequence, the $BR(\tau \to 3\mu)$ as 
$\tan^6 \beta)$, and also because the SUSY one-loop 
contributions 
do not decouple in these couplings. These large $\tan\beta$ enhacement and 
SUSY non-decoupling behaviour were also found in the 
LFV Higgs boson decays, 
$H_0,h_0,A_0 \to l_i\bar l_j$~\cite{Brignole1,Arganda:2004bz}. 
A more exhaustive study of the $\tau \to 3\mu$ and other Higgs-mediated 
LFV $\tau$ decays, 
including 
an
estimate of the Higgs contributions, were done in~\cite{Brignole2}.
However, these previous numerical estimates of the Higgs contributions to 
LFV $\tau$ and $\mu$
decays were done in the context of a generic MSSM (see also~\cite{Paradisi:2005tk}), where the Higgs boson mass, or
equivalently 
$m_A^0$, 
is an
input parameter and can take small values of the order of 100 GeV which 
produces 
larger rates. A more recent study on the LFV Higgs decays has been done
in~\cite{Parry:2005fp} in the SUSY-GUT $SU(5)$ context. We instead work here in the mSUGRA context where all the MSSM 
particle masses are
quantities derived from the mSUGRA parameters. We will see here that
this and the requirement of compatibility with
the present experimental lower bounds for all the SUSY particle 
masses~\cite{pdg2004} do indeed constraint
the contribution from the Higgs penguins.     

In the present work we also include the predictions for the $l_j\to l_i \gamma$
channels which, for the context we work with, are interestingly correlated with 
the $l_j\to 3 l_i$ rates. This correlation has been studied previously in the
generic MSSM context in~\cite{Brignole2} and in a similar mSUGRA 
context in~\cite{Ellis:2002fe}, but in this later the dominant photon 
penguin
approximation was used. We will 
update this comparative  analysis of the $l_j\to l_i \gamma$ and 
$l_j\to 3 l_i$ rates, in the mSUGRA context, including the full contributions, 
and considering the 
very recent upper bounds for $\tau\to \mu \gamma$~\cite{Aubert:2005ye} 
and $\tau\to e \gamma$~\cite{Aubert:2005wa}.  
In addition, we also require the input seesaw
parameters to be compatible with the present neutrino data. For this
comparison with the neutrino data we use the parametrization first
introduced in~\cite{Casas:2001sr} for the study of the $\mu \to e
\gamma$ decay.
  
Our final goal will be to use the SUSY contributions to all the above
LFV $\tau$ and $\mu$ decays 
as an efficient way to test the mSUGRA and seesaw parameters. 
With this goal in mind we will analize here the size of the branching ratios 
in terms of the mSUGRA and seesaw parameters and will explore 
in detail the restrictions imposed from the present experimental bounds. 
We
will find that for some plausible choices of the seesaw parameters,
being compatible with neutrino data, there are indeed large excluded 
regions in the 
mSUGRA parameter space.

The present work is organized as follows. In section~\ref{MSSMnuR} 
we will review the basic aspects of the MSSM extended with three
 RH neutrinos,
their SUSY partners and the seesaw mechanism for neutrino mass
generation. The lepton flavor mixing in the slepton sector and in the mSUGRA
context will be explained in section~\ref{LFV}. There we also include the
 exact diagonalization of the sfermion mass matrices, both in the slepton 
 and in the
 sneutrino sectors. The analytical results of the LFV
$l_j \to 3 l_i$ decays will be presented in section~\ref{analytical}. 
The numerical results for all the LFV $\tau$ and $\mu$ decays will be presented in
section~\ref{numerical}. Finally, section~\ref{conclu} will be devoted
to the conclusions.

\section{\label{MSSMnuR} The MSSM extended with RH neutrinos and sneutrinos}

In this section we briefly review the additional basic ingredients that are 
needed to
extend the MSSM in order to include three right handed neutrinos, their 
corresponding SUSY partners, i.e. the sneutrinos,
and the generation of neutrino
masses by the seesaw mechanism.  We follow closely 
the notation of 
refs.~\cite{Casas:2001sr,Arganda:2004bz} to describe the SUSY-seesaw scenario 
and the connection with neutrino data. For the other sectors of the MSSM we
assume here the standard conventions as defined, for instance, in
~\cite{Haber:1985rc,Gunion:1986yn}.

We start with the Yukawa-sector of the MSSM-seesaw that contains the three 
left handed (LH) SM neutrinos $\nu_{L, i}^o$ and three extra right handed (RH) 
massive 
neutrinos $\nu_{R, i}^o$, whose Yukawa interactions provide, after 
spontaneous electroweak symmetry breaking, together with the right 
handed neutrino masses, the following mass Lagrangian containing the Dirac and Majorana mass terms,
\begin{equation}
-L^\nu_{mass} = \frac{1}{2} (\overline{\nu^0_L}, (\overline{\nu^0_R})^C)
M^\nu \left(\begin{array}{c} (\nu^{0}_L)^C\\ \nu^0_R \end{array} \right)\  + h.c.,
\end{equation} 

where,
\begin{equation}
M^\nu\ =\ \left( \begin{array}{cc} 0 & m_D\\ m_D^T & m_M \end{array} \right).
\label{mass6x6}
\end{equation}  

Here $m_D$ is the $3 \times 3$ Dirac mass matrix that is related to 
the $3 \times 3 $ Yukawa coupling matrix $Y_{\nu}$ and the MSSM Higgs vacum
expectation value, $<H_2>=v_2=v\sin\beta$ with $v=174$ GeV, 
by $m_D=Y_{\nu} <H_2>$. The other MSSM Higgs doublet gives masses to 
the charged leptons by $m_l=Y_l <H_1>$, where $Y_l$ are the Yukawa couplings of
the charged leptons and  $<H_1>=v_1=v\cos\beta$.
The remaining $3 \times 3 $ mass matrix involved in the seesaw mechanism, $m_M$,
is real, non singular and symmetric, and provides the masses for the three RH neutrinos 

The mass matrix $M^{\nu}$ is a $6 \times 6$ complex symmetric matrix that can be diagonalized by a $6 \times 6$ unitary matrix $U^{\nu}$ in the following way:

\begin{equation}
U^{\nu T}M^\nu U^\nu =\hat{M}^\nu = diag (m_{\nu_1},m_{\nu_2},m_{\nu_3},m_{N_1},m_{N_2},m_{N_3}).
\label{matrizU}
\end{equation}
This gives 3 light Majorana neutrino mass eigenstates $\nu_i$, 
with masses $m_{\nu_i}$ (i=1,2,3), and three heavy ones $N_i$, 
with masses $m_{N_i}$ (i=1,2,3), which are related to the weak eigenstates 
via,

\begin{equation}
\left(\begin{array}{c} \nu^0_L \\ (\nu^{0}_R)^C \end{array} \right)\ =\ 
U^{\nu\ast}\ \left(\begin{array}{c} \nu_L \\ N_L \end{array} \right)\quad \mbox{and}\quad
\left(\begin{array}{c} (\nu^{0}_L)^C \\ \nu^0_R \end{array} \right)\ =\ 
 U^\nu\ \left(\begin{array}{c} \nu_R \\ N_R \end{array} \right).
\end{equation}
The seesaw mechanism for neutrino mass generation assumes 
a large separation between the two mass scales involved in $m_D$ and $m_M$ 
matrices. More specifically, we shall assume here that all matrix elements 
of $m_D$ are much smaller than those of $m_M$, $m_D<<m_M$, and 
the predictions of the seesaw model are then given in power series 
of a matrix defined as,
\begin{eqnarray}
\xi &\equiv &m_D m_M^{-1}.
\end{eqnarray}
In particular, the previous diagonalization of the mass matrix 
$M^{\nu}$ can be solved in power series of $\xi$. 
For simplicity, we choose to work here and in the rest of this paper, 
in a flavor basis where the RH Majorana mass matrix, $m_M$, and 
the charged lepton mass matrix, $m_l$, are flavor diagonal. 
This means that all flavor mixing of the LH sector is included in the 
mixing matrix $U_{MNS}$. By working to the lowest orders of 
these power series expansions one finds, in the flavor basis, 
the following neutrino $3 \times 3$ matrices, 

\begin{eqnarray}
m_{\nu}&=&-m_D \xi^T + \mathcal{O}(m_D \xi^3) \simeq -m_D m_M^{-1}m_D^T\\ \nonumber
m_N &=& m_M + \mathcal{O}(m_D \xi) \simeq m_M.
\end{eqnarray}
Here, $m_N$ is already diagonal, but $m_{\nu}$ is not yet diagonal. The rotation from this flavor basis to the mass eigenstate basis is finally given by the MNS unitary matrix, $U_{MNS}$. Thus,
\begin{eqnarray}
m_{\nu}^{diag}&=&U_{MNS}^T m_{\nu} U_{MNS}= diag(m_{\nu_1},m_{\nu_2},m_{\nu_3}),\\ \nonumber
m_N^{diag} &=& m_N = diag(m_{N_1},m_{N_2},m_{N_3}),
 \end{eqnarray}
and the diagonalization of $M^{\nu}$ in eqs.~(\ref{mass6x6}) and (\ref{matrizU}) can be performed by the following unitary $6 \times 6$ matrix:

\begin{equation}
U^\nu\ =\ \left( \begin{array}{cc} (1-\frac{1}{2} \xi^* \xi^T) U_{MNS} & \xi^* (1-\frac{1}{2} \xi^T \xi^*)\\ -\xi^T (1- \frac{1}{2} \xi^* \xi^T)U_{MNS} & (1-\frac{1}{2} \xi^T \xi^*) \end{array} \right) + \mathcal{O}(\xi ^4).
\label{eq8}
\end{equation}
As for the $U_{MNS}$ matrix, we use the standard parametrization given by,

\begin{equation}
U_{MNS}\ =\ \left( \begin{array}{ccc} c_{12} c_{13} & s_{12} c_{13}& s_{13} e^{-i \delta}\\ -s_{12} c_{23}-c_{12}s_{23}s_{13}e^{i \delta} & c_{12} c_{23}-s_{12}s_{23}s_{13}e^{i \delta} & s_{23}c_{13} \\ s_{12} s_{23}-c_{12}c_{23}s_{13}e^{i \delta} & -c_{12} s_{23}-s_{12}c_{23}s_{13}e^{i \delta} & c_{23}c_{13}\end{array} \right) diag(1,e^{i \alpha},e^{i \beta}).
\label{Umns}
\end{equation}
where $c_{ij} \equiv \cos \theta_{ij}$ and $s_{ij} \equiv \sin \theta_{ij}$.

Regarding the sneutrino sector, and because of SUSY, the introduction of three
RH neutrinos, $\nu_R$, leads to the addition of the three corresponding SUSY 
partners, ${\tilde \nu_R}$. Thus, there are     
 two complex scalar fields $\tilde \nu_L$ and $\tilde \nu_R$ 
 per generation, as in the   
 charged slepton case where there are $\tilde l_L$ and $\tilde l_R$. 
 The difference is that in the sneutrino sector, 
 the seesaw matrix $\xi$ is involved, as in the neutrino sector,   
 and gives rise to a natural suppression of the RH sneutrino 
 components in the relevant mass eigenstates. 
 This fact makes the diagonalization procedure simpler in the sneutrino sector 
 than in the charged slepton one. 
 In order to understand properly this feature of the MSSM-seesaw model, 
 we will illustrate in the following the simplest case of one generation,
 where this suppression already manifests. For this, we follow 
 closely~\cite{Grossman:1997is}.  The generalization of this 
 decoupling behaviour of the ${\tilde \nu_R}$ components to the  
 three generations case is straightforward and we omit to show it here, 
 for brevity.

One starts by adding the new terms in the MSSM Lagrangian that involve the 
$\nu_R$ and/or ${\tilde \nu_R}$. In particular, the usual MSSM
 soft SUSY breaking potential must be modified to include 
 new mass and coupling terms for the right handed sneutrinos, 
 which for the one generation case are the following,
\begin{eqnarray}
V_{soft}^{\tilde \nu}= m_{\tilde M}^2 \tilde \nu_R^* \tilde \nu_R - \left(
\frac{g}{\sqrt{2}m_W}\epsilon_{ij} \frac{m_D A_{\nu}}{\sin \beta} H_2^i \tilde l_L^j
\tilde \nu_R^* + h.c. \right) + \left( m_M B_M \tilde \nu_R^* \tilde \nu_R + h.c. \right) \nonumber \\
\end{eqnarray}
where $m_{\tilde M}$, $A_{\nu}$ and $B_M$ are the new soft breaking parameters.
These are in addition to the usual soft  parameters of the slepton sector,
$m_{\tilde L}$, $m_{\tilde E}$ and $A_{l}$. 
The sneutrino mass terms of the MSSM-seesaw model can then be written in the one 
generation case as,
\begin{equation}
-\mathcal{L}_{mass}^{\nu}=\left(\begin{array}{c} Re (\tilde{\nu}_L) \, Re (\tilde{\nu}_R) \, 
Im(\tilde{\nu}_L) \, Im (\tilde{\nu}_R)\end{array} \right)
\left( \begin{array}{cc} M_+^2 & 0\\ 0 & M_{-}^2 \end{array} \right)
 \left(\begin{array}{c} Re(\tilde{\nu}_L) \\  Re(\tilde{\nu}_R) \\ Im(\tilde{\nu}_L)\\ Im(\tilde{\nu}_R) \end{array} \right)
\end{equation}
with,
\begin{equation}
M_{\pm}^2=
\left( \begin{array}{cc} m_{\tilde{L}}^2 + m_D^2 + \frac{1}{2} m_Z^2 \cos 2 \beta & 
m_D (A_{\nu}- \mu \cot \beta \pm m_M)\\  m_D (A_{\nu}- \mu \cot \beta \pm m_M) & m_{\tilde{M}}^2+m_D^2+m_M^2 \pm 2 B_M m_M \end{array} \right)
\end{equation}
Notice that, in the sneutrino sector, there are several mass scales involved, the soft SUSY-breaking parameters, 
$m_{\tilde L}$, $m_{\tilde M}$, $B_M$ and $A_{\nu}$, 
the Dirac mass $m_D$, the $\mu$-mass parameter, the Z boson mass $m_Z$ and 
the Majorana neutrino mass $m_M$. Our basic assumption in all this work 
is that $m_M$ is much heavier than the other mass scales involved (except $M_X$), 
$m_M>>m_D, m_Z, \mu, m_{\tilde{L}}, m_{\tilde M}, A_{\nu}, B_M$. 
The size of $B_M$ has been discussed in the literature~\cite{Grossman:1997is} 
and seems more controversial. For simplicity, 
we shall assume here that this is also smaller than $m_M$. 
In this large $m_M$ limit, the diagonalization of the previous sneutrino 
squared mass matrix is simpler and leads to four mass eigenstates, 
two of which are light, $\xi_1^l$, $\xi_2^l$ and two heavy, 
$\xi_1^h$, $\xi_2^h$. In the leading orders of the series expansion 
in powers of $\xi$ the mass eigenstates and their corresponding mass eigenvalues are given by (We correct in the definitions of $M_{\pm}^2$ and $\xi_2^l$ some typos with wrong signs of ref.~\cite{Arganda:2004bz}),
\begin{eqnarray}
\xi_1^l &=& \sqrt{2} \left( Re(\tilde{\nu}_L) - \xi Re(\tilde{\nu}_R)\right) \,\,;
\xi_2^l = \sqrt{2} \left( Im(\tilde{\nu}_L) + \xi Im(\tilde{\nu}_R)\right) \nonumber \\
\xi_1^h &=& \sqrt{2} \left( Re(\tilde{\nu}_R) + \xi Re(\tilde{\nu}_L)\right) \,\,;
\xi_2^h = \sqrt{2} \left( Im(\tilde{\nu}_R) - \xi Im(\tilde{\nu}_L)\right) \nonumber \\
m_{\xi_{1,2}^l}^2 &=& m_{\tilde{L}}^2 + 
\frac{1}{2} m_Z^2 \cos 2 \beta \mp 2 m_D (A_{\nu} -\mu \cot \beta-B_N)\xi \nonumber \\
m_{\xi_{1,2}^h}^2 &=& m_M^{2} \pm 2 B_M m_M + m_{\tilde M}^2 + 2 m_D^2
\end{eqnarray}  
Here we can see that the heavy states $\xi_{1,2}^h$ will couple very
weakly to the rest of particles of the MSSM via their $\tilde{\nu}_L$
component, which is highly suppresed by the small factor $\xi$ and,
therefore, it is a good approximation to ignore them and keep just the
light states $\xi_{1,2}^l$, which are made mainly of $\tilde{\nu}_L$
and its complex conjugate $\tilde{\nu}_L^*$. One says then that the
heavy sneutrinos decouple from the low energy physics.

The generalization of the previous argument to the three generations case leads to
the conclusion that,  in the
seesaw limit, $\xi \ll 1$, the physical sneutrino eigenstates,
$\tilde{\nu}_{\beta}$ ($\beta = 1,2,3$) are made mainly of the 
$\tilde \nu_{L,\,l}$ states with
$l=e,\,\mu,\,\tau$ respectively, and their corresponding complex
conjugates. The process from the weak eigenstates to the 
mass eigenstates is simplified to the diagonalization of a $3 \times 3$ sneutrino
mass matrix. This is to be compared with the more complex case of charged
sleptons where the corresponding process requires the diagonalization of a 
$6 \times 6$ slepton mass matrix. This will be presented in the next 
section, where the most 
general case with lepton flavor mixing is considered.

To end this section, we shortly comment on the parameterization that we use to make 
contact with the neutrino data. It was first introduced in~\cite{Casas:2001sr} 
to study the $\mu \to e \gamma$ decay and used later by many other authors. The
advantage of this parameterization is that instead of using as input parameters the
seesaw mass matrices $m_D$ and $m_M$ it uses the three physical light neutrino 
masses, $m_{\nu_i}$, the three physical heavy neutrino masses, 
$m_{N_i}$, the $U_{MNS}$ matrix, and a general complex $3 \times 3$ orthogonal 
matrix $R$. With our signs and matrix conventions, the relation between the seesaw
mass matrices and these other more physical quantities is given by,
\begin{equation}
m_D^T =i \,m_N^{diag \, 1/2}\, R \,m_{\nu}^{diag \, 1/2}\, U_{MNS}^+
\label{Rcasas}
\end{equation}
where $R^T R=1$ and, as we have said, $m_{N_i} \simeq m_{M_i}$. 
Thus, instead of proposing directly possible 
textures for $m_D$, or  $Y_{\nu}$, one proposes possible values for 
$m_{N_1} \, ,m_{N_2} \, ,m_{N_3} $ and $R$, and sets 
$m_{\nu_1} \, ,m_{\nu_2} \, ,m_{\nu_3} $ and $U_{MNS}$ to their 
suggested values from the experimental data. Notice that any hypothesis for 
$R$ different from the unit matrix will lead to an additional lepton flavor 
mixing, besides the one introduced by the $U_{MNS}$. 
Notice also that the previous relation holds at the energy scale $m_M$, 
and to use it properly one must apply the Renormalization Group Equations to 
run the input experimental data $m_{\nu}^{diag}$ and $U_{MNS}$ from the 
low energies $m_W$ up to $m_M$. Therefore, we will also include these running 
effects in the numerical results for all the branching ratios presented 
in this work.

Regarding the matrix $R$, we will consider the following parameterization:
\begin{equation}
R =\ \left( \begin{array}{ccc} c_{2} c_{3} 
& -c_{1} s_{3}-s_1 s_2 c_3& s_{1} s_3- c_1 s_2 c_3\\ c_{2} s_{3} & c_{1} c_{3}-s_{1}s_{2}s_{3} & -s_{1}c_{3}-c_1 s_2 s_3 \\ s_{2}  & s_{1} c_{2} & c_{1}c_{2}\end{array} \right).
\end{equation}
where $c_i\equiv \cos \theta_i$, $s_i\equiv \sin\theta_i$ and $\theta_1$, 
$\theta_2$ and $\theta_3$ are arbitrary complex angles. This parameterization 
was proposed in ref.~\cite{Casas:2001sr} for the study of 
$\mu \rightarrow e \gamma$ decays. It has also been considered 
in ref.~\cite{Chankowski:2004jc,Bi:2003ea} with specific values for the 
$\theta_i$ angles to study the implications for baryogenesis 
in the case of hierarchical neutrinos. And it has also been considered 
by~\cite{Arganda:2004bz} to study the LFV Higgs boson 
decays into $l_i \bar{l}_j$.

Finally, for the numerical estimates in this work,  
we will consider the following two plausible scenarios, at the low energies, 
being compatible with data:
\begin{itemize}
\item Scenario A:
with quasi-degenerate light and degenerate heavy neutrinos,
\begin{eqnarray}
m_{\nu_1}&=&0.2 \, eV \, , m_{\nu_2}=m_{\nu_1}+\frac{\Delta m_{sol}^2}{2 m_{\nu_1}} \, , m_{\nu_3}=m_{\nu_1}+\frac{\Delta m_{atm}^2}{2 m_{\nu_1}}, \\ \nonumber
m_{N_1} &=& m_{N_2}= m_{N_3}= m_N
\end{eqnarray}
\item Scenario B:
with hierarchical light and hierarchical heavy neutrinos,
\begin{eqnarray}
m_{\nu_1} &\simeq& 0 \, eV \, , m_{\nu_2}= \sqrt{\Delta m_{sol}^2} \, , m_{\nu_3}=\sqrt{\Delta m_{atm}^2}, \\ \nonumber
m_{N_1} &\leq & m_{N_2} < m_{N_3}
\end{eqnarray}

\end{itemize}

In the two above scenarios, we will fix the input low energy data to 
the following values, $\sqrt{\Delta m_{sol}^2}=0.008$ eV, 
$\sqrt{\Delta m_{atm}^2}=0.05$ eV, $\theta_{12}=\theta_{sol}=30^o$, 
$\theta_{23}=\theta_{atm}=45^o$, $\theta_{13}=0^o$ and $\delta = \alpha= \beta =0$ 
(See for instance, ref.~\cite{review}). Some results will also be presented 
for the alternative choice of small but non-vanishing $\theta_{13}$.

\section{\label{LFV} Generation of flavor mixing in the slepton sector}

Once the three $\nu_R$ and the three $\tilde{\nu}_R$ are
 added to the MSSM particle content, lepton flavor mixing is generated in the
 slepton sector. This can be seen as the result of a misalignment 
 between the rotations leading to the mass eigenstate 
 basis of sleptons with respect to the one of leptons, which is generically 
 present in 
 the SUSY-seesaw models. 
 This misalignment 
 is radiatively generated from the Yukawa couplings of the Majorana neutrinos 
 and can be sizable, in both, the charged slepton and sneutrino sectors. 
  Usually, it is
implemented via the Renormalization Group Equations (RGEs), which we take 
within the context of mSUGRA extended with three right-handed neutrinos and their 
SUSY partners. In consequence, we assume here universal soft-SUSY-breaking 
parameters 
at the large energies $M_X >> m_M$, which must now include 
the corresponding parameters of the neutrino and sneutrino sectors, namely, 
\begin{eqnarray}
(m_{\tilde{L}})_{ij}^2 &=& M_0^2 \delta_{ij}, \, (m_{\tilde{E}})_{ij}^2 = M_0^2 \delta_{ij}, \, (m_{\tilde{M}})_{ij}^2 = M_0^2 \delta_{ij} \nonumber \\
(A_{l})_{ij}&=& A_0 (Y_l)_{ij}, \, (A_{\nu})_{ij}= A_0 (Y_{\nu})_{ij},\,i,j=1,2,3
\label{univ_cond}
\end{eqnarray}
Here, $M_0$ and $A_0$ are the usual universal soft SUSY breaking parameters in
mSUGRA,  
$(Y_{l})_{ij}=Y_{l_i} \delta_{ij}$ with $Y_{l_i}= m_{l_i}/v_1$, and
$(Y_{\nu})_{ij}=(m_D)_{ij}/v_2$. Notice that we have used the $3 \times 3$ 
matrix form with $i,j=1,2,3$ or equivalently $i,j=e,\mu,\tau$.

The effects of the running from $M_X$ down to $m_M$ on the soft mass matrices 
of the slepton sector are found then by solving the 
RGEs which now include the corresponding terms and equations for the 
Yukawas of the neutrinos and soft breaking parameters of the sneutrino sector, 
as they are active particles in this energy range.
Below the energy scales $m_M$, the right handed neutrinos decouple and 
the effects of running from $m_M$ down to the electroweak scale on the 
various parameters are obtained by solving the RGEs but now without the terms 
and equations containing the Yukawas and soft breaking neutrino parameters. 
The obtained values at the electroweak scale of the various SUSY parameters are 
the relevant ones in order to build the slepton and sneutrino mass matrices that
will be presented below. 

To solve numerically the RGEs we use 
the Fortran code SPheno~\cite{Porod:2003um} 
that we have adapted to include the full flavor structure of the 
$3 \times 3$ soft SUSY breaking mass and trilinear coupling matrices and of the 
Yukawa coupling matrices. This program solves the full RGEs 
(i.e. including the commented extra equations and neutrino terms) in 
the two loops approximation, computes the MSSM spectra at low energies, and 
uses as inputs the universal mSUGRA parameters, $M_0$, $A_0$, $M_{1/2}$; 
the value of $\tan\beta$ at the electroweak scale,
and the sign of the $\mu$ mass parameter. The value of $M_X$ is derived from the
unification condition for the $SU(2)$ and $U(1)$ couplings,  $g_1 = g_2$. 
For all the numerical analysis performed in this work, 
we have got very close values to $M_X = 2 \times 10^{16}$ GeV. The value of $|\mu|$
is derived from the requirement of the proper radiative electroweak symmetry 
breaking. 

We present next the slepton mass matrices, relevant to low energies, 
that include the lepton mixing generated from the neutrino Yukawa couplings 
by the RGEs. For the charged slepton case and referred to the  
$(\tilde{e}_L, \tilde{e}_R, \tilde{\mu}_L, \tilde{\mu}_R, \tilde{\tau}_L, 
\tilde{\tau}_R)$ basis, the squared mass matrix can be written as follows, 
\begin{equation}
M_{\tilde{l}}^2\ =\ \left( \begin{array}{cccccc} M_{LL}^{ee \, 2} & M_{LR}^{ee \, 2} & M_{LL}^{e \mu \, 2} & M_{LR}^{e \mu \, 2} & M_{LL}^{e \tau \, 2} & M_{LR}^{e \tau \, 2} \\ M_{RL}^{ee \, 2} & M_{RR}^{ee \, 2} & M_{RL}^{e \mu \, 2} & M_{RR}^{e \mu \, 2} & M_{RL}^{e \tau \, 2} & M_{RR}^{e \tau \, 2} \\ M_{LL}^{\mu e \, 2} &  M_{LR}^{\mu e \, 2}& M_{LL}^{\mu \mu \, 2} & M_{LR}^{\mu \mu \, 2} & M_{LL}^{\mu \tau \, 2} & M_{LR}^{\mu \tau \, 2} \\ M_{RL}^{\mu e \, 2} & M_{RR}^{\mu e \, 2} & M_{RL}^{\mu \mu \, 2} & M_{RR}^{\mu \mu \, 2} & M_{RL}^{\mu \tau \, 2} & M_{RR}^{\mu \tau \, 2} \\ M_{LL}^{\tau e \, 2} & M_{LR}^{\tau e \, 2} & M_{LL}^{\tau \mu \, 2} & M_{LR}^{\tau \mu \, 2} & M_{LL}^{\tau \tau \, 2} & M_{LR}^{\tau \tau \, 2}\\ M_{RL}^{\tau e \, 2} & M_{RR}^{\tau e \, 2} & M_{RL}^{\tau \mu \, 2} & M_{RR}^{\tau \mu \, 2} & M_{RL}^{\tau \tau \, 2} & M_{RR}^{\tau \tau \, 2} \end{array} \right)
\label{sleptonmatrix}
\end{equation}
where,
\begin{eqnarray}
M_{LL}^{ij \, 2} &=& m_{\tilde{L}, ij}^2 + v_1^2 \left( Y_l^{\dagger} Y_l \right)_{ij} + m_Z^2 \cos 2 \beta \left(-\frac{1}{2}+ \sin^2 \theta_{W} \right) \delta_{ij} \nonumber \\
M_{RR}^{ij \, 2} &=& m_{\tilde{E}, ij}^2 + v_1^2 \left( Y_l^{\dagger} Y_l \right)_{ij} - m_Z^2 \cos 2 \beta \sin^2 \theta_{W} \delta_{ij} \nonumber \\
M_{LR}^{ij \, 2} &=& v_1 \left(A_l^{ij}\right)^{\ast} -\mu Y_l^{ij} v_2 \nonumber \\
M_{RL}^{ij \, 2} &=& \left(M_{LR}^{ij \, 2}\right)^{\ast} \nonumber \\
\end{eqnarray}
The soft SUSY breaking mass matrices  and trilinear coupling matrices above, 
$m_{\tilde{L}, ij}$, 
$m_{\tilde{E}, ij}$ and  $A_l^{ij}$, with $i,j= e\,,\,\mu \,,\,\tau$, refer to 
their corresponding values at the electroweak scale which we get with the SPheno
program. 

After numerical diagonalization of the $M_{\tilde{l}}^2$ matrix one gets the 
physical slepton masses and the six mass eigenstates 
($\tilde{l_1},.....,\tilde{l_6}$)$\equiv \tilde{l}$ which are related 
to the previous weak eigenstates 
($\tilde{e}_L$,....$\tilde{\tau}_R$)$\equiv \tilde{l}'$ 
by  
$\tilde{l}' = R^{(l)}\tilde{l}$, where $R^{(l)}$ is a
 $6 \times 6$ rotation matrix
 such that,
\begin{eqnarray}
M_{\tilde l_{diag}}^2 &=& R^{(l)} M_{\tilde l}^2 R^{(l)\,\dag} =
diag(m_{\tilde l_1}^2,..,m_{\tilde l_6}^2).
\end{eqnarray}
For the sneutrino sector, 
the $3 \times 3$ squared mass matrix, 
referred to the $\tilde \nu'=
(\tilde \nu_{e,\,L}, \,\tilde \nu_{\mu,\,L}, \,\tilde \nu_{\tau,\,L})$ 
basis can be written as follows,
\begin{equation}
M_{\tilde{\nu}}^2\ =\ \left( \begin{array}{ccc} m_{\tilde{L}, ee}^2 + \frac{1}{2}
m_Z^2 \cos 2 \beta  & m_{\tilde{L}, e \mu}^2 & m_{\tilde{L}, e \tau}^2 \\
m_{\tilde{L}, \mu e}^2  &  m_{\tilde{L}, \mu \mu}^2 + \frac{1}{2} m_Z^2 \cos 2
\beta & m_{\tilde{L}, \mu \tau}^2 \\ m_{\tilde{L}, \tau e}^2 & m_{\tilde{L}, \tau
\mu}^2 &  m_{\tilde{L}, \tau \tau}^2 + \frac{1}{2} m_Z^2 \cos 2 \beta  \end{array} \right)
\end{equation}
where $m_{\tilde{L}, ij}^2$ are the same as in the previous charged 
slepton squared mass matrix. After diagonalization of the $M_{\tilde{\nu}}^2$ 
matrix one gets the relevant physical sneutrino masses and eigenstates, 
$\tilde{\nu}_{\beta}$ ($\beta = 1,2,3$) which are related to the 
previous states $\tilde{\nu}_{\alpha}'$ by the corresponding $3 \times 3$ 
rotation matrix, $\tilde{\nu}'=R^{(\nu)} \tilde{\nu}$, and is such that,
\begin{eqnarray}
M_{\tilde \nu_{diag}}^2 &=& R^{(\nu)} M_{\tilde \nu}^2 R^{(\nu)\,\dag} =
diag(m_{\tilde \nu_1}^2,m_{\tilde \nu_2}^2,m_{\tilde \nu_3}^2).
\end{eqnarray}

Finally, in order to illustrate later the size of the misalignment effects 
in the slepton sector we define the following dimesionless parameters,
\begin{eqnarray}
\delta_{LL}^{ij} = \frac{M_{LL}^{ij 2}}{\tilde{m}^2} \label{deltaLL}\\
\delta_{LR}^{ij} = \frac{M_{LR}^{ij 2}}{\tilde{m}^2} \label{deltaLR}\\
\delta_{RR}^{ij} = \frac{M_{RR}^{ij 2}}{\tilde{m}^2} \label{deltaRR}
\end{eqnarray}
where 
\begin{equation}
\tilde{m}^2 = \left( m_{\tilde{l}_1}^2 m_{\tilde{l}_2}^2 m_{\tilde{l}_3}^2 m_{\tilde{l}_4}^2 m_{\tilde{l}_5}^2 m_{\tilde{l}_6}^2 \right)^{1/6}
\end{equation}
is an average slepton squared mass.
These parameters have also been considered by other authors in a more model
independent approach and  
with the purpose of getting bounds from experimental data. Some of these
bounds can be found in~\cite{9604387,Chankowski:2005jh,Paradisi:2005fk}.

For all the numerical results presented in this paper, we will set
values for all the following input parameters and physical quantities:
\begin{itemize}
\item {mSUGRA parameters}: $M_0$, $M_{1/2}$, $A_0$, $\mbox{sign}(\mu)$ and $\tan{\beta}$.
\item {seesaw parameters}: $m_{N_1}$, $m_{N_2}$, $m_{N_3}$ and $R$ 
(or equivantly $\theta_1$, $\theta_2$, $\theta_3$).
\item {physical quantities}: $m_{\nu_1}$, $m_{\nu_2}$, $m_{\nu_3}$, $U_{MNS}$
\end{itemize}


\section{\label{analytical} Analytical results for the 
$l_j^- \to l_i^- l_i^- l_i^+$ decays }

In this section we present the analytical results for the
 LFV $\tau$ and $\mu$ decays into three leptons 
 with the same flavor, within the mSUGRA-seesaw context that we have 
 presented in the
 previous sections. We perform a complete one-loop computation 
 of the $\tau$ and $\mu$ decay widths for all the three possible 
 channels, $\tau^- \to \mu^- \mu^- \mu^+$, $\tau \to e^- e^- e^+$ and 
 $\mu \to e^- e^- e^+$, and include all the contributing SUSY loops. We present 
  each contribution separately, $\gamma$-penguin, $Z$-penguin, 
 Higgs-penguin and boxes. The contributions from the Higgs-penguin diagrams 
 are, to our knowledge, computed exactly by the first time here. 
 We have also reviewed the analytical results in~\cite{Hisano:1995cp} and 
 correct their results
  for the Z-penguin contributions.
 Notice that we make the computation in the physical mass eigenstate basis.  That
 is, we consider the one-loop contributions from charged sleptons, $\tilde l_X$
 ($X=1,..,6$), sneutrinos $\tilde \nu_X$ ($X=1,2,3$), charginos 
 ${\tilde{\chi}_A^-}$ ($A=1,2$), and neutralinos ${\tilde{\chi}_A^0}$ 
 ($A=1,..,4$).
 In all this section we follow closely the notation and way of presentation of
~\cite{Hisano:1995cp}. The interactions in the physical mass eigenstate basis that
 are needed for this computation are collected 
 in the form of Feynman rules in Appendix~\ref{apendice1}.
 
 First, we define the amplitudes for the 
 $l_j^-(p) \to l_i^-(p_1) l_i^-(p_2) l_i^+(p_3)$
 decays
 as the sum of the various contributions,
 \begin{equation}
T(l_j^- \to l_i^- l_i^- l_i^+) = T_{\gamma-{\rm penguin}} + T_{Z-{\rm penguin}} +
T_{\rm H-penguin} + T_{\rm boxes}.
\end{equation}
  In the following we present the results for these contributions in terms 
  of some convenient 
 form factors.

\subsection{The $\gamma$-penguin contributions}

\begin{figure}[h]
\begin{center}
\includegraphics[width=10cm]{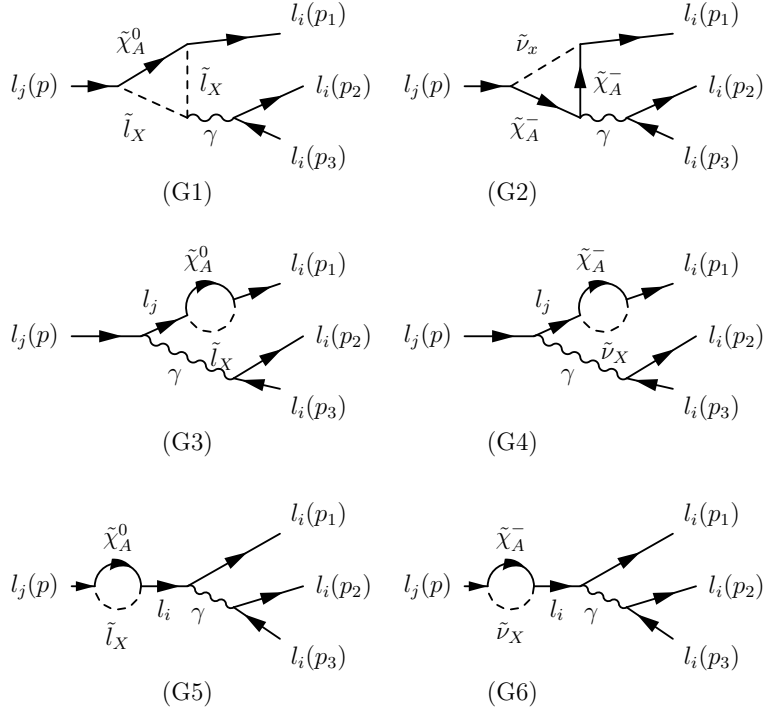}
\caption{$\gamma$-penguin diagrams contributing to the $l_j^- \to l_i^- l_i^- l_i^+$ decay}
\label{GammaPenguin}
\end{center}
\end{figure}

The diagrams where a photon is exchanged are called $\gamma$-penguin
diagrams and are shown in fig.~\ref{GammaPenguin}.
The result for the $\gamma$-penguin amplitude contributing to the 
$l_j^- \to l_i^- l_i^- l_i^+$
 decays is usually written as,
\begin{eqnarray}
T_{\gamma-{\rm penguin}} &=& \bar{u}_i(p_1)\left[q^2 \gamma_{\mu} (A_1^L P_L + A_1^R P_R) + i m_{l_j} \sigma_{\mu \nu} q^{\nu} \left( A_2^L P_L + A_2^R P_R \right) \right] u_j(p) \nonumber \\
&\times& \frac{e^2}{q^2}  \bar{u}_i(p_2) \gamma^ {\mu} v_i(p_3) - (p_1 \leftrightarrow p_2)
\end{eqnarray}
where $q$ is the photon momentum and $e$ is the electric charge. 
 The photon-penguin amplitude has two contributions in the MSSM-seesaw 
 from the chargino and neutralino sectors respectively, as can be seen 
 in the structure of the form factors,
\begin{equation}
A_a^{L,R} = A_a^{(n)L.R} + A_a^{(c)L,R}, \quad a = 1, 2
\end{equation}

The neutralino contributions are given by
\begin{eqnarray}
A_1^{(n)L} &=& \frac{1}{576 \pi^2} N_{iAX}^R N_{jAX}^{R \ast} \frac{1}{m_{\tilde{l}_X}^2} \frac{2 - 9 x_{AX} + 18 x_{AX}^2 - 11 x_{A}^3 + 6 x_{AX}^3 \log{x_{AX}}}{\left( 1 - x_{AX} \right)^4} \nonumber \\
\\
A_2^{(n)L} &=& \frac{1}{32 \pi^2} \frac{1}{m_{\tilde{l}_X}^2} \left[ N_{iAX}^L N_{jAX}^{L \ast} \frac{1 - 6 x_{AX} + 3 x_{AX}^2 + 2 x_{AX}^3 - 6 x_{AX}^2 \log{x_{AX}}}{6 \left( 1 - x_{AX} \right)^4} \right. \nonumber \\
&+& N_{iAX}^R N_{jAX}^{R \ast} \frac{m_{l_i}}{m_{l_j}} \frac{1 - 6 x_{AX} + 3 x_{AX}^2 + 2 x_{AX}^3 - 6 x_{AX}^2 \log{x_{AX}}}{6 \left( 1 - x_{AX} \right)^4} \nonumber \\
&+& \left. N_{iAX}^L N_{jAX}^{R \ast} \frac{m_{\tilde{\chi}_A^0}}{m_{l_j}} \frac{1 -
x_{AX}^2 +2 x_{AX} \log{x_{AX}}}{\left( 1 - x_{AX} \right)^3} \right] \label{A2Lneut}\\
A_a^{(n)R} &=& \left. A_a^{(n)L} \right|_{L \leftrightarrow R}\label{ARneut}
\end{eqnarray}
where $x_{AX} = m_{\tilde{\chi}_A^0}^2/m_{\tilde{l}_X}^2$. On the other hand, the chargino contributions are
\begin{eqnarray}
A_1^{(c)L} &=& -\frac{1}{576 \pi^2} C_{iAX}^R C_{jAX}^{R \ast} \frac{1}{m_{\tilde{\nu}_X}^2} \frac{16 - 45 x_{AX} + 36 x_{AX}^2 - 7 x_{A}^3 + 6 (2 - 3 x_{AX}) \log{x_{AX}}}{\left( 1 - x_{AX} \right)^4} \nonumber \\
& & \\ \cr
A_2^{(c)L} &=& -\frac{1}{32 \pi^2} \frac{1}{m_{\tilde{\nu}_X}^2} \left[ C_{iAX}^L C_{jAX}^{L \ast} \frac{2 + 3 x_{AX} - 6 x_{AX}^2 + x_{AX}^3 + 6 x_{AX} \log{x_{AX}}}{6 \left( 1 - x_{AX} \right)^4} \right. \nonumber \\
&+& C_{iAX}^R C_{jAX}^{R \ast} \frac{m_{l_i}}{m_{l_j}} \frac{2 + 3 x_{AX} - 6 x_{AX}^2 + x_{AX}^3 + 6 x_{AX} \log{x_{AX}}}{6 \left( 1 - x_{AX} \right)^4} \nonumber \\
&+& \left. C_{iAX}^L C_{jAX}^{R \ast} \frac{m_{\tilde{\chi}_A^-}}{m_{l_j}} \frac{-3 +
4 x_{AX} - x_{AX}^2 - 2 \log{x_{AX}}}{\left( 1 - x_{AX} \right)^3} \right]
\label{A2Lchar} \\
A_a^{(c)R} &=& \left. A_a^{(c)L} \right|_{L \leftrightarrow R}\label{ARchar}
\end{eqnarray}
where $x_{AX} = m_{\tilde{\chi}_A^-}^2/m_{\tilde{\nu}_X}^2$. Notice that in 
both neutralino and chargino contributions a summation over the indices 
$A$ and $X$ is understood. Notice also that we have not neglected any of 
the fermion masses. If we neglect these masses in the previous
formulas we get the same result as in~\cite{Hisano:1995cp}. 
The expressions for the $N$ and $C$ couplings are given in the 
Appendix~\ref{apendice1}.

\subsection{The $Z$-penguin contributions}
\begin{figure}[h]
\begin{center}
\includegraphics[width=10cm]{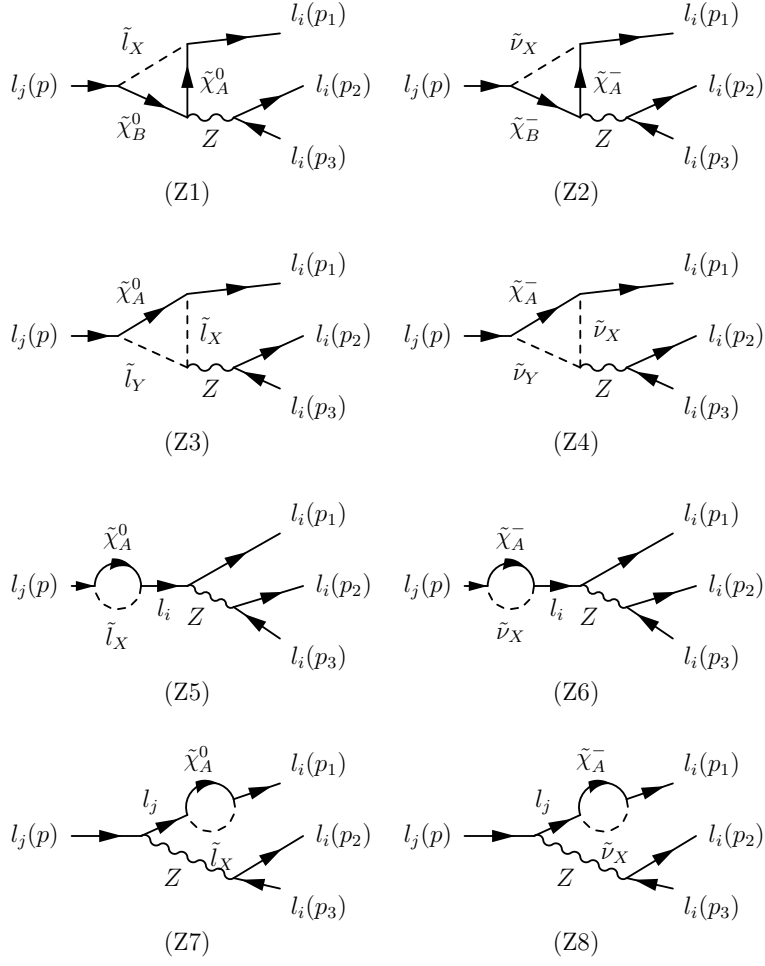}
\caption{$Z$-penguin diagrams contributing to the $l_j^- \to l_i^- l_i^- l_i^+$ decay}
\label{ZPenguin}
\end{center}
\end{figure}

 The diagrams where a Z boson is exchanged are called the $Z$-penguin
 diagrams and
are shown in fig.~\ref{ZPenguin}. The amplitude in this case is
\begin{eqnarray}
T_{Z-\mbox{\rm penguin}} &=& \frac{1}{m_Z^2} \bar{u}_i(p_1) \left[ \gamma_{\mu} \left( F_L P_L + F_R P_R \right) \right] u_j(p) \nonumber \\
&\times& \bar{u}_i(p_2) \left[ \gamma^{\mu} \left( Z_L^{(l)} P_L + Z_R^{(l)} P_R \right) \right] v_i(p_3) - (p_1 \leftrightarrow p_2)
\end{eqnarray}
where, as before, $F_{L(R)} = F_{L(R)}^{(n)} + F_{L(R)}^{(c)}$. 
The expressions for these form factors are the following:
\begin{eqnarray}
F_L^{(n)} &=& -\frac{1}{16 \pi^2} \left\{ N_{iBX}^R N_{jAX}^{R \ast} \left[ 2 E_{BA}^{R(n)} C_{24}(m_{\tilde{l}_X}^2, m_{\tilde{\chi}_A^0}^2, m_{\tilde{\chi}_B^0}^2) - E_{BA}^{L(n)} m_{\tilde{\chi}_A^0} m_{\tilde{\chi}_B^0} C_0(m_{\tilde{l}_X}^2, m_{\tilde{\chi}_A^0}^2, m_{\tilde{\chi}_B^0}^2) \right] \right. \nonumber \\
&+& \left. N_{iAX}^R N_{jAY}^{R \ast} \left[ 2 Q_{XY}^{\tilde{l}} C_{24}(m_{\tilde{\chi}_A^0}^2, m_{\tilde{l}_X}^2, m_{\tilde{l}_Y}^2) \right] + N_{iAX}^R N_{jAX}^{R \ast} \left[ Z_L^{(l)} B_1(m_{\tilde{\chi}_A^0}^2, m_{\tilde{l}_X}^2) \right] \right\} \\
F_R^{(n)} &=& \left. F_L^{(n)} \right|_{L \leftrightarrow R} \\
F_L^{(c)} &=& -\frac{1}{16 \pi^2} \left\{ C_{iBX}^R C_{jAX}^{R \ast} \left[ 2 E_{BA}^{R(c)} C_{24}(m_{\tilde{\nu}_X}^2, m_{\tilde{\chi}_A^-}^2, m_{\tilde{\chi}_B^-}^2) - E_{BA}^{L(c)} m_{\tilde{\chi}_A^-} m_{\tilde{\chi}_B^-} C_0(m_{\tilde{\nu}_X}^2, m_{\tilde{\chi}_A^-}^2, m_{\tilde{\chi}_B^-}^2) \right] \right. \nonumber \\
&+& \left. C_{iAX}^R C_{jAY}^{R \ast} \left[ 2 Q_{XY}^{\tilde{\nu}} C_{24}(m_{\tilde{\chi}_A^-}^2, m_{\tilde{\nu}_X}^2, m_{\tilde{\nu}_Y}^2) \right] + C_{iAX}^R C_{jAX}^{R \ast} \left[ Z_L^{(l)} B_1(m_{\tilde{\chi}_A^-}^2, m_{\tilde{\nu}_X}^2) \right] \right\} \\
F_R^{(c)} &=& \left. F_L^{(c)} \right|_{L \leftrightarrow R}
\end{eqnarray}

Notice that all the loop functions are evaluated at zero external 
momenta which is a very good approximation in these decays. That is,
\begin{eqnarray}
B(m_1^2, m_2^2) &=& B(0, m_1^2, m_2^2) \\
C(m_1^2, m_2^2, m_3^2) &=& C(0, 0, m_1^2, m_2^2, m_3^2)
\end{eqnarray}
 The expressions for the couplings are collected in 
 Appendix~\ref{apendice1}
 and the loop functions~\cite{Hollik} are 
given in the Appendix~\ref{apendice2}. 
Notice that our result for the $Z$-penguin contributions differs 
significantly from the result in~\cite{Hisano:1995cp}. In fact, these
authors did not consider all the diagrams in these 
$Z$-penguin contributions, which we think is not justified.

\subsection{The box contributions}

\begin{figure}[h]
\begin{center}
\includegraphics[width=6cm]{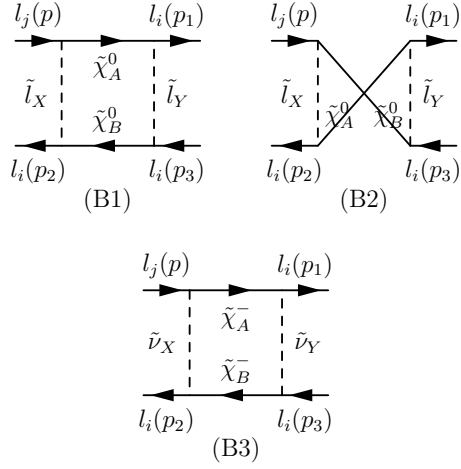}
\caption{Box-type diagrams contributing to the $l_j^- \to l_i^- l_i^- l_i^+$ decay}
\label{Boxes}
\end{center}
\end{figure}
The box-type diagrams are shown  in fig.~\ref{Boxes}. We have computed
these diagrams and found a result in agreement 
with~\cite{Hisano:1995cp}.
The amplitude for these box contributions can be written as,
\begin{eqnarray}
T_{\rm boxes} &=& e^2 B_1^L \left[ \bar{u}_i(p_1) \left( \gamma^{\mu} P_L \right) u_j(p) \right] \left[ \bar{u}_i(p_2) \left( \gamma_{\mu} P_L \right) v_i(p_3) \right] \nonumber \\
&+& e^2 B_1^R \left[ \bar{u}_i(p_1) \left( \gamma^{\mu} P_R \right) u_j(p) \right] \left[ \bar{u}_i(p_2) \left( \gamma_{\mu} P_R \right) v_i(p_3) \right] \nonumber \\
&+& e^2 B_2^L \left\{ \left[ \bar{u}_i(p_1) \left( \gamma^{\mu} P_L \right) u_j(p) \right] \left[ \bar{u}_i(p_2) \left( \gamma_{\mu} P_R \right) v_i(p_3) \right] - (p_1 \leftrightarrow p_2) \right\} \nonumber \\
&+& e^2 B_2^R \left\{ \left[ \bar{u}_i(p_1) \left( \gamma^{\mu} P_R \right) u_j(p) \right] \left[ \bar{u}_i(p_2) \left( \gamma_{\mu} P_L \right) v_i(p_3) \right] - (p_1 \leftrightarrow p_2) \right\} \nonumber \\
&+& e^2 B_3^L \left\{ \left[ \bar{u}_i(p_1) P_L u_j(p) \right] \left[ \bar{u}_i(p_2) P_L v_i(p_3) \right] - (p_1 \leftrightarrow p_2) \right\} \nonumber \\
&+& e^2 B_3^R \left\{ \left[ \bar{u}_i(p_1) P_R u_j(p) \right] \left[ \bar{u}_i(p_2) P_R v_i(p_3) \right] - (p_1 \leftrightarrow p_2) \right\} \nonumber \\
&+& e^2 B_4^L \left\{ \left[ \bar{u}_i(p_1) \left( \sigma_{\mu \nu} P_L \right) u_j(p) \right] \left[ \bar{u}_i(p_2) \left( \sigma^{\mu \nu} P_L \right) v_i(p_3) \right] - (p_1 \leftrightarrow p_2) \right\} \nonumber \\
&+& e^2 B_4^R \left\{ \left[ \bar{u}_i(p_1) \left( \sigma_{\mu \nu} P_R u_j(p) \right) \right] \left[ \bar{u}_i(p_2) \left( \sigma^{\mu \nu} P_R \right) v_i(p_3) \right] - (p_1 \leftrightarrow p_2) \right\} \nonumber \\
\end{eqnarray}
where
\begin{equation}
B_a^{L,R} = B_a^{(n)L,R} + B_a^{(c)L,R} \quad a = 1, ..., 4
\end{equation}
  The different neutralino contributions are,
\begin{eqnarray}
e^2 B_1^{(n)L} &=& \frac{1}{16 \pi^2} \left[ \frac{\tilde{D}_0}{2} N_{iAY}^R N_{jAX}^{R \ast} N_{iBX}^R N_{iBY}^{R \ast} + D_0 m_{\tilde{\chi}_A^0} m_{\tilde{\chi}_B^0} N_{iBY}^R N_{iBX}^R N_{jAX}^{R \ast} N_{iAY}^{R \ast} \right] \nonumber \\
\\
e^2 B_2^{(n)L} &=& \frac{1}{16 \pi^2} \left[ \frac{\tilde{D}_0}{4} N_{iAY}^R N_{jAX}^{R \ast} N_{iBX}^L N_{iBY}^{L \ast} - \frac{D_0}{2} m_{\tilde{\chi}_A^0} m_{\tilde{\chi}_B^0} N_{iAY}^L N_{jAX}^{R \ast} N_{iBX}^R N_{iBY}^{L \ast} \right. \nonumber \\
&-& \left. \frac{\tilde{D}_0}{4} N_{iBY}^L N_{iBX}^R N_{jAX}^{R \ast} N_{iAY}^{L \ast} + \frac{\tilde{D}_0}{4} N_{iBY}^R N_{iBX}^L N_{jAX}^{R \ast} N_{iAY}^{L \ast} \right] \\
e^2 B_3^{(n)L} &=& \frac{1}{16 \pi^2} \left[ D_0 m_{\tilde{\chi}_A^0} m_{\tilde{\chi}_B^0} N_{iAY}^L N_{jAX}^{R \ast} N_{iBX}^L N_{iBY}^{R \ast} + \frac{D_0}{2} m_{\tilde{\chi}_A^0} m_{\tilde{\chi}_B^0} N_{iBY}^L N_{iBX}^L N_{jAX}^{R \ast} N_{iAY}^{R \ast} \right] \nonumber \\
\\
e^2 B_4^{(n)L} &=& \frac{1}{16 \pi^2} \left[ \frac{D_0}{8} m_{\tilde{\chi}_A^0} m_{\tilde{\chi}_B^0} N_{jAX}^{R \ast} N_{iAY}^{R \ast} N_{iBY}^L N_{iBX}^L \right] \\
B_a^{(n)R} &=& \left. B_a^{(n)L} \right|_{L \leftrightarrow R} \quad a = 1, ..., 4
\end{eqnarray}
where
\begin{eqnarray}
D_0 &=& D_0(m_{\tilde{\chi}_A^0}^2, m_{\tilde{\chi}_B^0}^2, m_{\tilde{l}_X}^2, m_{\tilde{l}_Y}^2) \\
\tilde{D}_0 &=& \tilde{D}_0(m_{\tilde{\chi}_A^0}^2, m_{\tilde{\chi}_B^0}^2, m_{\tilde{l}_X}^2, m_{\tilde{l}_Y}^2)
\end{eqnarray}
 The chargino contributions read,
\begin{eqnarray}
e^2 B_1^{(c)L} &=& \frac{1}{16 \pi^2} \left[ \frac{\tilde{D}_0}{2} C_{iAY}^R C_{jAX}^{R \ast} C_{iBX}^R C_{iBY}^{R \ast} \right] \\
e^2 B_2^{(c)L} &=& \frac{1}{16 \pi^2} \left[ \frac{\tilde{D}_0}{4} C_{iAY}^R C_{jAX}^{R \ast} C_{iBX}^L C_{iBY}^{L \ast} - \frac{D_0}{2} m_{\tilde{\chi}_A^-} m_{\tilde{\chi}_B^-} C_{iAY}^L C_{jAX}^{R \ast} C_{iBX}^R C_{iBY}^{L \ast} \right] \nonumber \\
\\
e^2 B_3^{(c)L} &=& \frac{1}{16 \pi^2} \left[ D_0 m_{\tilde{\chi}_A^-} m_{\tilde{\chi}_B^-} C_{iAY}^L C_{jAX}^{R \ast} C_{iBX}^L C_{iBY}^{R \ast} \right] \\
e^2 B_4^{(c)L} &=& 0 \\
B_a^{(c)R} &=& \left. B_a^{(c)L} \right|_{L \leftrightarrow R} \quad a = 1, ..., 4
\end{eqnarray}
where 
\begin{eqnarray}
D_0 &=& D_0(m_{\tilde{\chi}_A^-}^2, m_{\tilde{\chi}_B^-}^2, m_{\tilde{\nu}_X}^2, m_{\tilde{\nu}_Y}^2) \\
\tilde{D}_0 &=& \tilde{D}_0(m_{\tilde{\chi}_A^-}^2, m_{\tilde{\chi}_B^-}^2, m_{\tilde{\nu}_X}^2, m_{\tilde{\nu}_Y}^2)
\end{eqnarray}

\subsection{The Higgs-penguin contributions}

\begin{figure}[h]
\begin{center}
\includegraphics[width=10cm]{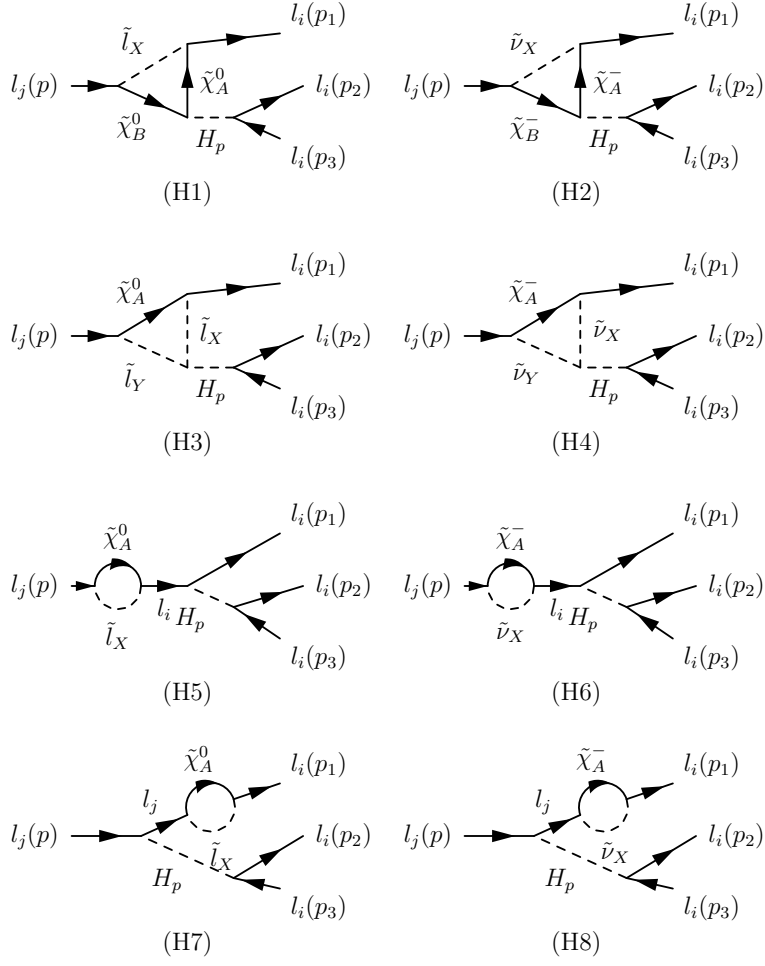}
\caption{Higgs-penguin diagrams contributing to the 
$l_j^- \to l_i^- l_i^- l_i^+$ decay. Here $H_p (p = 1, 2, 3) = h^0, H^0, A^0$.}
\label{HPenguin}
\end{center}
\end{figure}

The diagrams where a Higgs boson is exchanged are called the
Higgs-penguin diagrams. These are shown in fig.~\ref{HPenguin} and 
have been computed here by the first time. 
 These are usually not considered in the literature. In particular, in
 the most complete study so far of~\cite{Hisano:1995cp} these
 Higgs-penguin diagrams were not included.
However, they are expected to be relevant at large 
$\tan{\beta}$~\cite{Babu:2002et}. We will therefore include them here.
Specifically, we include the contributions
from the three neutral MSSM Higgs bosons, $h_0$, $H_0$ and $A_0$ and
consider all SUSY loops.

In this case, the amplitude can be written as,
\begin{eqnarray}
T_{\rm Higgs} &=& e^2 B_{2, \rm Higgs}^L \left\{ \left[ \bar{u}_i(p_1) \left( \gamma^{\mu} P_L \right) u_j(p) \right] \left[ \bar{u}_i(p_2) \left( \gamma_{\mu} P_R \right) v_i(p_3) \right] - (p_1 \leftrightarrow p_2) \right\} \nonumber \\
&+& e^2 B_{2, \rm Higgs}^R \left\{ \left[ \bar{u}_i(p_1) \left( \gamma^{\mu} P_R \right) u_j(p) \right] \left[ \bar{u}_i(p_2) \left( \gamma_{\mu} P_L \right) v_i(p_3) \right] - (p_1 \leftrightarrow p_2) \right\} \nonumber \\
&+& e^2 B_{3, \rm Higgs}^L \left\{ \left[ \bar{u}_i(p_1) P_L u_j(p) \right] \left[ \bar{u}_i(p_2) P_L v_i(p_3) \right] - (p_1 \leftrightarrow p_2) \right\} \nonumber \\
&+& e^2 B_{3, \rm Higgs}^R \left\{ \left[ \bar{u}_i(p_1) P_R u_j(p) \right] \left[ \bar{u}_i(p_2) P_R v_i(p_3) \right] - (p_1 \leftrightarrow p_2) \right\}
\end{eqnarray}
where
\begin{equation}
B_{a, \rm Higgs}^{L,R} = B_{a, \rm Higgs}^{(n)L,R} + B_{a, \rm Higgs}^{(c)L,R} \quad a = 2, 3
\end{equation} 
The first term represents the neutralino contribution, which we find to be
\begin{eqnarray}
e^2 B_{2, \rm Higgs}^{(n)L} &=& \sum_{p=1}^3 \left(-\frac{1}{2}\right) \frac{1}{m_{H_p}^2} H_{L, n}^{(p)} S_{R, i}^{(p)} \\
e^2 B_{3, \rm Higgs}^{(n)L} &=& \sum_{p=1}^3 \frac{1}{m_{H_p}^2} H_{L, n}^{(p)} S_{L, i}^{(p)} \\
B_{a, \rm Higgs}^{(n)R} &=& \left. B_{a, \rm Higgs}^{(n)L} \right|_{L \leftrightarrow R} \quad a = 2, 3
\end{eqnarray}
where $H_p (p = 1, 2, 3) = h^0, H^0, A^0$ and
\begin{eqnarray}
H_{L, n}^{(p)} &=& -\frac{1}{16 \pi^2} \left\{ \left[ B_0(m_{\tilde{\chi}_A^0}^2, m_{\tilde{\chi}_B^0}^2) + m_{\tilde{l}_X}^2 C_0(m_{\tilde{l}_X}^2, m_{\tilde{\chi}_A^0}^2, m_{\tilde{\chi}_B^0}^2) + m_{l_j}^2 C_{12}(m_{\tilde{l}_X}^2, m_{\tilde{\chi}_A^0}^2, m_{\tilde{\chi}_B^0}^2) \right. \right. \nonumber \\
&+& \left. m_{l_i}^2 (C_{11} - C_{12})(m_{\tilde{l}_X}^2, m_{\tilde{\chi}_A^0}^2, m_{\tilde{\chi}_B^0}^2) \right] N_{iAX}^L D_{R, AB}^{(p)} N_{jBX}^{R \ast} \nonumber \\
&+& m_{l_i} m_{l_j} (C_{11} + C_0)(m_{\tilde{l}_X}^2, m_{\tilde{\chi}_A^0}^2, m_{\tilde{\chi}_B^0}^2) N_{iAX}^R D_{L, AB}^{(p)} N_{jBX}^{L \ast} \nonumber \\
&+& m_{l_i} m_{\tilde{\chi}_B^0} (C_{11} - C_{12} + C_0)(m_{\tilde{l}_X}^2, m_{\tilde{\chi}_A^0}^2, m_{\tilde{\chi}_B^0}^2) N_{iAX}^R D_{L, AB}^{(p)} N_{jBX}^{R \ast} \nonumber \\
&+& m_{l_j} m_{\tilde{\chi}_B^0} C_{12}(m_{\tilde{l}_X}^2, m_{\tilde{\chi}_A^0}^2, m_{\tilde{\chi}_B^0}^2) N_{iAX}^L D_{R, AB}^{(p)} N_{jBX}^{L \ast} \nonumber \\
&+& m_{l_i} m_{\tilde{\chi}_A^0} (C_{11} - C_{12})(m_{\tilde{l}_X}^2, m_{\tilde{\chi}_A^0}^2, m_{\tilde{\chi}_B^0}^2) N_{iAX}^R D_{R, AB}^{(p)} N_{jBX}^{R \ast} \nonumber \\
&+& m_{l_j} m_{\tilde{\chi}_A^0} (C_{12} + C_0)(m_{\tilde{l}_X}^2, m_{\tilde{\chi}_A^0}^2, m_{\tilde{\chi}_B^0}^2) N_{iAX}^L D_{L, AB}^{(p)} N_{jBX}^{L \ast} \nonumber \\
&+& m_{\tilde{\chi}_A^0} m_{\tilde{\chi}_B^0} C_0(m_{\tilde{l}_X}^2, m_{\tilde{\chi}_A^0}^2, m_{\tilde{\chi}_B^0}^2) N_{iAX}^L D_{L, AB}^{(p)} N_{jBX}^{R \ast} \nonumber \\
&+& G_{XY}^{(p) \tilde{l}} \left[ - m_{l_i} (C_{11} - C_{12})(m_{\tilde{\chi}_A^0}^2, m_{\tilde{l}_X}^2, m_{\tilde{l}_Y}^2) N_{iAX}^R N_{jAY}^{R \ast} \right. \nonumber \\
&-& \left. m_{l_j} C_{12}(m_{\tilde{\chi}_A^0}^2, m_{\tilde{l}_X}^2, m_{\tilde{l}_Y}^2) N_{iAX}^L N_{jAY}^{L \ast} + m_{\tilde{\chi}_A^0} C_0(m_{\tilde{\chi}_A^0}^2, m_{\tilde{l}_X}^2, m_{\tilde{l}_Y}^2) N_{iAX}^L N_{jAY}^{R \ast} \right] \nonumber \\
&+& \frac{S_{L, j}^{(p)}}{m_{l_i}^2 - m_{l_j}^2} \left[ - m_{l_i}^2 B_1(m_{\tilde{\chi}_A^0}^2, m_{\tilde{l}_X}^2) N_{iAX}^L N_{jAX}^{L \ast} + m_{l_i} m_{\tilde{\chi}_A^0} B_0(m_{\tilde{\chi}_A^0}^2, m_{\tilde{l}_X}^2) N_{iAX}^R N_{jAX}^{L \ast} \right. \nonumber \\
&-& \left. m_{l_i} m_{l_j} B_1(m_{\tilde{\chi}_A^0}^2, m_{\tilde{l}_X}^2) N_{iAX}^R N_{jAX}^{R \ast} + m_{l_j} m_{\tilde{\chi}_A^0} B_0(m_{\tilde{\chi}_A^0}^2, m_{\tilde{l}_X}^2) N_{iAX}^L N_{jAX}^{R \ast} \right] \nonumber \\
&+& \frac{S_{L, i}^{(p)}}{m_{l_j}^2 - m_{l_i}^2} \left[ - m_{l_j}^2 B_1(m_{\tilde{\chi}_A^0}^2, m_{\tilde{l}_X}^2) N_{iAX}^R N_{jAX}^{R \ast} + m_{l_j} m_{\tilde{\chi}_A^0} B_0(m_{\tilde{\chi}_A^0}^2, m_{\tilde{l}_X}^2) N_{iAX}^R N_{jAX}^{L \ast} \right. \nonumber \\
&-& \left. \left. m_{l_i} m_{l_j} B_1(m_{\tilde{\chi}_A^0}^2, m_{\tilde{l}_X}^2) N_{iAX}^L N_{jAX}^{L \ast} + m_{l_i} m_{\tilde{\chi}_A^0} B_0(m_{\tilde{\chi}_A^0}^2, m_{\tilde{l}_X}^2) N_{iAX}^L N_{jAX}^{R \ast} \right] \right\} \\
H_{R, n}^{(p)} &=& \left. H_{L, n}^{(p)} \right|_{L \leftrightarrow R} \quad p = 1, 2, 3
\end{eqnarray}
The values of the couplings are given again in Appendix~\ref{apendice1} and the loop functions in Appendix~\ref{apendice2}. Correspondingly, the result for the chargino contribution is given by,
\begin{eqnarray}
e^2 B_{2, \rm Higgs}^{(c)L} &=& \sum_{p=1}^3 \left(-\frac{1}{2}\right) \frac{1}{m_{H_p}^2} H_{L, c}^{(p)} S_{R, i}^{(p)} \\
e^2 B_{3, \rm Higgs}^{(c)L} &=& \sum_{p=1}^3 \frac{1}{m_{H_p}^2} H_{L, c}^{(p)} S_{L, i}^{(p)} \\
B_{a, \rm Higgs}^{(c)R} &=& \left. B_{a, \rm Higgs}^{(c)L} \right|_{L
\leftrightarrow R} \quad a = 2, 3
\end{eqnarray}
where $H_{L (R), c}^{(p)}$ can be obtained from the previous $H_{L (R), n}^{(p)}$  by 
replacing everywhere 
\begin{eqnarray}
\tilde{l} &\to& \tilde{\nu} \nonumber \\
\tilde{\chi}^0 &\to& \tilde{\chi}^- \nonumber \\
N^{L(R)} &\to& C^{L(R)} \nonumber \\
D_{L(R)} &\to& W_{L(R)} \nonumber
\end{eqnarray}
Again the values of the couplings and the loop functions are given in 
Appendices~\ref{apendice1} and~\ref{apendice2} respectively.

\subsection{$l_j^- \to l_i^- l_i^- l_i^+$ decay width}

The decay width for $l_j^- \to l_i^- l_i^- l_i^+$ can be written in terms of 
the form factors given in the previous sections as~\cite{Hisano:1995cp}:
\begin{eqnarray}
\Gamma(l_j^- \to l_i^- l_i^- l_i^+) &=& \frac{e^4}{512 \pi^3} m_{l_j}^5 \left[ \left| A_1^L \right|^2 + \left| A_1^R \right|^2 - 2 \left( A_1^L A_2^{R \ast} + A_2^L A_1^{R \ast} + h.c. \right) \right. \nonumber \\
&+& \left( \left| A_2^L \right|^2 + \left| A_2^R \right|^2 \right) \left( \frac{16}{3} \log{\frac{m_{l_j}}{m_{l_i}}} - \frac{22}{3} \right) \nonumber \\
&+& \frac{1}{6} \left( \left| B_1^L \right|^2 + \left| B_1^R \right|^2 \right) + \frac{1}{3} \left( \left| \hat{B}_2^L \right|^2 + \left| \hat{B}_2^R \right|^2 \right) \nonumber \\
&+& \frac{1}{24} \left( \left| \hat{B}_3^L \right|^2 + \left| \hat{B}_3^R \right|^2 \right) + 6 \left( \left| B_4^L \right|^2 + \left| B_4^R \right|^2 \right) \nonumber \\
&-& \frac{1}{2} \left( \hat{B}_3^L B_4^{L \ast} + \hat{B}_3^R B_4^{R \ast} + h.c. \right) \nonumber \\
&+& \frac{1}{3} \left( A_1^L B_1^{L \ast} + A_1^R B_1^{R \ast} + A_1^L \hat{B}_2^{L \ast} + A_1^R \hat{B}_2^{R \ast} + h.c. \right) \nonumber \\
&-& \frac{2}{3} \left( A_2^R B_1^{L \ast} + A_2^L B_1^{R \ast} + A_2^L \hat{B}_2^{R \ast} + A_2^R \hat{B}_2^{L \ast} + h.c. \right) \nonumber \\
&+& \frac{1}{3} \left\{ 2 \left( \left| F_{LL} \right|^2 + \left| F_{RR} \right|^2 \right) + \left| F_{LR} \right|^2 + \left| F_{RL} \right|^2 \right. \nonumber \\
&+& \left( B_1^L F_{LL}^{\ast} + B_1^R F_{RR}^{\ast}  + \hat{B}_2^L F_{LR}^{\ast}  + \hat{B}_2^R F_{RL}^{\ast} + h.c. \right) \nonumber \\
&+& 2 \left( A_1^L F_{LL}^{\ast} + A_1^R F_{RR}^{\ast} + h.c. \right) + \left( A_1^L F_{LR}^{\ast} + A_1^R F_{RL}^{\ast} + h.c. \right) \nonumber \\
&-& 4 \left. \left. \left( A_2^R F_{LL}^{\ast} + A_2^L F_{RR}^{\ast} + h.c. \right) - 2 \left( A_2^L F_{RL}^{\ast} + A_2^R F_{LR}^{\ast} + h.c. \right) \right\} \right] \nonumber \\
\label{decay}
\end{eqnarray}
where
\begin{eqnarray}
F_{LL} &=& \frac{F_L Z_L^{(l)}}{g^2 \sin^2 \theta_W m_Z^2} \\
F_{RR} &=& \left. F_{LL} \right|_{L \leftrightarrow R} \\
F_{LR} &=& \frac{F_L Z_R^{(l)}}{g^2 \sin^2 \theta_W m_Z^2} \\
F_{RL} &=& \left. F_{LR} \right|_{L \leftrightarrow R}
\end{eqnarray}
Notice that we have put the Higgs contributions together with the box ones in order
to follow closely the way of presentation of~\cite{Hisano:1995cp}
\begin{eqnarray}
\hat{B}_2^{L,R} &=& B_2^{L,R} + B_{2, \rm Higgs}^{L,R} \\
\hat{B}_3^{L,R} &=& B_3^{L,R} + B_{3, \rm Higgs}^{L,R} 
\end{eqnarray}
Notice that we have corrected the result in ref.\cite{Hisano:1995cp} for the term that goes with $\left( \left|A_2^L\right| + \left|A_2^R\right| \right)$.

\section{\label{numerical} Numerical results for the LFV branching ratios}

We present in this section the numerical results for all the branching ratios 
of LFV $\tau$ and $\mu$ decays in the context 
of the mSUGRA-seesaw scenario that has been introduced in the previous 
sections. We focus on the following LFV decays, $\tau^- \to \mu^- \mu^- \mu^+$, 
$\tau^- \to e^- e^- e^+$ and $\mu^- \to e^- e^- e^+$, and the radiative decays
$\tau^-\to \mu^- \gamma$, $\tau^- \to e^- \gamma$ and 
$\mu^- \to e^- \gamma$. The reason to consider these radiative decays together 
with the decays into three leptons is that there are insteresting correlations among 
them that provide additional information in testing SUSY. Specifically, we  
show in this section  the correlations between the ratios of 
$\tau^- \to \mu^- \mu^- \mu^+$ and $\tau^- \to \mu^- \gamma$; between 
$\tau^- \to e^- e^- e^+$ and $\tau^- \to e^- \gamma$; and between $\mu^- \to e^- e^- e^+$
and $\mu^- \to e^- \gamma$. For the numerical estimates of the radiative decays we use 
the formula of~\cite{Hisano:1995cp}, 
which is given in terms of the $A_2^{L,R}$ as,
\begin{eqnarray}
\Gamma(l_j^- \to l_i^- \gamma) &=& \frac{e^2}{16 \pi} m^5_{l_j}(|A_2^L|^2+|A_2^R|^2) 
\end{eqnarray}
but we use our expressions for the form factors in eqs.(\ref{A2Lneut}),
(\ref{ARneut}), (\ref{A2Lchar}) and (\ref{ARchar})
 that include the 
lepton mass contributions. 
We explore here in full detail the 
size of the SUSY contributions to all these LFV $l_j \to 3l_i$ and $l_j \to l_i
\gamma$ decays
 as a function of all the mSUGRA 
parameters, $M_0$, $M_{1/2}$, $A_0$, $\tan\beta$ and sign($\mu$) and the 
seesaw parameters $m_{N_i}$, $i=1,2,3$ and $R$ or, equivalently, $\theta_1$, 
$\theta_2$ and $\theta_3$. 
In all this numerical analysis we require compatibility 
with the neutrino data and with the present upper experimental bounds for all
these branching ratios~\cite{Aubert:2003pc,Bellgardt:1987du,Aubert:2005wa,Aubert:2005ye,mue}, 
as given explicitely in the introduction. We also 
demand the complete set of SUSY
particle masses, which we derive with the SPheno program, 
to be above the present experimental lower bounds~\cite{pdg2004}. The 
numerical values of
the total $\tau$ and $\mu$ widths (lifetimes) are taken from~\cite{pdg2004}.
We show first the results for the scenario A with quasi-degenerate light and 
degenerate heavy neutrinos and next the
most interesting scenario B with hierarchical light and hierarchical heavy 
neutrinos.

\subsection{Degenerate case}

\begin{figure}
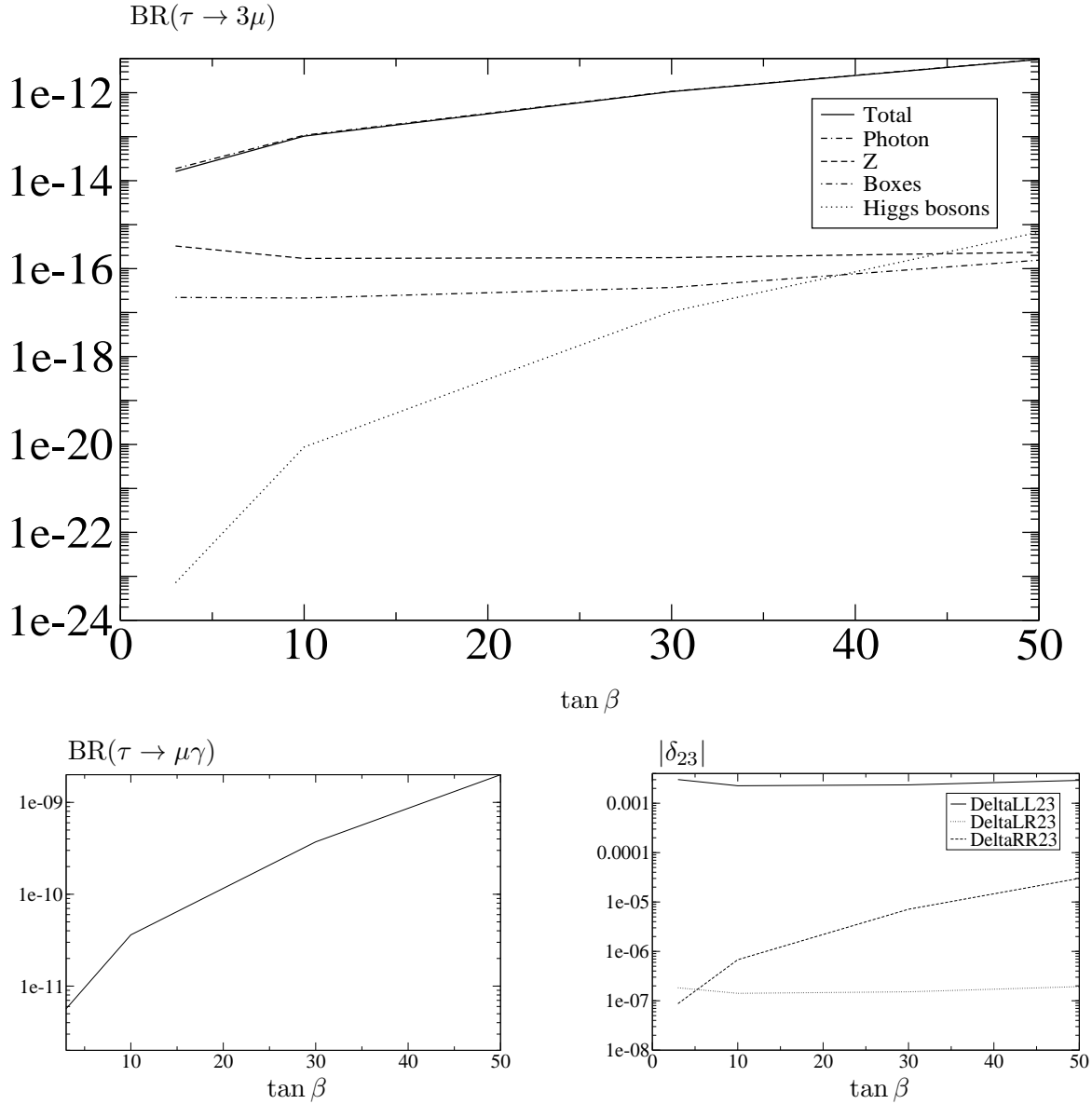

\hspace{-0.5cm}
\includegraphics[width=10.0cm,height=15.0cm,angle=-90]{fig1a_tau3mu.epsi}
\hspace{0.25cm}
\vspace{0.5cm}\\
\hspace{-0.5cm}
\includegraphics[width=5.0cm,height=7.0cm,angle=-90]{fig1a_taumugamma.epsi}
\hspace{1.0cm}
\includegraphics[width=5.0cm,height=7.0cm,angle=-90]{fig1a_deltas23.epsi}
\caption{Dependence of LVF $\tau$ decays with $\tan{\beta}$ in scenario A with degenerate heavy neutrinos and real R, for $m_N = 10^{14}$ GeV. (a) Upper panel, $BR(\tau \to \mu^- \mu^- \mu^+)$ and its different contributions, (b) lower-left panel, $BR(\tau \to \mu \gamma)$ and (c) lower-right panel, $|\delta^{23}_{LL, LR, RR}|$. The other input parameters are, $M_0=400 $ GeV, $M_{1/2}=300 $ GeV, $A_0 = 0$ and $\mbox{sign}(\mu) > 0$.
\label{fig:1a}}
\end{figure}

\begin{figure}
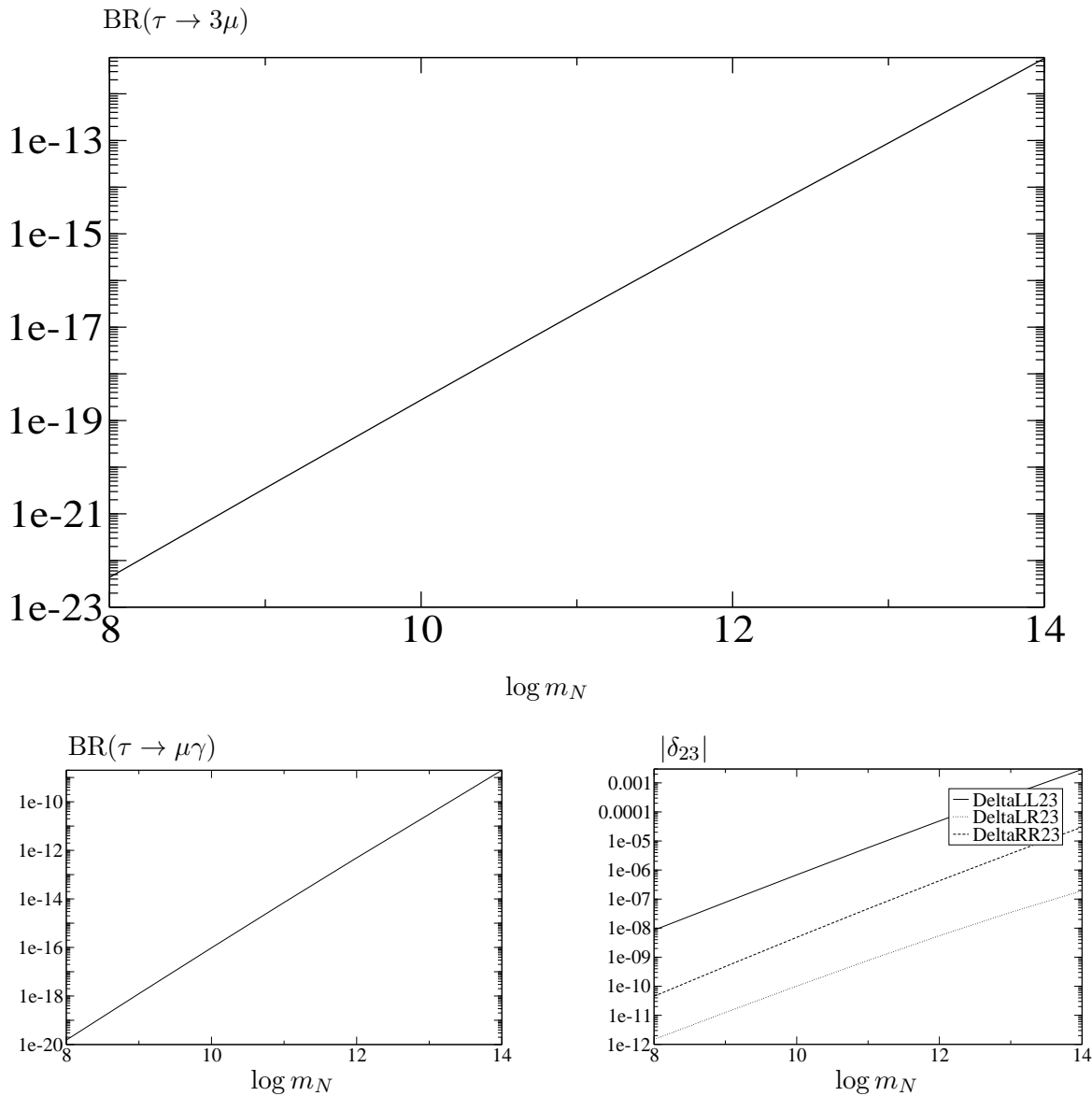

\hspace{-0.5cm}
\includegraphics[width=10.0cm,height=15.0cm,angle=-90]{fig1b_tau3mu.epsi}
\hspace{0.25cm}
\vspace{0.5cm}\\
\hspace{-0.5cm}
\includegraphics[width=5.0cm,height=7.0cm,angle=-90]{fig1b_taumugamma.epsi}
\hspace{1.0cm}
\includegraphics[width=5.0cm,height=7.0cm,angle=-90]{fig1b_deltas23.epsi}
\caption{Dependence of LFV $\tau$ decays with $m_N$ in scenario A with degenerate heavy neutrinos and real R, for $\tan{\beta} = 50$. (a) Upper panel, $BR(\tau \to \mu^- \mu^- \mu^+)$, (b) lower-left panel, $BR(\tau \to \mu \gamma)$ and (c) lower-right panel, $|\delta^{23}_{LL, LR, RR}|$. The other input parameters are, $M_0=400 $ GeV, $M_{1/2}=300 $ GeV, $A_0 = 0$ and $\mbox{sign}(\mu) > 0$.
\label{fig:1b}}
\end{figure}

\begin{figure}
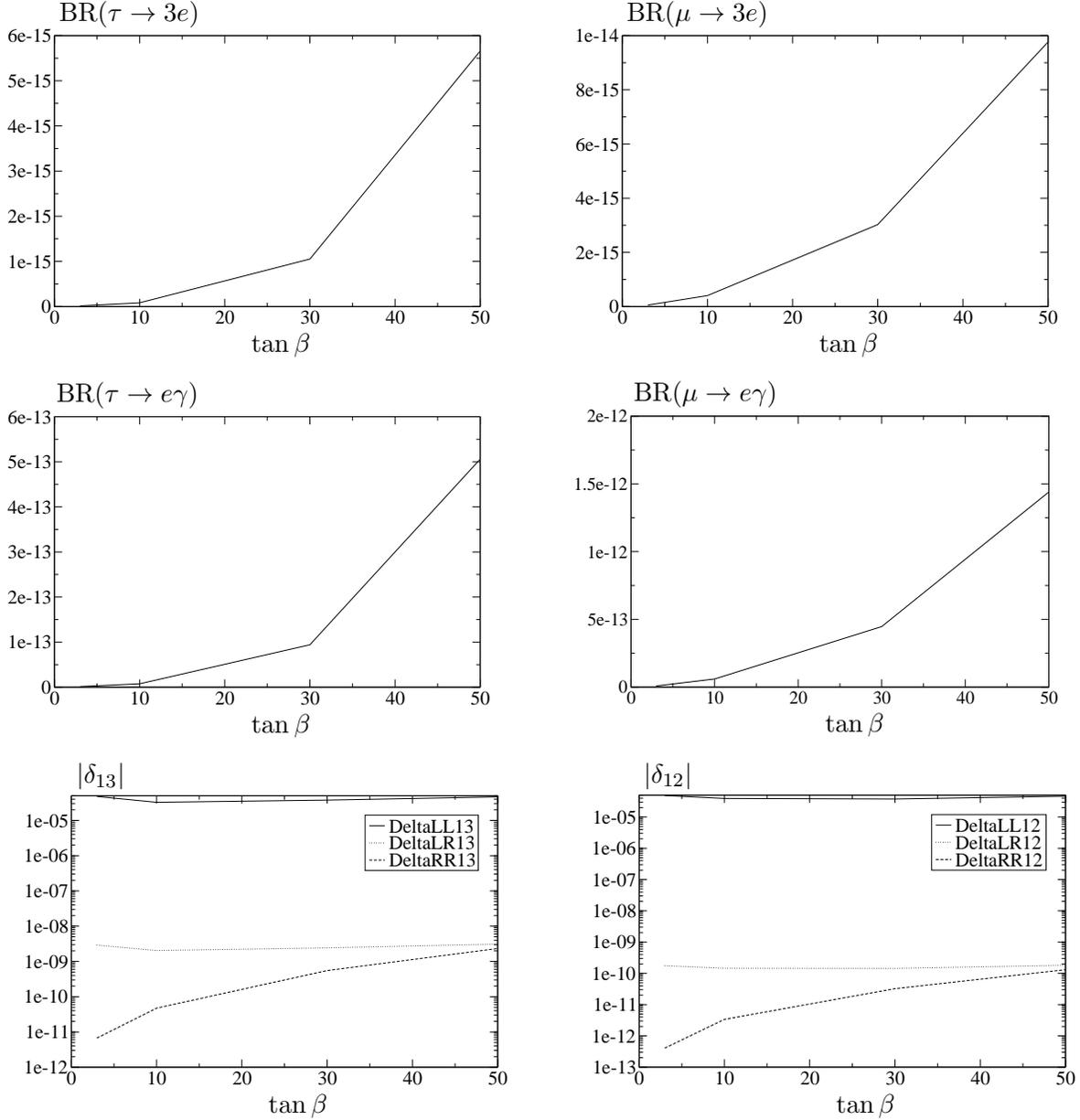

\hspace{-0.5cm}
\includegraphics[width=5.0cm,height=7.0cm,angle=-90]{fig1c_tau3e.epsi}
\hspace{1.0cm}
\includegraphics[width=5.0cm,height=7.0cm,angle=-90]{fig1d_mu3e.epsi}
\hspace{0.25cm}
\vspace{0.5cm}\\
\hspace{-0.5cm}
\includegraphics[width=5.0cm,height=7.0cm,angle=-90]{fig1c_tauegamma.epsi}
\hspace{1.0cm}
\includegraphics[width=5.0cm,height=7.0cm,angle=-90]{fig1d_muegamma.epsi}
\hspace{0.25cm}
\vspace{0.5cm}\\
\hspace{-0.5cm}
\includegraphics[width=5.0cm,height=7.0cm,angle=-90]{fig1c_deltas13.epsi}
\hspace{1.0cm}
\includegraphics[width=5.0cm,height=7.0cm,angle=-90]{fig1d_deltas12.epsi}
\caption{Dependence of LFV $\tau$ and $\mu$ decays with $\tan{\beta}$ in scenario A with degenerate heavy neutrinos and real R, for $m_N = 10^{14}$ GeV. (a) Upper-left panel, $BR(\tau \to e^- e^- e^+)$, (b) upper-right panel, $BR(\mu \to e^- e^- e^+)$, (c) middle-left panel, $BR(\tau \to e \gamma)$, (d) middle-right panel, $BR(\mu \to e \gamma)$, (e) lower-left panel, $|\delta^{13}_{LL, LR, RR}|$ and (f) lower-right panel, $|\delta^{12}_{LL, LR, RR}|$. The other input parameters are, $M_0=400 $ GeV, $M_{1/2}=300 $ GeV, $A_0 = 0$ and $\mbox{sign}(\mu) > 0$. 
\label{fig:1cd}}
\end{figure}

\begin{figure}
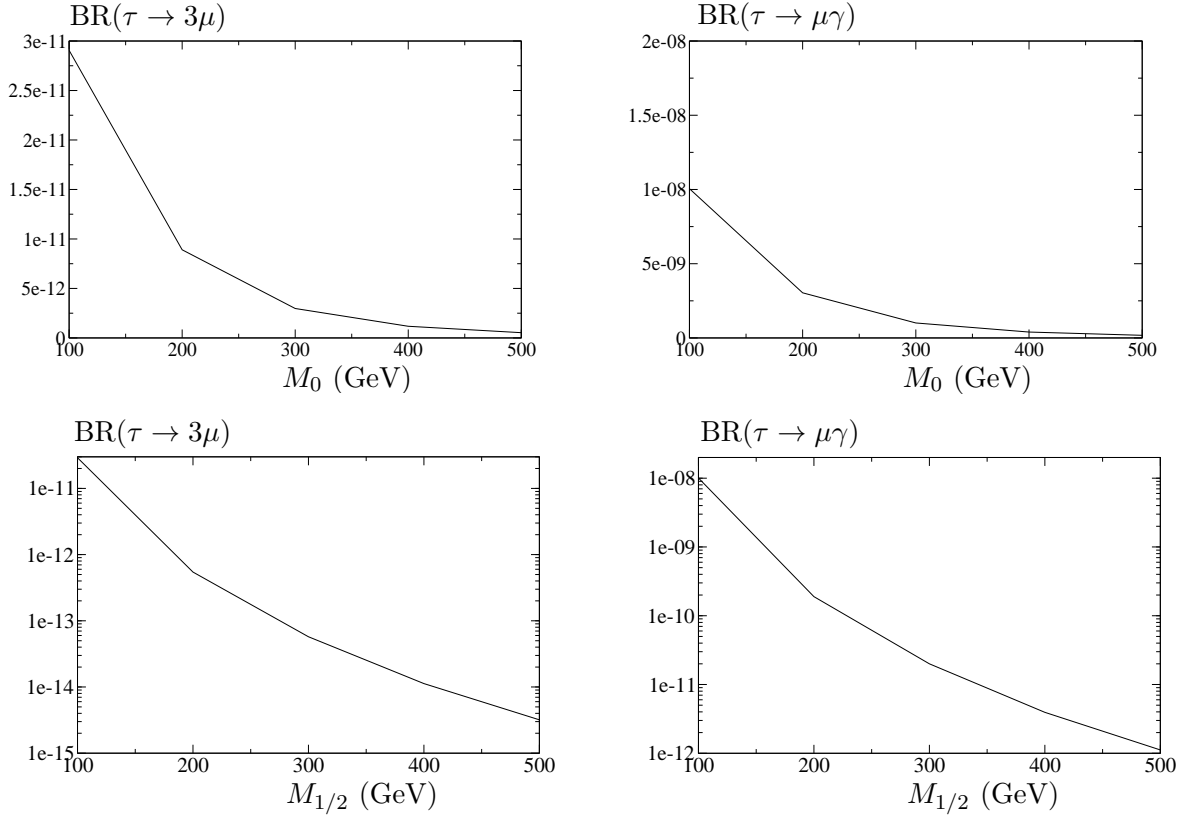

\hspace{-0.5cm}
\includegraphics[width=5.0cm,height=7.0cm,angle=-90]{fig1e_tau3mu.epsi}
\hspace{1.0cm}
\includegraphics[width=5.0cm,height=7.0cm,angle=-90]{fig1e_taumugamma.epsi}
\hspace{0.25cm}
\vspace{0.5cm}\\
\hspace{-0.5cm}
\includegraphics[width=5.0cm,height=7.0cm,angle=-90]{fig1f_tau3mu.epsi}
\hspace{1.0cm}
\includegraphics[width=5.0cm,height=7.0cm,angle=-90]{fig1f_taumugamma.epsi}
\caption{Dependence of $BR(\tau^- \to \mu^- \mu^- \mu^+)$ and $BR(\tau \to \mu \gamma)$ with $M_0$ and $M_{1/2}$ in scenario A with degenerate heavy neutrinos and real R, for $m_N = 10^{14}$ GeV and $\tan{\beta} = 50$. (a) Upper-left panel, $BR(\tau \to \mu^- \mu^- \mu^+)$ as a function of $M_0$ for $M_{1/2} = 100$ GeV, (b) upper-right panel, $BR(\tau \to \mu \gamma)$ as a function of $M_0$ for $M_{1/2} = 100$ GeV, (c) lower-left panel, $BR(\tau \to \mu^- \mu^- \mu^+)$ as a function of $M_{1/2}$ for $M_0 = 100$ GeV, (d) lower-right panel, $BR(\tau \to \mu \gamma)$ as a function of $M_{1/2}$ for $M_0 = 100$ GeV. In all the plots we take $A_0 = 0$ and $\mbox{sign}(\mu) > 0$.
\label{fig:1ef}}
\end{figure}

 We show in figs.~\ref{fig:1a} through~\ref{fig:1ef} the numerical results 
 of the branching  ratios for the LFV $\tau$ and $\mu$ decays 
 in scenario A with degenerate heavy neutrinos of mass $m_N$. 
 We show our predictions for the three channels, 
 $\tau^- \to \mu^- \mu^- \mu^+$, $\tau^- \to e^- e^- e^+$ and 
 $\mu^- \to e^- e^- e^+$, and similarly, for the comparison with the 
 leptonic radiative decays, $l_j\to l_i \gamma$, we also show in the 
 plots the correlated decay, $\tau^- \to \mu^- \gamma$, $\tau^- \to e^- \gamma$ and 
 $\mu^- \to e^- \gamma$, respectively. 

The results of the branching ratios for the $\tau^-\to \mu^- \mu^- \mu^+$ and 
$\tau^- \to \mu^- \gamma$ decays 
as a function of $\tan{\beta}$  are 
illustrated in fig.~\ref{fig:1a}. In these plots we set 
$m_N = 10^{14}$ GeV and
assume the matrix $R$ to be real. 
Notice that in the degenerate case with real $R$ 
these LFV ratios do not depend on the particular choice for $R$. 
This can be easily 
understood because the dependence on $R$ drops in the  
relevant factor, $(Y_\nu^*Y_\nu^T)_{ij}$, appearing in the dominant $\delta^{ij}_{LL}$ slepton mixing, 
and due to the property $R^TR=1$. 
From this figure we also see that the predicted rates for both channels are 
well below their respective experimental upper bounds for all 
$\tan \beta$ values, eventhough the total rates grow fast with $\tan \beta$.
We also see clearly the mentioned 
correlation between the  $\tau^-\to \mu^- \mu^- \mu^+$ and 
$\tau^- \to \mu^- \gamma$ rates. In fact, this correlation is an inmediate
consequence of the dominance of the $\gamma$-penguin contributions
which clearly governs the size of the $\tau^-\to \mu^- \mu^- \mu^+$ rates.
This dominance is illustrated in fig.~\ref{fig:1a}.(a). where the various
contributions are shown separately. In fact,
 the contributions  from the $\gamma$-penguin diagrams are 
  almost undistinguishable from the total rates for all $\tan \beta$ values. 
 For low $\tan \beta$ values the next dominant contribution is from 
 the $Z$-penguin diagrams, but this is still more than one order of magnitude
 smaller than the $\gamma$-penguin contribution. The contributions from the box
 diagrams are even smaller. We also learn that the $Z$ and boxes 
 contributions do not depend 
 significantly on $\tan{\beta}$, while the photon contribution 
 goes approximately as $(\tan{\beta})^2$ at large $\tan{\beta}$.
 In this large $\tan{\beta}$ region 
 it is interesting to note that  
 the total Higgs contribution becomes larger than the $Z$ contribution 
 and the boxes, due to the fact that it grows approximately 
 as $(\tan{\beta})^6$. In this total Higgs contribution the dominant penguins
 are those with $H_0$ and $A_0$ exchanged, which are several orders of 
 magnitude larger than the $h_0$-penguin contribution. However, in spite of
  this 
 huge enhacement of the total Higgs contribution occurring at large 
 $\tan{\beta}$, its relative size as compared to the photon-penguin 
 contribution is 
 still negligible. For instance, for the values set in this figure of 
  $M_0 = 400$ GeV, $M_{1/2} = 300$ GeV, $A_0 = 0$, 
 $\mbox{sign} \mu > 0$ and $m_{N} = 10^{14}$ GeV, the Higgs contribution 
  is still four orders of 
 magnitude smaller than the photon-penguin 
 contribution at $\tan{\beta}=50$. This set of values give rise to the
 following
 MSSM spectrum (we just specify here the relevant sectors),  
\begin{center}
\begin{tabular}{ccc}
$m_{\tilde{l}_1} = 247$ GeV & $m_{\tilde{\chi}_1^0} = 121$ GeV & $m_{h^0} = 114$ GeV \\
$m_{\tilde{l}_2} = 397$ GeV & $m_{\tilde{\chi}_2^0} = 232$ GeV & $m_{H^0} = 457$ GeV \\
$m_{\tilde{l}_3} = 413$ GeV & $m_{\tilde{\chi}_3^0} = 484$ GeV & $m_{A^0} = 457$ GeV \\
$m_{\tilde{l}_4} = 416$ GeV & $m_{\tilde{\chi}_4^0} = 493$ GeV & $m_{\tilde{\nu}_1} = 351$ GeV \\
$m_{\tilde{l}_5} = 417$ GeV & $m_{\tilde{\chi}_1^-} = 232$ GeV& $m_{\tilde{\nu}_2} = 409$ GeV \\
$m_{\tilde{l}_6} = 419$ GeV & $m_{\tilde{\chi}_2^-} = 495$ GeV& $m_{\tilde{\nu}_3} = 410$ GeV. 
\end{tabular}
\end{center} 
We have checked that other choices of parameters, specially lower $M_0$ and
$M_{1/2}$ lead to larger contributions from the Higgs penguins, since one gets
ligther SUSY spectra and more importantly lighter $H_0$ and $A_0$ bosons.
 However, the present experimental lower bounds on the MSSM particle masses,
 do not allow to decrease much these $M_0$ and $M_{1/2}$ values, so that in this 
 mSUGRA context the relevant $m_{H_0}$, and $m_{A_0}$ masses can never get
 low enough values such that their corresponding Higgs-penguin contributions be 
 competitive with the $\gamma$-penguin ones.
 From this figure we conclude then that the leading $\gamma$-penguin
 approximation works extremely well, for all $\tan{\beta}$ values.
 In this approximation one gets,
 \begin{eqnarray}
 \frac{BR(l_j \to 3 l_i)}{BR(l_j \to l_i \gamma)} &=& 
 \frac{\alpha}{3\pi}\left(\log\frac{m_{l_j}^2}{m_{l_i}^2}-\frac{11}{4}\right)
 \end{eqnarray}
 which leads to the approximate values of
 $\frac{1}{440}$, $\frac{1}{94}$ and $\frac{1}{162}$ for 
 $(l_jl_i)= (\tau \mu), (\tau e)$ and  $(\mu e)$, respectively. As will be seen later it also works extremely
 well in the other channels. These nearly constant values of the ratios of 
 branching ratios will be 
 showing along this work. Obviously, if these ratios could be measured 
 they could provide interesting information.     
 
In fig.~\ref{fig:1a}(c) 
we have included our predictions for $|\delta_{LL}^{23}|, |\delta_{LR}^{23}|$
and, 
$|\delta_{RR}^{23}|$, as defined in eqs.(\ref{deltaLL}),(\ref{deltaLR}) and
 (\ref{deltaRR}) respectively, 
as a function of $\tan{\beta}$. These are the flavor changing parameters 
that 
are the relevant ones for the $\tau$ decays having $\mu$ in the final state. 
It is also interesting to compare them with the predictions in the leading 
logarithmic
approximation where the generated mixing in the off-diagonal terms
$(i\neq j, i,j=1,2,3)$, through the running from $M_X$ down to $m_M$,
 is given by
\begin{eqnarray}
(\Delta m_{\tilde{L}}^2)_{ij}&=&-\frac{1}{8 \pi^2} (3 M_0^2+ A_0^2) (Y_{\nu}^* L Y_{\nu}^T)_{ij} \nonumber \\
(\Delta A_l)_{ij}&=&- \frac{3}{16 \pi^2} A_0 Y_{l_i} (Y_{\nu}^* L Y_{\nu}^T)_{ij}\nonumber \\
(\Delta m_{\tilde{E}}^2)_{ij}&=&0\,\,;\, L_{kl} \equiv \log \left( \frac{M_X}{m_{M_k}}\right) \delta_{kl}.
\label{misalignment_sleptons}
\end{eqnarray}
and, in consequence, it predicts the hierachy, 
$ |\delta_{LL}^{23}|>|\delta_{LR}^{23}|>|\delta_{RR}^{23}| $.

As expected from the leading-log approximation, we see in
fig.~\ref{fig:1a}(c) that $|\delta_{LL}^{23}|$ 
is much larger than $|\delta_{LR}^{23}|$ and $|\delta_{RR}^{23}|$. 
However, we get 
$|\delta_{RR}^{23}|$ larger than $|\delta_{LR}^{23}|$ and it can be 
indeed two orders of magnitude larger than $|\delta_{LR}^{23}|$ 
at large $\tan{\beta}$. It is clear that, at least for our choice here 
of $A_0=0$, the leading-log approximation does not fully work. 
We also learn from this figure that the size of the mixing is always small 
in the degenerate case, being the largest $|\delta_{LL}^{23}|$ 
about $3 \times 10^{-3}$.

We next comment on the relevance of the choice for the $m_N$ values.
In fig.~\ref{fig:1b} we have illustrated the 
$\tau \to \mu^- \mu^- \mu^+$ and $\tau \to \mu \gamma$ branching ratios 
as a function of $m_N$ for degenerate heavy neutrinos and 
$\tan{\beta} = 50$. The explored range in $m_N$ is from $10^8 $ GeV up to 
$10^{14} $ GeV which is favorable for baryogenesis. 
Both rates have the same behaviour with $m_N$ which 
corresponds approximately to 
$BR(\tau \to \mu^- \mu^- \mu^+)$, 
$BR(\tau \to \mu \gamma) \propto |m_N\log(m_N)|^2 $. 
As before, these two predicted branching ratios are well bellow their 
experimental upper bounds, even at the largest $m_N$ value of $10^{14}$ GeV. In the last plot of fig.~\ref{fig:1b} 
we inlude the dependence of 
$|\delta_{LL}^{23}|, |\delta_{LR}^{23}|, |\delta_{RR}^{23}|$ on $m_N$ which 
clearly show a correlated behaviour with the previous plots. 
Again, $|\delta_{LL}^{23}|$ is the dominant one reaching values up 
to about $3 \times 10^{-3}$, and $\delta_{RR}^{23}$ is larger than 
$\delta_{LR}^{23}$.

For completeness, we also include the results of the other four
LFV $\tau$ and $\mu$ decays in fig.~\ref{fig:1cd}, where the predictions
are shown as a function of $\tan{\beta}$. 
These behaviours are very similar to those 
in $BR(\tau^- \to \mu^- \mu^- \mu^+)$ and $BR(\tau \to \mu \gamma)$ decays
correspondingly. The main difference is in the lower plots, where now
$|\delta^{12(13)}_{LR}|$ is larger than $|\delta^{12(13)}_{RR}|$. 
The maximum reached values are very small in this case,
$|\delta^{12(13)}_{LL}| \sim 5 \times 10^{-5}$. 
We see again that the leading $\gamma$-penguin approximation works
extremely well for these channels, and the previously mentioned values of 
the ratios of branching
ratios give a pretty good answer. 
We also find that the rates for all these four decays are well  
below their corresponding experimental bounds, in the degenerate case, 
for all the explored values
of $\tan \beta$ and $m_N$.

To end up the study of the degenerate case, we have also explored the 
dependence of the largest ratios 
$BR(\tau^- \to \mu^- \mu^- \mu^+)$ and $BR(\tau^- \to \mu^- \gamma)$ 
with the 
mSUGRA parameters $M_0$ and $M_{1/2}$. These results are shown in 
fig~\ref{fig:1ef}. We see clearly a similar behaviour in the two 
channels and their rates decrease as expected when increasing
the soft SUSY breaking mass parameters. This 
implies that for large enough values of $M_0$ or $M_{1/2}$ 
the branching ratios are considerably suppresed, 
due to the decoupling of the heavy SUSY particles in the dominant 
loops which are common to both observables. 
Thus, looking at these plots we can obviously conclude that the 
lighter the SUSY spectrum is, the larger branching ratios we get. However, as
already said, the more interesting region of low $M_0$ and/or $M_{1/2}$ values, 
being close to $100$ GeV, is not allowed by the present experimental lower
bounds on the MSSM particle masses.   

In summary, 
in the case of degenerate heavy neutrinos, we get LFV 
$\tau$ and $\mu$ decay rates which are still below their 
 present experimental upper bounds, 
for all the explored values of the seesaw and mSUGRA parameters, which have
been required to provide a full MSSM spectrum with masses being 
compatible with the present experimental bounds.

\subsection{Hierarchical case}

\begin{figure}
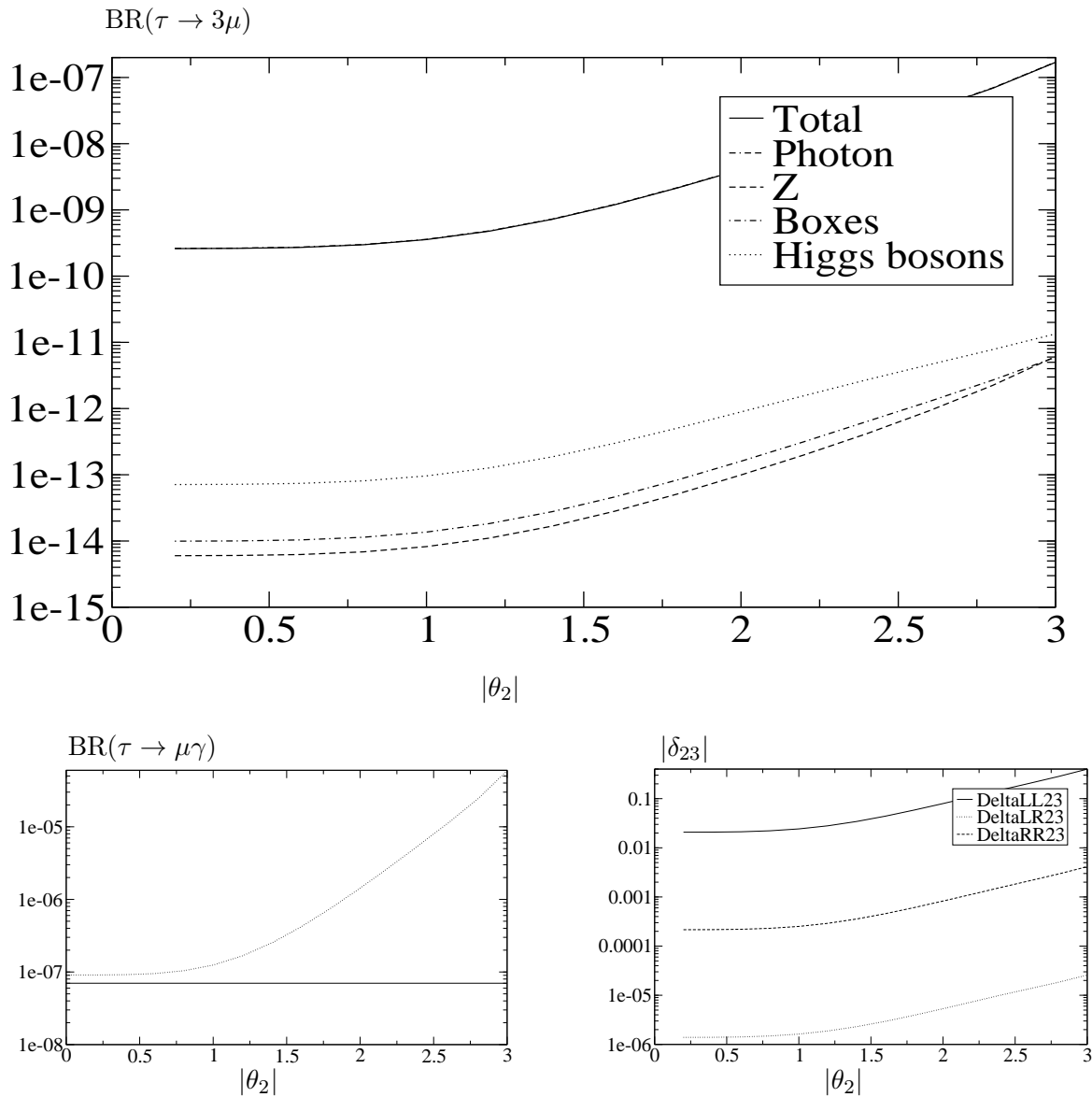

\hspace{-0.5cm}
\includegraphics[width=10.0cm,height=15.0cm,angle=-90]{fig2a_tau3mu.epsi}
\hspace{0.25cm}
\vspace{0.5cm}\\
\hspace{-0.5cm}
\includegraphics[width=5.0cm,height=7.0cm,angle=-90]{fig2a_taumugamma.epsi}
\hspace{1.0cm}
\includegraphics[width=5.0cm,height=7.0cm,angle=-90]{fig2a_deltas23.epsi}
\caption{(a) Upper panel: Dependence of $BR(\tau^- \to \mu^- \mu^- \mu^+)$ with  
$\vert \theta_2 \vert$, (b) lower-left panel: Dependence of $BR(\tau \to \mu \gamma)$ with $\vert \theta_2 \vert$, (c) lower right panel: $|\delta^{23}_{LL, LR, RR}|$ with $\vert \theta_2 \vert$. The horizontal line is the upper experimental bound. All panels are in scenario B, 
for $\arg(\theta_2) = \pi/4$, $(m_{N_1},m_{N_2},m_{N_3})=(10^8, 
2 \times 10^8, 10^{14})$ GeV, $\theta_1=\theta_3=0$, $\tan \beta = 50$, 
$M_0=400 $ GeV, $M_{1/2}=300 $ GeV, $A_0 = 0$ and $\mbox{sign}(\mu) > 0$.
\label{fig:2a}}
\end{figure}

\begin{figure}
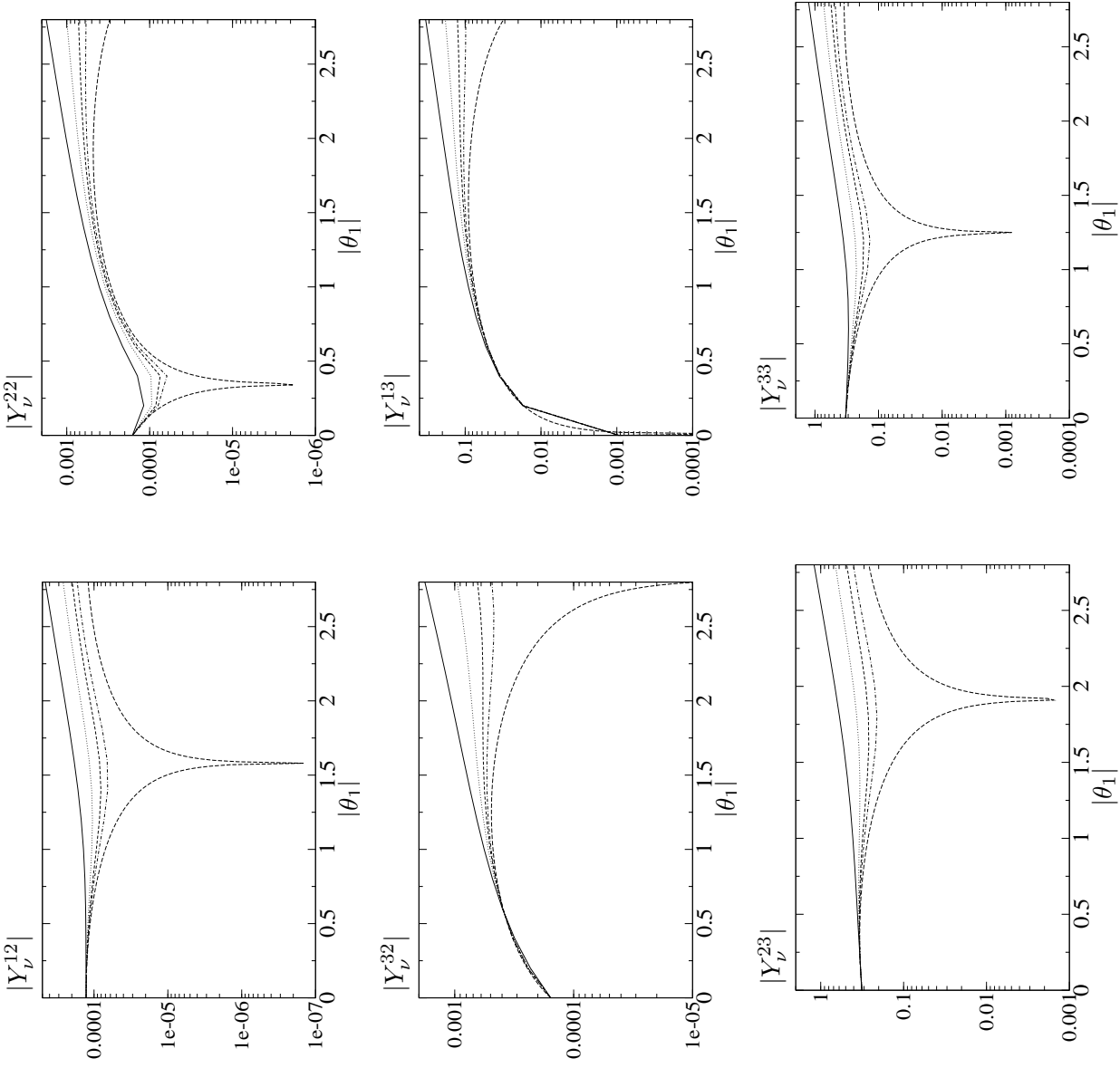

\hspace{-0.5cm}
\includegraphics[width=5.0cm,height=7.0cm,angle=-90]{Ynu21.epsi}
\hspace{1.0cm}
\includegraphics[width=5.0cm,height=7.0cm,angle=-90]{Ynu22.epsi}
\hspace{0.25cm}
\vspace{0.5cm}\\
\hspace{-0.5cm}
\includegraphics[width=5.0cm,height=7.0cm,angle=-90]{Ynu23.epsi}
\hspace{1.0cm}
\includegraphics[width=5.0cm,height=7.0cm,angle=-90]{Ynu31.epsi}
\hspace{0.25cm}
\vspace{0.5cm}\\
\hspace{-0.5cm}
\includegraphics[width=5.0cm,height=7.0cm,angle=-90]{Ynu32.epsi}
\hspace{1.0cm}
\includegraphics[width=5.0cm,height=7.0cm,angle=-90]{Ynu33.epsi}
\caption{Dependence of $|Y_{\nu}|$ with $\vert \theta_1 \vert$ in scenario B with hierarchical heavy neutrinos 
and complex R, for $\arg(\theta_1) = 0, \pi/10, \pi/8, \pi/6, \pi/4$ in radians
(lower to upper lines), $(m_{N_1},m_{N_2},m_{N_3})=(10^8, 
2 \times 10^8, 10^{14})$ GeV, $\theta_2=\theta_3=0$, $\tan \beta = 50$, 
$M_0=400 $ GeV, $M_{1/2}=300 $ GeV and $A_0 = 0$.
\label{fig:Ynu}}
\end{figure}

\begin{figure}
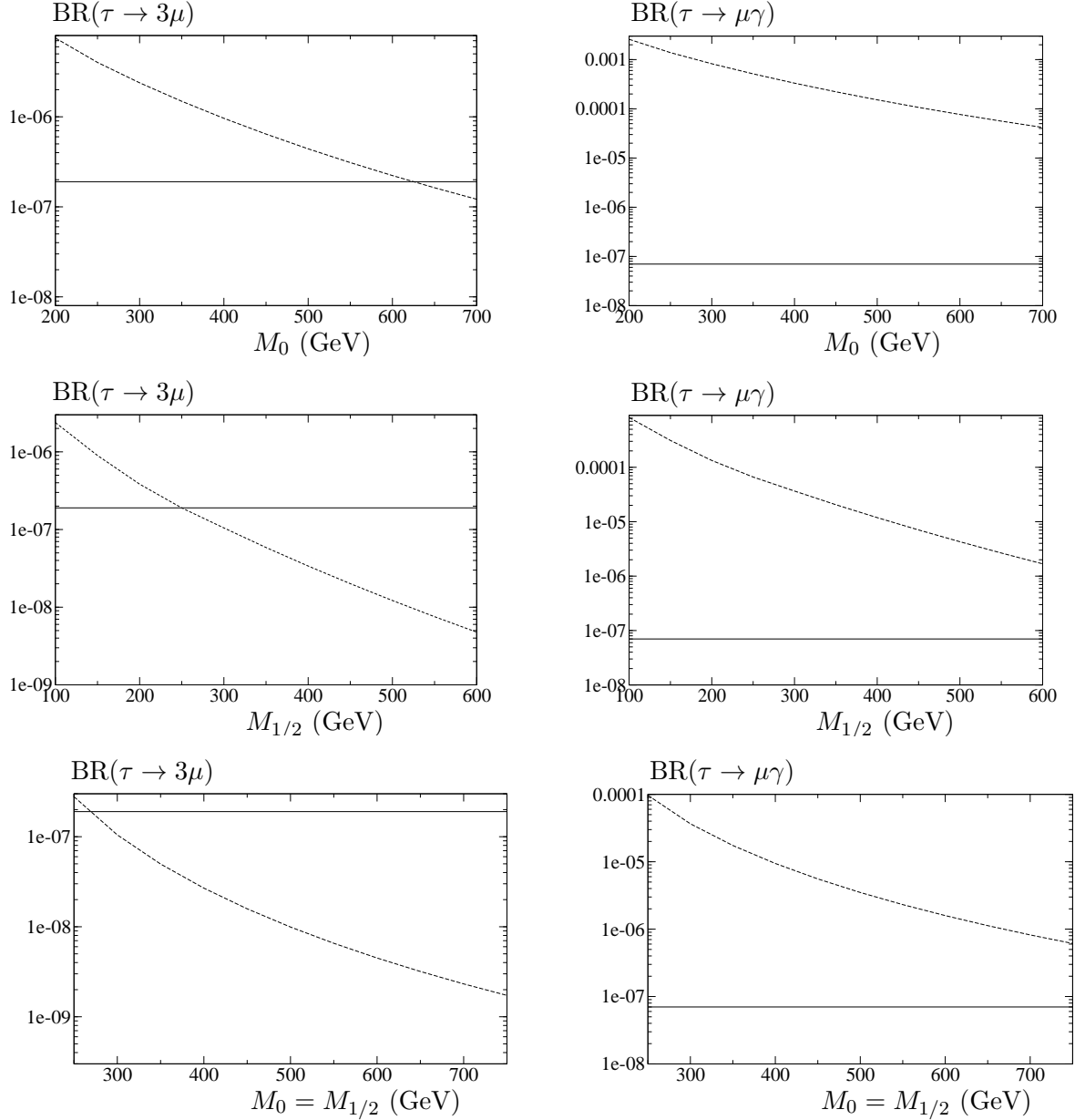

\hspace{-0.5cm}
\includegraphics[width=5.0cm,height=7.0cm,angle=-90]{fig2f_tau3mu_M0.epsi}
\hspace{1.0cm}
\includegraphics[width=5.0cm,height=7.0cm,angle=-90]{fig2f_taumugamma_M0.epsi}
\hspace{0.25cm}
\vspace{0.5cm}\\
\hspace{-0.5cm}
\includegraphics[width=5.0cm,height=7.0cm,angle=-90]{fig2f_tau3mu_Mhalf.epsi}
\hspace{1.0cm}
\includegraphics[width=5.0cm,height=7.0cm,angle=-90]{fig2f_taumugamma_Mhalf.epsi}
\hspace{0.25cm}
\vspace{0.5cm}\\
\hspace{-0.5cm}
\includegraphics[width=5.0cm,height=7.0cm,angle=-90]{fig2f_tau3mu_M0eqMhalf.epsi}
\hspace{1.0cm}
\includegraphics[width=5.0cm,height=7.0cm,angle=-90]{fig2f_taumugamma_M0eqMhalf.epsi}
\caption{Dependence of $BR(\tau^- \to \mu^- \mu^- \mu^+)$ and $BR(\tau \to \mu \gamma)$ with $M_0$ and $M_{1/2}$ in scenario B with hierarchical heavy neutrinos and complex R, for $(m_{N_1},m_{N_2},m_{N_3})=(10^8, 2 \times 10^8, 10^{14})$ GeV, $\tan \beta = 50$, $\vert \theta_2 \vert = 2.8$ and $\arg(\theta_2) = \pi/4$ ($\theta_1$ = $\theta_3$ = 0). (a) Upper-left panel, $BR(\tau \to \mu^- \mu^- \mu^+)$ as a function of $M_0$ for $M_{1/2} = 100$ GeV, (b) upper-right panel, $BR(\tau \to \mu \gamma)$ as a function of $M_0$ for $M_{1/2} = 100$ GeV, (c) middle-left panel, $BR(\tau \to \mu^- \mu^- \mu^+)$ as a function of $M_{1/2}$ for $M_0 = 200$ GeV, (d) middle-right panel, $BR(\tau \to \mu \gamma)$ as a function of $M_{1/2}$ for $M_0 = 200$ GeV, (e) lower-left panel, $BR(\tau \to \mu^- \mu^- \mu^+)$ as a function of $M_0 = M_{1/2}$, (f) lower-right panel, $BR(\tau \to \mu \gamma)$ as a function of $M_0 = M_{1/2}$. In all the plots we take $A_0 = 0$ and $\mbox{sign}(\mu) > 0$. The horizontal lines are the experimental bounds. 
\label{fig:2f}}
\end{figure}

We next present the results for hierarchical neutrinos, scenario B, which are
much more promissing. In this case the choice for $R$ is very relevant. 
The results for the general complex $R$ case and for the particular 
mass hierarchy  $(m_{N_1},m_{N_2},m_{N_3})=(10^8, 2 \times 10^8, 10^{14})$ GeV,
 are shown in figs.~\ref{fig:2a} through~\ref{fig:2f}. This particular 
 choice for the heavy neutrino masses seems to generate a proper rate for 
 baryogenesis via leptogenesis in the hierarchical 
 case~\cite{Chankowski:2004jc}. 
 We will later explore other choices as well. 
 
 From these figures we first confirm that the 
 LFV $\tau$ and $\mu$ decay rates are much larger in the hierarchical case
  than in the degenerate one. This is true even for the case of real $R$, 
 which  corresponds in our plots to the predictions at 
 $\arg(\theta_1)=\arg(\theta_2)= \arg(\theta_3)=0$. Furthermore, we get severe 
 restrictions on the maximum allowed decay rates coming from the 
 experimental upper bounds.

The predictions for  $BR(\tau^- \to \mu^- \mu^- \mu^+)$ and 
$BR(\tau \to \mu \gamma)$ as a function of $\vert \theta_2 \vert$  are 
depicted in fig~\ref{fig:2a}. Here $\theta_1$ 
and $\theta_3$ are set to zero, and $\arg(\theta_2) = \pi/4$. From now on the arguments of $\theta_1$, $\theta_2$ and $\theta_3$ are written in radians. The other
parameters are set to 
$\tan \beta = 50$, 
$M_0=400 $ GeV, $M_{1/2}=300 $ GeV, $A_0 = 0$ and $\mbox{sign}(\mu) > 0$. 
In fig.~\ref{fig:2a}(a) we show separately the various contributions to 
$BR(\tau^- \to \mu^- \mu^- \mu^+)$.
The dominant one is again the photon-penguin contribution (which is
undistinguisible from the total in this figure) and the 
others are several orders of magnitude smaller. 
We also see that the relative size of the subdominant contributions have 
changed respect to the previously studied degenerate case. 
Now the  Higgs contribution is larger than the boxes one and this
is larger than the $Z$ 
one. This is so because the largest $\tan{\beta} = 50$ value has been set. 
All the rates for $\tau^- \to \mu^- \mu^- \mu^+$ in this plot are within 
the allowed range by the experimental bound, which is placed just at the
upper line of the rectangle. In contrast, one can see in fig~.\ref{fig:2a}(b) 
that, for the chosen mSUGRA and seesaw parameters, the predicted 
$BR(\tau \to \mu \gamma)$ are clearly 
above the experimental bound. 
The dependence of $|\delta^{23}_{LL, LR, RR}|$ with $\vert \theta_2 \vert$ 
is shown in fig.~\ref{fig:2a}(c). We see that $|\delta^{23}_{LL}|$ can 
reach very large values, up to 0.4, for $\vert \theta_2 \vert = 3$ and 
$\arg(\theta_2) = \pi/4$. We have checked that this particular choice 
of $\theta_2 = 3 e^{i \pi/4}$ gives rise to large neutrino Yukawa matrix 
elements $|Y_{\nu}^{33}|$ and $|Y_{\nu}^{23}|$ of the order of 1, which are the
responsible for this large mixing in the slepton sector. 

It is also interesting
to compare the MSSM spectrum for this hierarchical case with the previous 
degenerate case. For the input values of fig.~\ref{fig:2a} but with 
$\theta_2$ set to the extreme value $\theta_2 = 2.8e^{i\frac{\pi}{4}}$ 
we get the following
masses,
\begin{center}
\begin{tabular}{ccc}
$m_{\tilde{l}_1} = 230$ GeV & $m_{\tilde{\chi}_1^0} = 122$ GeV & $m_{h^0} = 114$ GeV \\
$m_{\tilde{l}_2} = 356$ GeV & $m_{\tilde{\chi}_2^0} = 232$ GeV & $m_{H^0} = 455$ GeV \\
$m_{\tilde{l}_3} = 413$ GeV & $m_{\tilde{\chi}_3^0} = 481$ GeV & $m_{A^0} = 455$ GeV \\
$m_{\tilde{l}_4} = 417$ GeV & $m_{\tilde{\chi}_4^0} = 490$ GeV & $m_{\tilde{\nu}_1} = 296$ GeV \\
$m_{\tilde{l}_5} = 436$ GeV & $m_{\tilde{\chi}_1^-} = 232$ GeV& $m_{\tilde{\nu}_2} = 422$ GeV \\
$m_{\tilde{l}_6} = 448$ GeV &$m_{\tilde{\chi}_2^-} = 492$ GeV & $m_{\tilde{\nu}_3} = 441$ GeV \\
\end{tabular}
\end{center}
It is obvious that the complex $R$ affects significantly the predictions 
of the MSSM masses,
specially in the slepton sector. In general, the slepton mixing generated by 
the complex 
$\theta_i$, lower the lightest charged slepton and the lightest sneutrino
masses and increases the heaviest charged slepton and sneutrino masses.

\begin{figure}
\hspace{-0.5cm}
\includegraphics[width=5.0cm,height=7.0cm,angle=-90]{fig2b_tau3mu.epsi}
\hspace{1.0cm}
\includegraphics[width=5.0cm,height=7.0cm,angle=-90]{fig2b_taumugamma.epsi}
\hspace{0.25cm}
\vspace{0.5cm}\\
\hspace{-0.5cm}
\includegraphics[width=5.0cm,height=7.0cm,angle=-90]{fig2b_tau3e.epsi}
\hspace{1.0cm}
\includegraphics[width=5.0cm,height=7.0cm,angle=-90]{fig2b_tauegamma.epsi}
\hspace{0.25cm}
\vspace{0.5cm}\\
\hspace{-0.5cm}
\includegraphics[width=5.0cm,height=7.0cm,angle=-90]{fig2b_mu3e.epsi}
\hspace{1.0cm}
\includegraphics[width=5.0cm,height=7.0cm,angle=-90]{fig2b_muegamma.epsi}
\caption{Dependence of LVF $\tau$ and $\mu$ decays with 
$\vert \theta_2 \vert$ in scenario B with hierarchical heavy neutrinos 
and complex R, for $\arg(\theta_2) = 0, \pi/10, \pi/8, \pi/6, \pi/4$ in radians 
(lower to upper lines), $(m_{N_1},m_{N_2},m_{N_3})=(10^8, 2 \times 10^8, 10^{14})$ GeV, 
$\theta_1=\theta_3=0$, $\tan \beta = 50$, $M_0=400 $ GeV, $M_{1/2}=300 $ GeV, 
$A_0 = 0$ and $\mbox{sign}(\mu) > 0$. The horizontal lines are the upper experimental bounds.
\label{fig:2b}}
\end{figure}

In fig.~\ref{fig:2b} we show the predictions of
 $BR(l_j^- \to l_i^- l_i^- l_i^+)$ and $BR(l_j \to l_i \gamma)$ as 
 functions of $\vert \theta_2 \vert$, for all the channels and for the 
 different
 values
 of $\arg(\theta_2)=0, \pi/10, \pi/8, \pi/6, \pi/4$. 
 In all these plots we set again $\tan{\beta} = 50$, $M_0=400 $ GeV, 
 $M_{1/2}=300 $ GeV, $A_0 = 0$, 
 $\mbox{sign}(\mu) > 0$ and 
 $(m_{N_1},m_{N_2},m_{N_3})=(10^8, 2 \times 10^8, 10^{14})$ GeV.
 The upper lines correspond to $\arg(\theta_2) = \pi/4$ and the lower ones 
 to $\arg(\theta_2) = 0$. These lower lines are therefore the corresponding predictions for real $R$. 
 It is clear that all the branching ratios have a soft behaviour 
 with $\vert \theta_2 \vert$ except for the case of real 
 $\theta_2$ where appears a narrow dip in each plot. 
 In this fig.~\ref{fig:2b} we see that 
  all the rates obtained are below their experimental upper bounds, 
  except for the processes $\tau \to \mu \gamma$ and $\mu \to e \gamma$, 
  where the predicted rates for complex $\theta_2$ with large 
  $\vert \theta_2 \vert$ are clearly above the allowed region. The most
  restrictive channel in this case is $\tau \to \mu \gamma$ where
  compatibility with data occurs just for real $\theta_2$ and for complex 
  $\theta_2$ but with $\vert \theta_2 \vert$ values near the region of
  the narrow dip. We also see that the rates for $BR(\mu \to 3 e)$ enter in
  conflict with experiment at the upper corner of large 
  $\vert \theta_2 \vert$ and large $\arg(\theta_2) = \pi/4$.
  

\begin{figure}
\hspace{-0.5cm}
\includegraphics[width=5.0cm,height=7.0cm,angle=-90]{fig2c_tau3mu.epsi}
\hspace{1.0cm}
\includegraphics[width=5.0cm,height=7.0cm,angle=-90]{fig2c_taumugamma.epsi}
\hspace{0.25cm}
\vspace{0.5cm}\\
\hspace{-0.5cm}
\includegraphics[width=5.0cm,height=7.0cm,angle=-90]{fig2c_tau3e.epsi}
\hspace{1.0cm}
\includegraphics[width=5.0cm,height=7.0cm,angle=-90]{fig2c_tauegamma.epsi}
\hspace{0.25cm}
\vspace{0.5cm}\\
\hspace{-0.5cm}
\includegraphics[width=5.0cm,height=7.0cm,angle=-90]{fig2c_mu3e.epsi}
\hspace{1.0cm}
\includegraphics[width=5.0cm,height=7.0cm,angle=-90]{fig2c_muegamma.epsi}
\caption{Dependence of LFV $\tau$ and $\mu$ decays with 
$\vert \theta_1 \vert$ in scenario B with hierarchical heavy neutrinos 
and complex R, for $\arg(\theta_1) = 0, \pi/10, \pi/8, \pi/6, \pi/4$ in radians
(lower to upper lines), $(m_{N_1},m_{N_2},m_{N_3})=(10^8, 
2 \times 10^8, 10^{14})$ GeV, $\theta_2=\theta_3=0$, $\tan \beta = 50$, 
$M_0=400 $ GeV, $M_{1/2}=300 $ GeV and $A_0 = 0$. The horizontal lines are 
the upper experimental bounds.
\label{fig:2c}}
\end{figure}

\begin{figure}
\hspace{-0.5cm}
\includegraphics[width=5.0cm,height=7.0cm,angle=-90]{fig2c_bis_tau3mu.epsi}
\hspace{1.0cm}
\includegraphics[width=5.0cm,height=7.0cm,angle=-90]{fig2c_bis_taumugamma.epsi}
\hspace{0.25cm}
\vspace{0.5cm}\\
\hspace{-0.5cm}
\includegraphics[width=5.0cm,height=7.0cm,angle=-90]{fig2c_bis_tau3e.epsi}
\hspace{1.0cm}
\includegraphics[width=5.0cm,height=7.0cm,angle=-90]{fig2c_bis_tauegamma.epsi}
\hspace{0.25cm}
\vspace{0.5cm}\\
\hspace{-0.5cm}
\includegraphics[width=5.0cm,height=7.0cm,angle=-90]{fig2c_bis_mu3e.epsi}
\hspace{1.0cm}
\includegraphics[width=5.0cm,height=7.0cm,angle=-90]{fig2c_bis_muegamma.epsi}
\caption{Dependence of LFV $\tau$ and $\mu$ decays with $\vert \theta_1 \vert$ 
in scenario B with hierarchical heavy neutrinos and complex R, for 
$\arg(\theta_1) = 0, \pi/10, \pi/8, \pi/6, \pi/4$ in radians (lower to upper lines), 
$(m_{N_1},m_{N_2},m_{N_3})=(10^8, 2 \times 10^8, 10^{14})$ GeV, 
$\theta_2=\theta_3=0$, $\tan \beta = 50$, $M_0=250 $ GeV, $M_{1/2}=150 $ GeV 
and $A_0 = 0$. The horizontal lines are the upper experimental bounds.
\label{fig:2d}}
\end{figure}

Even more interesting are the predictions for 
$BR(l_j^- \to l_i^- l_i^- l_i^+)$ and $BR(l_j \to l_i \gamma)$ as functions of
 $\vert \theta_1 \vert$, due
to the large values of the relevant entries of the $Y_{\nu}$ coupling
matrix, which are illustrated in fig.~\ref{fig:Ynu}. Concretely,
$|Y_{\nu}^{13}|$ can be as large as $\sim 0.2$ for $|\theta_1|
\sim 2.5$ and $\arg{(\theta_1)} = \pi/4$, and $|Y_{\nu}^{23}|$ and
$|Y_{\nu}^{33}|$ are in the range $0.1 - 1$ for all studied complex $\theta_1$ values. The results for $BR(l_j^- \to l_i^- l_i^- l_i^+)$ and $BR(l_j \to l_i \gamma)$ as  functions of
 $\vert \theta_1 \vert$, for different values of $\arg{(\theta_1)}$, are 
 illustrated in fig.~\ref{fig:2c}. Here $\theta_2$ and $\theta_3$ are set 
 to zero. The same set of mSUGRA parameters and heavy neutrino masses as in
 fig.~\ref{fig:2b} are
 taken for comparison. We see clearly that 
 the restrictions are 
 more severe in this case than in the previous one. In fact, 
 all the rates cross the horizontal lines of 
 the experimental bounds except for $BR(\tau^- \to \mu^- \mu^- \mu^+)$ and 
 $BR(\tau^- \to e^- e^- e^+)$. The most restrictive channel is now the
 $\mu \to e \gamma$ decay. 
 More specifically, we see that 
 all the points in the plot of $BR(\mu \to e \gamma)$, except for the particular
 values $\theta_1= 0$ and real $\theta_1$ at the dip, are   
 excluded by the experimental upper bound. Also the predictions for 
 $BR(\mu \to 3e)$ are mostly excluded, except again for the region close to 
 zero and
 the dip. Notice that the qualitative behaviour of these all branching
 ratios with $|\theta_1|$ in fig.~\ref{fig:2c} and the locations of the dips can be explained from the Yukawa
 coupling matrix behaviour in fig.~\ref{fig:Ynu}.
 
The scenario most seriously in conflict with experiment is shown in 
 fig.~\ref{fig:2d} where the predictions for 
$BR(l_j^- \to l_i^- l_i^- l_i^+)$ and $BR(l_j \to l_i \gamma)$ are again plotted
as a function of  $\vert \theta_1 \vert$ and for the same choices of 
$\arg(\theta_1)$ as in the previous case, but now the mSUGRA mass parameters are
set to the lower values, $M_0=250 $ GeV, and $M_{1/2}=150 $ GeV. These lead 
to a
lighter MSSM spectrum and and in consequence to higher rates. For comparison
with the previous cases, we include below the predicted masses of the 
relevant MSSM
particles, for the particular value $\theta_1=2.8e^{i \frac{\pi}{4}}$,
\begin{center}
\begin{tabular}{ccc}
$m_{\tilde{l}_1} = 94$ GeV & $m_{\tilde{\chi}_1^0} = 58$ GeV & $m_{h^0} = 108$ GeV \\
$m_{\tilde{l}_2} = 218$ GeV & $m_{\tilde{\chi}_2^0} = 107$ GeV & $m_{H^0} = 269$ GeV \\
$m_{\tilde{l}_3} = 259$ GeV & $m_{\tilde{\chi}_3^0} = 284$ GeV & $m_{A^0} = 269$ GeV \\
$m_{\tilde{l}_4} = 259$ GeV & $m_{\tilde{\chi}_4^0} = 296$ GeV & $m_{\tilde{\nu}_1} = 143$ GeV \\
$m_{\tilde{l}_5} = 273$ GeV & $m_{\tilde{\chi}_1^-} = 107$ GeV& $m_{\tilde{\nu}_2} = 247$ GeV \\
$m_{\tilde{l}_6} = 273$ GeV &$m_{\tilde{\chi}_2^-} = 300$ GeV & $m_{\tilde{\nu}_3} = 261$ GeV \\
\end{tabular}
\end{center}
Notice that the lightest slepton, neutralino, chargino and Higgs boson have masses
 close to their
experimental lower bounds. 

We conclude from this fig.~\ref{fig:2d} that the predictions for  
$BR(\mu \to e \gamma)$ and $BR(\mu \to 3 e)$ are totally excluded by present 
data and the predictions for $BR(\tau \to \mu \gamma)$ are practically excluded,
with the exception of the two narrow dips. The predictions for  
$BR(\tau \to e \gamma)$ get severe restrictions for complex $\theta_1$ with 
large $\vert \theta_1 \vert$ and/or large  $\arg(\theta_1)$, and the rates 
for $BR(\tau \to 3 \mu)$ start being sensitive to the present experimental 
bounds for large complex $\theta_1$ values in the upper corner of the plot.

We have also explored the dependence with the complex $\theta_3$ angle, and it 
turns out that the predictions for all rates are nearly constant with this
angle. For instance, for $\tan \beta = 50$, 
$M_0=400 $ GeV, $M_{1/2}=300 $ GeV, $A_0 = 0$ and $\mbox{sign}(\mu) > 0$, 
we get
$BR(\tau \to 3 \mu) =2.6 \times 10^{-10}$,
$BR(\tau \to 3 e) =8.8 \times 10^{-15}$,
$BR(\mu \to 3 e) =1.8 \times 10^{-14}$,
$BR(\tau \to \mu \gamma)= 9.1 \times 10^{-8}$, 
$BR(\tau \to e \gamma)= 7.8 \times 10^{-13}$ and
$BR(\mu \to e \gamma)= 2.6 \times 10^{-12}$.   
In this case only the prediction for  $BR(\tau \to \mu \gamma)$ is in conflict 
with the experiment.
 
\begin{figure}
\hspace{-0.5cm}
\includegraphics[width=5.0cm,height=7.0cm,angle=-90]{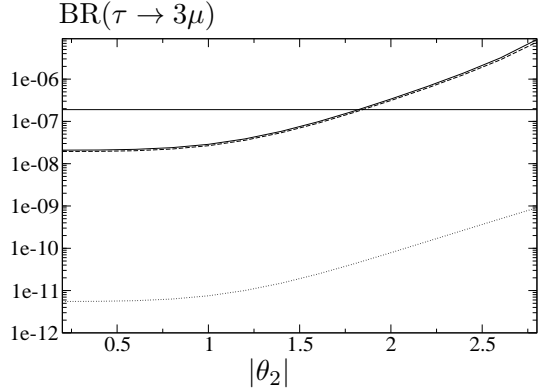}
\caption{Dependence of LFV $\tau \to 3 \mu$  with $|\theta_{2}|$ in
  scenario B with hierarchical heavy neutrinos, for different
  $m_{N_i}$ choices. Solid line is for
  $(m_{N_1},m_{N_2},m_{N_3})=(10^8, 2 \times 10^8, 10^{14})$ GeV,
  dashed line is for $(m_{N_1},m_{N_2},m_{N_3})=(10^{10}, 2 \times
  10^{10}, 10^{14})$ GeV, and dotted line is for
  $(m_{N_1},m_{N_2},m_{N_3})=(10^8, 2 \times 10^8, 10^{12})$ GeV. The
  rest of parameters are set to $\tan \beta = 50$, $M_0=200 $ GeV,
  $M_{1/2}=100 $ GeV, $A_0 = 0$, $\mbox{sign}(\mu) > 0$ and
  $\mbox{arg}(\theta_2) = \pi/4$. The horizontal line is the experimental bound.
\label{fig:2g}}
\end{figure}

\begin{figure}
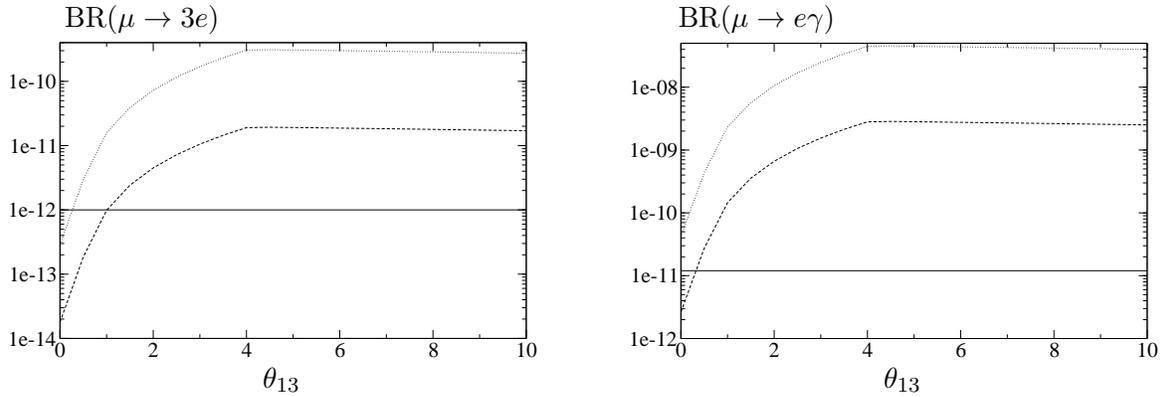

\hspace{-0.5cm}
\includegraphics[width=5.0cm,height=7.0cm,angle=-90]{fig2e_mu3e.epsi}
\hspace{1.0cm}
\includegraphics[width=5.0cm,height=7.0cm,angle=-90]{fig2e_muegamma.epsi}
\caption{Dependence of LFV $\mu$ decays with $\theta_{13}$ in degrees in scenario B 
with hierarchical heavy neutrinos and $R = 1$, for 
$(m_{N_1},m_{N_2},m_{N_3})=(10^8, 2 \times 10^8, 10^{14})$ GeV, 
$\tan \beta = 50$, $A_0 = 0$ and 
$\mbox{sign}(\mu) > 0$. The upper lines are for $M_0=250 $ GeV, $M_{1/2}=150 $ GeV 
and the lower lines
are for $M_0=400 $ GeV, $M_{1/2}=300 $ GeV. 
The horizontal lines are the experimental bounds.
\label{fig:2e}}
\end{figure}

The dependence of $BR(\tau^- \to \mu^- \mu^- \mu^+)$ and 
$BR(\tau \to \mu \gamma)$ with the mSUGRA parameters $M_0$ and $M_{1/2}$ 
is illustrated in fig.~\ref{fig:2f}. We see a similar behaviour as in the 
degenerate case, where a suppresion of the branching ratios occurs for large 
values of $M_0$ and/or $M_{1/2}$. Whereas the ratios for $BR(\tau \to
3 \mu)$ enter in to the allowed region by the experimental bound for
large enough $M_0$ and/or $M_{1/2}$, the ratios for $B(\tau \to \mu
  \gamma)$ are well above their bound for all $M_0$ and $M_{1/2}$
  values explored. The main point again is the particular value of $\theta_2$ with large 
$|\theta_2|$ and large $\arg(\theta_2)$, which generates large rates.

With the purpose of exploring other choices of the mSUGRA parameters, 
we have also generated results for the specific value $A_0=-100$ and 
found very close predictions to the $A_0 = 0$ case, the lines in the plots being
nearly undistinguisable respect to this case. 
We have also run the alternative case of $\mbox{sign}(\mu)< 0$,  
and found again very close predictions to the $\mbox{sign}(\mu)> 0$ case,
with the lines in the plots being undistinguisable from this case. 

Finally, we have also tried another input values for the heavy neutrino masses.
The results for $BR(\tau \to 3 \mu)$ are shown in fig.~\ref{fig:2g}. Here we
compare the predictions for the three following set of values,
$(m_{N_1},m_{N_2},m_{N_3})=(10^8, 2 \times 10^8, 10^{14})$ GeV,
$(10^{10}, 2 \times 10^{10}, 10^{14})$ GeV and 
$(10^8, 2 \times 10^8, 10^{12})$ GeV. We conclude, that the relevant mass is the
heaviest one, $m_{N_3}$, and the scaling with this mass is approximately as 
the scaling with
the common mass $m_N$ in the  degenerate case. Because of this, the rates for 
the two first sets are nearly undistinguisable, and the rates for the third set
are about four orders of magnitude below.

Last but not least, we consider the very interesting case where the 
angle $\theta_{13}$ of the $U_{MNS}$ is non vanishing. It is known that the
present neutrino
data still allows for small values of this angle, $\theta_{13}<10^o$. 
The dependence of 
$BR(\mu^- \to e^- e^- e^+)$ and $BR(\mu \to e \gamma)$ with this 
$\theta_{13}$ is shown in fig.~\ref{fig:2e} where we explore values in the
$0 < \theta_{13} < 10^o$ range. We choose these two channels
because they are the most sensitive ones to this angle. For this study we 
assume the most conservative choice of $R = 1$, and set the other parameters 
to the following values: 
$\tan \beta = 50$, $A_0 = 0$, $\mbox{sign}(\mu) > 0$, and 
$(m_{N_1},m_{N_2},m_{N_3})=(10^8, 2 \times 10^8, 10^{14})$ GeV. The upper lines
are for $M_0=250 $ GeV, $M_{1/2}=150 $ GeV and the lower ones for
$M_0=400 $ GeV, $M_{1/2}=300 $ GeV.
We conclude that, for this choice of parameters, values of $\theta_{13}$ 
larger than 1 degree are totally excluded by the data on LFV $\mu$ decays.
It is a quite stricking result.

In summary, we obtain in the hierachical case much 
larger rates than in the degenerate one, 
and one must pay attention to these values, 
because the rates in several channels do get in conflict with the 
experimental bounds. More specifically, the choice of a complex $R$ matrix 
 with large modules and/or large arguments of $\theta_1$ and/or $\theta_2$ 
and a light SUSY spectrum is very constrained by data. 
We also confirm that the experimental upper bounds of the 
processes 
$l_j \to l_i \gamma$ are more restrictive than the 
$l_j^- \to l_i^- l_i^- l_i^+$ ones but all together will allow to 
extract large excluded regions of the mSUGRA and seesaw parameter space. 
A more precise conclusion on the excluded regions of this parameter space 
deserves a more devoted study.

\section{\label{conclu} Conclusions}

We have shown in this paper that the LFV $\tau$ and $\mu$ decays do provide a 
very efficient tool to look for indirect SUSY signals. Whereas the predicted
rates for these processes are negligible within the SM, the SUSY scenario 
considered here provides in contrast
significant rates which are at the present experimental reach for some of the
studied channels. This scenario consists of 
the well known mSUGRA extended with three right handed neutrinos 
and their
SUSY partners, and with the needed neutrino masses being generated via the
seesaw mechanism. The reason for these significant rates is because of the
important lepton flavor mixing that is generated in the slepton sector 
due to large Yukawa neutrino
couplings, which is transmited via the RGE running from the large energies 
down the electroweak
scale.  

With the motivation in mind of testing SUSY we have studied exhaustively 
in this work the particular decays
 $\tau \to 3\mu$, $\tau \to 3e$ and $\mu \to 3e$, and the correlated 
radiative decays $\tau \to \mu \gamma$, $\tau \to e \gamma$ and 
$\mu \to e \gamma$. All of these channels have quite challenging 
experimental bounds 
and they are expected to improve in the future . We have 
explored the dependence of the branching ratios
 for these LFV processes with the various parameters
involved, namely, the mSUGRA and seesaw parameters. 
We have computed and analyzed in full detail all the contributions from 
the SUSY
loops to the $l_j^- \to l_i^- l_i^- l_i^+$ 
decays. Our analytical results for these decays correct and complete previous 
results in the literature. In particular we have presented the results for
the separate contributions from the $\gamma$-penguin, the $Z$-penguin, the
Higgs-penguin and the box diagrams and shown explicitely the $\gamma$-penguin
dominance.
In the numerical estimates we have presented results for 
both the $l_j^- \to l_i^- l_i^- l_i^+$ and the correlated radiative decays 
$l_j \to l_i \gamma$. 

For the degenerate heavy neutrinos case, 
we have got rates for all the studied LFV $\tau$ and $\mu$ decays that
are below the present experimental upper bounds. 
The largest rates we get, within the explored range of the seesaw and 
mSUGRA parameter space, are for the $\tau$ decays. Specifically, 
$BR(\tau \to \mu \gamma) \sim 10^{-8}$ and 
$BR(\tau^- \to \mu^- \mu^- \mu^+) \sim 3 \times 10^{-11}$, 
corresponding to the extreme values of $\tan{\beta} = 50$ and 
$m_{N} = 10^{14}$ GeV and for the lowest values of $M_0$ and $M_{1/2}$ 
explored. 
The case of hierarchical heavy neutrinos turns out to be much more interesting.
 First of all, we get much larger branching ratios than in the previous case and
 secondly they are in many cases above the present experimental bounds.       
 
 We have analyzed in detail the behaviour of the branching ratios 
 with the mSUGRA and seesaw parameters also in the hierarchical case. 
 The largest ratios found are again 
 for $\tau \to \mu \gamma$ and $\tau^- \to \mu^- \mu^- \mu^+$ decays. 
 All the LFV $\tau$ and $\mu$ decay rates are mainly sensitive 
 to $\tan{\beta}$, 
 the heaviest neutrino mass $m_{N_3}$, which we have set to 
 $m_{N_3} = 10^{14}$ GeV, and the complex angles in the $R$ matrix 
 $\theta_1$ and $\theta_2$, which have been taken in the range 
 $3 < \tan{\beta} < 50$, $0 < |\theta_i|< 3$ and $0 < \arg(\theta_i) < \pi/4$. 
 For the values of these parameters at the upper limit 
 of this studied interval
  we have found that some of the predicted branching ratios are clearly 
  above 
  the corresponding experimental upper bounds. The most restrictive channels
  being  $\mu \to e \gamma$, $\mu \to 3e$ and $\tau \to \mu \gamma$.
  Therefore, we get in this region important restrictions on the 
  posible values of the mSUGRA and seesaw parameters. 
  In particular, for $\theta_2 = 2.8 e^{i \pi/4}$, we get that 
  the whole studied range of 
  $100 \, \mbox{GeV} < M_0, M_{1/2} < 800 \, \mbox{GeV}$ 
  with $\tan{\beta} = 50$ is totally excluded by $\tau \to \mu \gamma$. 
  Values of $M_0$ and $M_{1/2}$ in the low region below $250$ GeV are also 
  excluded by $\tau^- \to \mu^- \mu^- \mu^+$ data. The case of $\theta_1$
  is even more restrictive, because the predictions for $\mu \to e \gamma$, 
  $\mu \to 3e$ and $\tau \to \mu \gamma$ totally exclude a light SUSY
  scenario, for practicaly all $\theta_1$ values. 
  
Perhaps, the most striking result is that even for the most conservative 
choice of $R = 1$, that is $\theta_1 = \theta_2 = \theta_3 = 0$, 
there are also important restrictions at low $M_0$, $M_{1/2}$ and 
large $\tan{\beta}$ values. In particular, for $\tan{\beta} = 50$, 
values lower or equal than $M_0 = 250$ GeV and $M_{1/2} = 150$ GeV 
are totally excluded 
by $\tau \to \mu \gamma$, $\mu \to e \gamma$ and 
$\mu^- \to e^- e^- e^+$ data. 

For this conservative choice of $R=1$ we have also found the surprising
result that both $\mu \to e \gamma$ and $\mu^- \to e^- e^- e^+$ place 
important
restrictions on the allowed values for the $U_{MNS}$ angle 
$\theta_{13}$. For values lower or equal than  $M_0 = 250$ GeV and 
$M_{1/2} = 150$ GeV and for $\tan\beta=50$ and $m_{N_3 }=10^{14}$ GeV, 
we get that values
of $\theta_{13}$ larger than 1 degree are not allowed by these LFV data.

In conclusion,  it is clear from these results that the 
LFV $\tau$ and $\mu$
decays studied here do restrict significantly the mSUGRA and seesaw 
parameter space. A more refined analysis of the restrictions on 
this multidimensional parameter space,  
deserves a further study.

\begin{acknowledgments}
 We thank the kind hospitality of the SLAC theory group members, the valuable 
 enviroment for discussions there and the
 facilities provided at SLAC, where 
 most of this work was done. We also thank K. Tobe for some clarifications regarding some results in~\cite{Hisano:1995cp}.  M.J. Herrero acknowledges the finantial support
 from the Spanish Ministery of Science and Education (MEC) by the grant 
 under the name
 "Estancias de Profesores de Universidad en Centros Extranjeros", ref:
 PR2005-0069. 
 E. Arganda acknowledges
the Spanish MEC for finantial support by his FPU grant 
AP2003-3776. This work was also supported by the Spanish MEC under project 
FPA2003-04597. 
\end{acknowledgments}

\appendix
\label{apendices}

\section{}
\label{apendice1}

In this appendix we collect the Feynman rules for the interactions 
that are relevant in this work. They are expressed in the physical 
eigenstate basis, for all the MSSM sectors involved: sleptons ${\tilde l_X}$ 
$(X=1,..,6)$, sneutrinos ${\tilde{\nu_X}}$ $(X=1,2,3)$, neutralinos 
${\tilde \chi^0_A}$ $(A=1,..,4)$, charginos ${\tilde \chi^-_A}$ $(A=1,2)$ and 
the neutral Higgs bosons $H_p \,(p=1,2,3)\,=h^0, H^0, A^0$.

\subsection{Photon interactions}

The Feynman rules for the photon interactions that are used in this work are given by,

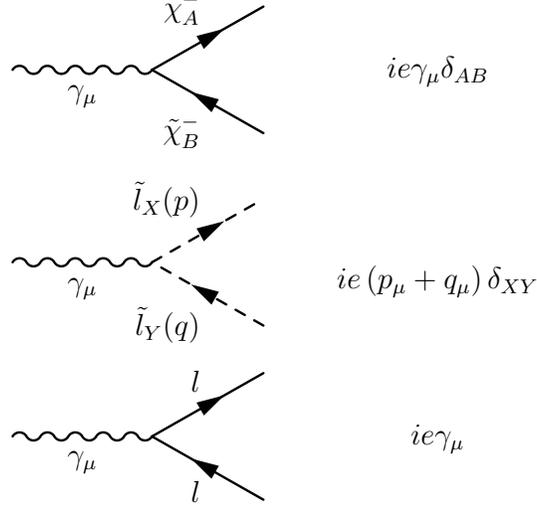
\begin{figure}[h!]
\begin{center}
\includegraphics[width=7cm]{photon.epsi}
\label{neutralinos_feyn}
\end{center}
\end{figure}


\subsection{Neutralino interactions}

The Feynman rules for neutralinos that take part in the one-loop diagrams computed here are the following:

\begin{figure}[h!]
\begin{center}
\includegraphics[width=9.5cm]{neutralinos.epsi}
\label{neutralinos_feyn}
\end{center}
\end{figure}

where
\begin{eqnarray}
N_{iAX}^L &=& -g \sqrt{2} \left\{ \frac{m_{l_i}}{2 M_W \cos{\beta}} N_{A3}^{\ast} R_{(1, 3, 5) X}^{(l)} + \tan{\theta_W} N_{A1}^{\ast} R_{(2, 4, 6) X}^{(l)} \right\} \\
N_{iAX}^R &=& -g \sqrt{2} \left\{ -\frac{1}{2} \left( \tan{\theta_W} N_{A1} + N_{A2} \right) R_{(1, 3, 5) X}^{(l)} + \frac{m_{l_i}}{2 M_W \cos{\beta}} N_{A3} R_{(2, 4, 6) X}^{(l)} \right\} \nonumber \\
\end{eqnarray}
$C$ is the charge conjugation matrix and $P_{L, R} = \frac{1 \mp \gamma_5}{2}$. 
Here $R^{(l)}$ and $N$ are the rotation matrices in the charge slepton and 
neutralino sectors, respectively. The definition of $N$ can be found in
~\cite{Haber:1985rc,Gunion:1986yn}. 

\newpage

\subsection{Chargino interactions}

The Feynman rules for the chargino interactions are given by

\begin{figure}[h!]
\begin{center}
\includegraphics[width=8.5cm]{charginos.epsi}
\label{charginos_feyn}
\end{center}
\end{figure}

where
\begin{eqnarray}
C_{iAX}^L &=& g \frac{m_{l_i}}{\sqrt{2} M_W \cos{\beta}} U_{A2}^{\ast} R_{(1, 2, 3) X}^{(\nu)} \\
C_{iAX}^R &=& -g V_{A1} R_{(1, 2, 3) X}^{(\nu)}
\end{eqnarray}
and $R^{(\nu)}$, $U$ and $V$ are the rotation matrices in the sneutrino 
and chargino sectors, respectively. The definitions of $U$ and $V$ can be 
found in
~\cite{Haber:1985rc,Gunion:1986yn}.

\subsection{$Z$ boson interactions}

The Feynman rules for $Z$ boson interactions are given by,
\begin{figure}[h!]
\begin{center}
\includegraphics[width=9cm]{Zneutralinos.epsi}
\label{charginos_feyn}
\end{center}
\end{figure}

where
\begin{eqnarray}
E_{AB}^{L(n)} &=& \frac{g}{\cos{\theta_W}} O_{AB}^{\prime \prime L} = \frac{g}{c_W} \left( -\frac{1}{2} N_{A3} N_{B3}^{\ast} + \frac{1}{2} N_{A4} N_{B4}^{\ast} \right) \\
E_{AB}^{R(n)} &=& \frac{g}{\cos{\theta_W}} O_{AB}^{\prime \prime R} = -\frac{g}{c_W} \left( -\frac{1}{2} N_{A3}^{\ast} N_{B3} + \frac{1}{2} N_{A4}^{\ast} N_{B4} \right)
\end{eqnarray}


\begin{figure}[h!]
\begin{center}
\includegraphics[width=9cm]{Zcharginos.epsi}
\label{charginos_feyn}
\end{center}
\end{figure}

where
\begin{eqnarray}
E_{AB}^{L(c)} &=& -\frac{g}{\cos{\theta_W}} O_{AB}^{\prime R} = -\frac{g}{c_W} \left[ -\left( \frac{1}{2} - s_W^2 \right) U_{A2}^{\ast} U_{B2} - c_W^2 U_{A1}^{\ast} U_{B1} \right] \\
E_{AB}^{R(c)} &=& -\frac{g}{\cos{\theta_W}} O_{AB}^{\prime L} = -\frac{g}{c_W} \left[ -\left( \frac{1}{2} - s_W^2 \right) V_{A2} V_{B2}^{\ast} - c_W^2 V_{A1} V_{B1}^{\ast} \right]
\end{eqnarray}

\begin{figure}[h!]
\begin{center}
\includegraphics[width=7cm]{Zslpetons.epsi}
\label{charginos_feyn}
\end{center}
\end{figure}

where
\begin{equation}
Q_{XY}^{(\tilde{l})} = -\frac{g}{c_W} \sum_{k=1}^3 \left[ \left( -\frac{1}{2} + s_W^2 \right) R_{2k-1, X}^{(l) \ast} R_{2k-1, Y}^{(l)} + s_W^2 R_{2k, X}^{(l) \ast} R_{2k, Y}^{(l)}  \right]
\end{equation}

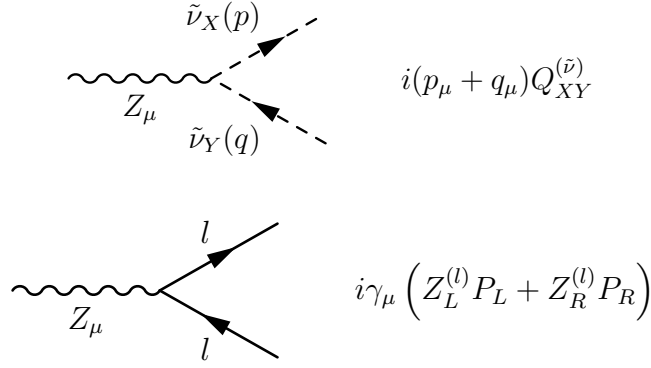
\vspace{2cm}

\begin{figure}[h!]
\begin{center}
\includegraphics[width=7cm]{Zsneutrinos.epsi}
\label{charginos_feyn}
\end{center}
\end{figure}

where
\begin{equation}
Q_{XY}^{(\tilde{\nu})} = -\frac{g}{2 c_W} \delta_{XY}
\end{equation}

\begin{figure}[h!]
\begin{center}
\includegraphics[width=8.5cm]{Zleptons.epsi}
\label{charginos_feyn}
\end{center}
\end{figure}

where
\begin{eqnarray}
Z_L^{(l)} &=& -\frac{g}{c_W} \left[ -\frac{1}{2} + s_W^2 \right] \\
Z_R^{(l)} &=& -\frac{g}{c_W} s_W^2
\end{eqnarray}
 We have used here and everywhere the short notation $s_W = \sin{\theta_W}$ and $c_W = \cos{\theta_W}$.

\subsection{Higgs boson interactions}

The Feynman rules for the three neutral Higgs bosons read as,

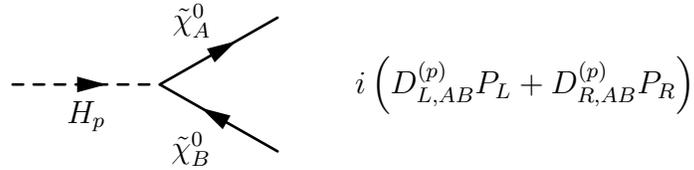
\begin{figure}[h!]
\begin{center}
\includegraphics[width=9cm]{Hneutralinos.epsi}
\label{charginos_feyn}
\end{center}
\end{figure}

where
\begin{eqnarray}
D_{L, AB}^{(p)} &=& -\frac{g}{\sin{\beta}} \left[ Q_{BA}^{'' \ast} \sigma_5^{(p)} - R_{BA}^{'' \ast} \sigma_2^{(p)} + \frac{m_{\chi_A^0}}{2 M_W} \sigma_2^{(p)} \delta_{BA} \right] \\
D_{R, AB}^{(p)} &=& -\frac{g}{\sin{\beta}} \left[ Q_{BA}^{''} \sigma_5^{(p) \ast} - R_{BA}^{''} \sigma_2^{(p) \ast} + \frac{m_{\chi_A^0}}{2 M_W} \sigma_2^{(p) \ast} \delta_{BA} \right]
\end{eqnarray}
and
\begin{eqnarray}
Q_{AB}^{''} &=& \frac{1}{2} \left[ N_{A3} \left( N_{B2} - \tan{\theta_W} N_{B1} \right) + N_{B3} \left( N_{A2} - \tan{\theta_W} N_{A1} \right)  \right] \\
R_{AB}^{''} &=& \frac{1}{2 M_W} \left[ M_2^* N_{A2} N_{B2} + M_1^* N_{A1} N_{B1} - \mu^* \left( N_{A3} N_{B4} + N_{A4} N_{B3} \right) \right]
\end{eqnarray}

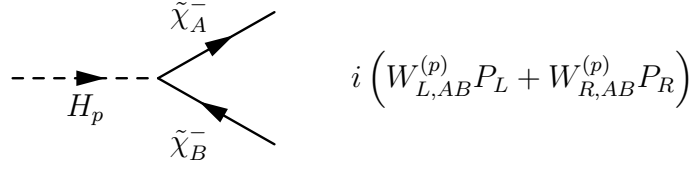
\begin{figure}[h!]
\begin{center}
\includegraphics[width=9cm]{Hcharginos.epsi}
\label{charginos_feyn}
\end{center}
\end{figure}

where
\begin{eqnarray}
W_{L, AB}^{(p)} &=& -\frac{g}{\sin{\beta}} \left[ Q_{BA}^{\ast} \sigma_5^{(p)} - R_{BA}^{\ast} \sigma_2^{(p)} + \frac{m_{\chi_A^-}}{2 M_W} \sigma_2^{(p)} \delta_{BA} \right] \\
W_{R, AB}^{(p)} &=& -\frac{g}{\sin{\beta}} \left[ Q_{AB} \sigma_5^{(p) \ast} - R_{AB} \sigma_2^{(p) \ast} + \frac{m_{\chi_A^-}}{2 M_W} \sigma_2^{(p) \ast} \delta_{AB} \right]
\end{eqnarray}
and
\begin{eqnarray}
Q_{AB} &=& \frac{1}{\sqrt{2}} U_{A2} V_{B1} \\
R_{AB} &=& \frac{1}{2 M_W} \left[ M_2^* U_{A1} V_{B1} + \mu^* U_{A2} V_{B2} \right]
\end{eqnarray}

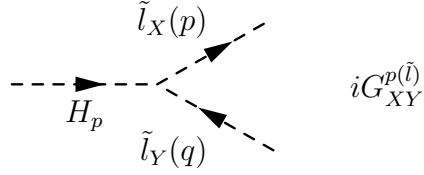
\begin{figure}[h!]
\begin{center}
\includegraphics[width=5.5cm]{Hsleptons.epsi}
\label{charginos_feyn}
\end{center}
\end{figure}

where
\begin{eqnarray}
G_{XY}^{p (\tilde{l})} &=& -g \left[ g_{LL, e}^{(p)} R_{1X}^{*(l)} R_{1Y}^{(l)} + g_{RR, e}^{(p)} R_{2X}^{*(l)} R_{2Y}^{(l)} + g_{LR, e}^{(p)} R_{1X}^{*(l)} R_{2Y}^{(l)} + g_{RL, e}^{(p)} R_{2X}^{*(l)} R_{1Y}^{(l)}\right. \nonumber \\
&+& g_{LL, \mu}^{(p)} R_{3X}^{*(l)} R_{3Y}^{(l)} + g_{RR, \mu}^{(p)} R_{4X}^{*(l)} R_{4Y}^{(l)} + g_{LR, \mu}^{(p)} R_{3X}^{*(l)} R_{4Y}^{(l)} + g_{RL, \mu}^{(p)} R_{4X}^{*(l)} R_{3Y}^{(l)} \nonumber \\
&+& \left. g_{LL, \tau}^{(p)} R_{5X}^{*(l)} R_{5Y}^{(l)} + g_{RR, \tau}^{(p)} R_{6X}^{*(l)} R_{6Y}^{(l)} + g_{LR, \tau}^{(p)} R_{5X}^{*(l)} R_{6Y}^{(l)} + g_{RL, \tau}^{(p)} R_{6X}^{*(l)} R_{5Y}^{(l)} \right]
\end{eqnarray}

with
\begin{eqnarray}
g_{LL, l}^{(p)} &=&  \frac{M_Z}{\cos{\theta_W}} \sigma_3^{(p)} \left( \frac{1}{2}- \sin^2{\theta_W} \right) + \frac{m_{l}^2}{M_W \cos{\beta}} \sigma_4^{(p)} \\
g_{RR, l}^{(p)} &=&  \frac{M_Z}{\cos{\theta_W}} \sigma_3^{(p)} \left(  \sin^2{\theta_W} \right) + \frac{m_{l}^2}{M_W \cos{\beta}}  \sigma_4^{(p)} \\
g_{LR, l}^{(p)} &=& \left(-\sigma_1^{(p)}A_l-\sigma_2^{(p)*}\mu\right) \frac{m_{l}}{2 M_W \cos{\beta}} \\
g_{RL, l}^{(p)} &=& g_{LR, l}^{(p)*}
\end{eqnarray}
with $A_l = (A_l)^{ii}/(Y_l)^{ii}, i = 1, 2, 3$ for $l = e, \mu, \tau$, respectively.

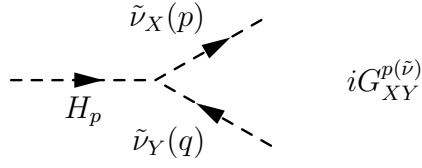
\begin{figure}[h!]
\begin{center}
\includegraphics[width=5.5cm]{Hsneutrinos.epsi}
\label{charginos_feyn}
\end{center}
\end{figure}

where
\begin{equation}
G_{XY}^{p (\tilde{\nu})} = -g \left[ g_{LL, \nu}^{(p)} R_{1X}^{*(\nu)} R_{1Y}^{(\nu)} + g_{LL, \nu}^{(p)} R_{2X}^{*(\nu)} R_{2Y}^{(\nu)} + g_{LL, \nu}^{(p)} R_{3X}^{*(\nu)} R_{3Y}^{(\nu)} \right]
\end{equation}

with
\begin{equation}
g_{LL, \nu}^{(p)} = -\frac{M_Z}{2\cos{\theta_W}} \sigma_3^{(p)}
\end{equation}


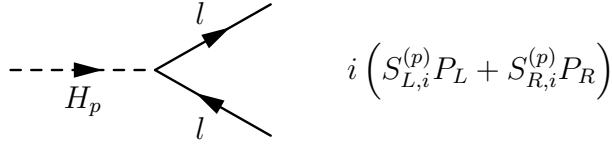
\begin{figure}[h!]
\begin{center}
\includegraphics[width=8cm]{Hleptons.epsi}
\label{charginos_feyn}
\end{center}
\end{figure}

where
\begin{eqnarray}
S_{L, i}^{(p)} &=& g \frac{m_{l_i}}{2 M_W \cos{\beta}} \sigma_1^{(p) \ast} \\
S_{R, i}^{(p)} &=& S_{L, i}^{(p) \ast} 
\end{eqnarray}

In all the above equations,
\begin{eqnarray}
\sigma_1^{(p)} &=& \left( \begin{array}{c} \sin{\alpha} \\ -\cos{\alpha} \\ i \sin{\beta} \end{array} \right) \\ 
\sigma_2^{(p)} &=& \left( \begin{array}{c} \cos{\alpha} \\ \sin{\alpha} \\ -i \cos{\beta} \end{array} \right) \\
\sigma_3^{(p)} &=& \left( \begin{array}{c} \sin{(\alpha + \beta)} \\ -\cos{(\alpha + \beta)} \\ 0 \end{array} \right) \\
\sigma_4^{(p)} &=& \left( \begin{array}{c} -\sin{\alpha} \\ \cos{\alpha} \\ 0 \end{array} \right) \\
\sigma_5^{(p)} &=& \left( \begin{array}{c} -\cos{(\beta - \alpha)} \\ \sin{(\beta - \alpha)} \\  i \cos{2\beta} \end{array} \right)
\end{eqnarray}
and $H_p (p = 1, 2, 3) = h^0, H^0, A^0$. We have also used here
the standard notation for the MSSM soft-gaugino-mass parameters $M_{1,2}$ and
the $\mu$ parameter.

\section{}
\label{apendice2}

In this appendix we present the analytical expressions of the loop-functions for the calculations of the $l_j^- \to l_i^- l_i^- l_j^+$ decays. In these expressions we neglect the external fermion momenta/masses which for the present computation works extremely well. That is,
\begin{eqnarray}
B(k^2, m_1^2, m_2^2) &\simeq& B(0, m_1^2, m_2^2) = B(m_1^2, m_2^2) \\
C(k_1^2, k_2^2, m_1^2, m_2^2, m_3^2) &\simeq& C(0, 0, m_1^2, m_2^2, m_3^2) = C(m_1^2, m_2^2, m_3^2) \\
D(k_1^2, k_2^2, k_3^2, m_1^2, m_2^2, m_3^2, m_4^2) &\simeq& D(0, 0, 0, m_1^2, m_2^2, m_3^2, m_4^2) \nonumber \\
&=& D(m_1^2, m_2^2, m_3^2, m_4^2)
\end{eqnarray}

\subsection{Two-points functions}

The analytical expressions for $B_0$ and $B_1$ functions are the following:
\begin{eqnarray}
B_0(m_1^2, m_2^2) &=& -\log{m_2^2} + \frac{m_2^2 - m_1^2 + m_1^2 \log{\left(\frac{m_1^2}{m_2^2}\right)}}{m_2^2 - m_1^2} \\
B_1(m_1^2, m_2^2) &=& -\frac{1}{2} + \frac{1}{2} \log{m_2^2} - \frac{m_1^2 - m_2^2 + 2 m_1^2 \log{\left(\frac{m_2^2}{m_1^2}\right)}}{4 \left( m_1^2 - m_2^2 \right)^2}
\end{eqnarray}

\subsection{Three-points functions}

The expressions for the three-points functions used in this work are given by,
\begin{eqnarray}
C_0(m_1^2, m_2^2, m_3^2) &=& -\frac{1}{m_2^2 - m_3^2} \left( \frac{m_1^2 \log{m_1^2} - m_2^2 \log{m_2^2}}{m_1^2 - m_2^2} - \frac{m_1^2 \log{m_1^2} - m_3^2 \log{m_3^2}}{m_1^2 - m_3^2} \right) \\
\tilde{C}_0(m_1^2, m_2^2, m_3^2) &=& 1 - \frac{1}{m_2^2 - m_3^2} \left( \frac{m_1^4 \log{m_1^2} - m_2^4 \log{m_2^2}}{m_1^2 - m_2^2} - \frac{m_1^4 \log{m_1^2} - m_3^4 \log{m_3^2}}{m_1^2 - m_3^2} \right) \\
C_{11}(m_1^2, m_2^2, m_3^2) &=& \frac{m_1^2}{2 (m_1^2 - m_2^2)^2 (m_1^2 - m_3^2)^2 (m_2^2 - m_3^2)} \nonumber \\
&\times& \left[ -(m_1^2 - m_2^2) (m_1^2 - m_3^2) (m_2^2 - m_3^2) + m_1^2 m_2^2 (2 m_1^2 - m_2^2) \log{\frac{m_1^2}{m_2^2}} \right. \nonumber \\
&+& \left. m_1^2 m_3^2 (-2 m_1^2 + m_3^2) \log{\frac{m_1^2}{m_3^2}} + m_2^2 m_3^2 (-2 m_1^2 + m_2^2) (-2 m_1^2 + m_3^2) \log{\frac{m_2^2}{m_3^2}} \right] \nonumber \\
\\
C_{12}(m_1^2, m_2^2, m_3^2) &=& \frac{1}{2 (m_1^2 - m_2^2) (m_1^2 - m_3^2)^2 (m_2^2 - m_3^2)^2} \nonumber \\
&\times& \left[ (m_1^2 - m_2^2) (m_1^2 - m_3^2) (m_2^2 - m_3^2) m_3^2 + m_2^4 m_3^2 (2 m_1^2 - m_3^2) \log{\frac{m_2^2}{m_3^2}} \right. \nonumber \\
&+& m_1^4 \left. \left( m_2^4 \log{\frac{m_1^2}{m_2^2}} + m_3^2 (-2 m_2^2 + m_3^2) \log{\frac{m_1^2}{m_3^2}} \right) \right] \\
C_{24}(m_1^2, m_2^2, m_3^2) &=& \frac{1}{4} \tilde{C}_0(0, 0, m_1^2, m_2^2, m_3^2)
\end{eqnarray}

\subsection{Four-points functions}

Finally, the four-points functions have the following expressions,
\begin{eqnarray}
D_0(m_1^2, m_2^2, m_3^2, m_4^2) &=& -\frac{m_1^2 \log{m_1^2}}{(m_1^2 - m_2^2) (m_1^2 - m_3^2) (m_1^2 - m_4^2)} + \frac{m_2^2 \log{m_2^2}}{(m_1^2 - m_2^2) (m_2^2 - m_3^2) (m_2^2 - m_4^2)} \nonumber \\
&-& \frac{m_3^2 \log{m_3^2}}{(m_1^2 - m_3^2) (m_2^2 - m_3^2) (m_3^2 - m_4^2)} + \frac{m_4^2 \log{m_4^2}}{(m_1^2 - m_4^2) (m_2^2 - m_4^2) (m_3^2 - m_4^2)} \nonumber \\
\\
\tilde{D}_0(m_1^2, m_2^2, m_3^2, m_4^2) &=& -\frac{m_1^4 \log{m_1^2}}{(m_1^2 - m_2^2) (m_1^2 - m_3^2) (m_1^2 - m_4^2)} + \frac{m_2^4 \log{m_2^2}}{(m_1^2 - m_2^2) (m_2^2 - m_3^2) (m_2^2 - m_4^2)} \nonumber \\
&-& \frac{m_3^4 \log{m_3^2}}{(m_1^2 - m_3^2) (m_2^2 - m_3^2) (m_3^2 - m_4^2)} + \frac{m_4^4 \log{m_4^2}}{(m_1^2 - m_4^2) (m_2^2 - m_4^2) (m_3^2 - m_4^2)} \nonumber \\
\end{eqnarray}




\begin{thebibliography}{10}

\bibitem{neutrinodata}
B.~T.~Cleveland {\it et al.},
Astrophys.\ J.\  {\bf 496} (1998) 505;
W.~Hampel {\it et al.},
Phys. \ lett.\ B {\bf 447} (1999) 127;
Q.~R.~Ahmad {\it et al.}  [SNO Collaboration],
Phys.\ Rev.\ Lett.\  {\bf 87} (2001) 071301
[arXiv:nucl-ex/0106015];
Q.~R.~Ahmad {\it et al.}  [SNO Collaboration],
Phys.\ Rev.\ Lett.\  {\bf 89} (2002) 011302
[arXiv:nucl-ex/0204009].
R.~Becker-Szendy {\it et al.},
  Nucl.\ Phys.\ Proc.\ Suppl.\  {\bf 38}, 331 (1995).
Y.~Fukuda {\it et al.}  [Kamiokande Collaboration],
  Phys.\ Lett.\ B {\bf 335}, 237 (1994);
Y.~Ashie {\it et al.}  [Super-Kamiokande Collaboration],
  Phys.\ Rev.\ Lett.\  {\bf 93}, 101801 (2004)
  [arXiv:hep-ex/0404034].
T.~Araki {\it et al.}  [KamLAND Collaboration],
  Phys.\ Rev.\ Lett.\  {\bf 94}, 081801 (2005)
  [arXiv:hep-ex/0406035];
E.~Aliu {\it et al.}  [K2K Collaboration],
  Phys.\ Rev.\ Lett.\  {\bf 94}, 081802 (2005)
  [arXiv:hep-ex/0411038];
T.~Araki {\it et al.}  [KamLAND Collaboration],
  Phys.\ Rev.\ Lett.\  {\bf 94}, 081801 (2005)
  [arXiv:hep-ex/0406035].

\bibitem{MNS}
Z.~Maki, M.~Nakagawa and S.~Sakata,
Prog. \ Theor. \ Phys.\ {\bf 28} (1962) 870;
B.~Pontecorvo, 
Zh. \ Eksp. \ Teor.\ Fiz. {\bf 33} (1957) 549 and {\bf 34} (1957) 247.

\bibitem{seesaw}
M.~Gell-Mann, P.~Ramond, and R.~Slansky, \emph{Complex spinors and unified
 theories}, in \emph{Supergravity} (P.~van Nieuwenhuizen and
D.~Z. Freedman,  eds.), North Holland, Amsterdam, 1979, p.~315;
Pierre Ramond, CALT-68-709, Feb 1979. 21pp.
Invited talk given at Sanibel Symposium, Palm Coast, Fla., Feb 25 - Mar
2, 1979; hep-ph/9809459;
T.~Yanagida, in \emph{Proceedings of the Workshop on the Unified Theory
and the  Baryon Number in the Universe} (O.~Sawada and A.~Sugamoto, eds.),
KEK,  Tsukuba, Japan, 1979, p.~95;
S.~L. Glashow, \emph{The future of elementary particle physics}, in
 \emph{Proceedings of the 1979 Carg{\`e}se Summer Institute on Quarks and
 Leptons} (M.~L{\'e}vy, J.-L. Basdevant, D.~Speiser, J.~Weyers,
R.~Gastmans,
 and M.~Jacob, eds.), Plenum Press, New York, 1980, pp.~687--713;
R.~N. Mohapatra and G.~Senjanovi{\'c}, Phys. Rev. Lett. \textbf{44}, 912
(1980).

\bibitem{borzumati}
F.~Borzumati and A.~Masiero, 
Phys.\ Rev.\ Lett.\  {\bf 57}, 961 (1986).

\bibitem{Aubert:2003pc}
  B.~Aubert {\it et al.}  [BABAR Collaboration],
  Phys.\ Rev.\ Lett.\  {\bf 92}, 121801 (2004)
  [arXiv:hep-ex/0312027].

\bibitem{Bellgardt:1987du}
  U.~Bellgardt {\it et al.}  [SINDRUM Collaboration],
  Nucl.\ Phys.\ B {\bf 299}, 1 (1988).

\bibitem{Aubert:2005ye}
  B.~Aubert {\it et al.}  [BABAR Collaboration],
  Phys.\ Rev.\ Lett.\  {\bf 95}, 041802 (2005)
  [arXiv:hep-ex/0502032].

\bibitem{Aubert:2005wa}
  B.~Aubert {\it et al.}  [BABAR Collaboration],
  arXiv:hep-ex/0508012.

\bibitem{mue}
M.~L.~Brooks {\it et al.}  [MEGA Collaboration],
Phys.\ Rev.\ Lett.\  {\bf 83} (1999) 1521
[arXiv:hep-ex/9905013].

\bibitem{Hisano:1995cp}
J.~Hisano, T.~Moroi, K.~Tobe and M.~Yamaguchi,
Phys.\ Rev.\ D {\bf 53} (1996) 2442
[arXiv:hep-ph/9510309].
\bibitem{todos}
  J.~Hisano, D.~Nomura and T.~Yanagida,
  Phys.\ Lett.\ B {\bf 437}, 351 (1998)
  [arXiv:hep-ph/9711348];
  J.~Hisano and D.~Nomura,
  Phys.\ Rev.\ D {\bf 59}, 116005 (1999)
  [arXiv:hep-ph/9810479];
  W.~Buchmuller, D.~Delepine and F.~Vissani,
  Phys.\ Lett.\ B {\bf 459}, 171 (1999)
  [arXiv:hep-ph/9904219];
  J.~R.~Ellis, M.~E.~Gomez, G.~K.~Leontaris, S.~Lola and D.~V.~Nanopoulos,
  Eur.\ Phys.\ J.\ C {\bf 14}, 319 (2000)
  [arXiv:hep-ph/9911459];
  X.~J.~Bi, Y.~B.~Dai and X.~Y.~Qi,
  Phys.\ Rev.\ D {\bf 63}, 096008 (2001)
  [arXiv:hep-ph/0010270];
  J.~Hisano and K.~Tobe,
  Phys.\ Lett.\ B {\bf 510}, 197 (2001)
  [arXiv:hep-ph/0102315];
  X.~J.~Bi and Y.~B.~Dai,
  Phys.\ Rev.\ D {\bf 66}, 076006 (2002)
  [arXiv:hep-ph/0112077];
  D.~F.~Carvalho, J.~R.~Ellis, M.~E.~Gomez, S.~Lola and J.~C.~Romao,
  Phys.\ Lett.\ B {\bf 618}, 162 (2005)
  [arXiv:hep-ph/0206148];
  S.~Lavignac, I.~Masina and C.~A.~Savoy,
  Phys.\ Lett.\ B {\bf 520}, 269 (2001)
  [arXiv:hep-ph/0106245];
  Y.~Kuno and Y.~Okada,
  Rev.\ Mod.\ Phys.\  {\bf 73}, 151 (2001)
  [arXiv:hep-ph/9909265];
  J.~R.~Ellis, J.~Hisano, M.~Raidal and Y.~Shimizu,
  Phys.\ Rev.\ D {\bf 66}, 115013 (2002)
  [arXiv:hep-ph/0206110];
  F.~Deppisch, H.~Pas, A.~Redelbach, R.~Ruckl and Y.~Shimizu,
  Eur.\ Phys.\ J.\ C {\bf 28}, 365 (2003)
  [arXiv:hep-ph/0206122];
  A.~Dedes, J.~R.~Ellis and M.~Raidal,
  Phys.\ Lett.\ B {\bf 549}, 159 (2002)
  [arXiv:hep-ph/0209207];
  J.~Hisano,
  arXiv:hep-ph/0209005;
  S.~Pascoli, S.~T.~Petcov and C.~E.~Yaguna,
  Phys.\ Lett.\ B {\bf 564}, 241 (2003)
  [arXiv:hep-ph/0301095];
  T.~Fukuyama, T.~Kikuchi and N.~Okada,
  Phys.\ Rev.\ D {\bf 68}, 033012 (2003)
  [arXiv:hep-ph/0304190];
  J.~I.~Illana and M.~Masip,
  Eur.\ Phys.\ J.\ C {\bf 35}, 365 (2004)
  [arXiv:hep-ph/0307393];
  A.~Masiero, S.~K.~Vempati and O.~Vives,
  New J.\ Phys.\  {\bf 6}, 202 (2004)
  [arXiv:hep-ph/0407325].

\bibitem{Casas:2001sr}
J.~A.~Casas and A.~Ibarra,
Nucl.\ Phys.\ B {\bf 618} (2001) 171
[arXiv:hep-ph/0103065].

\bibitem{Babu:2002et}
K.~S.~Babu and C.~Kolda,
Phys.\ Rev.\ Lett.\  {\bf 89}, 241802 (2002)
[arXiv:hep-ph/0206310].

\bibitem{Ellis:2002fe}
J.~R.~Ellis, J.~Hisano, M.~Raidal and Y.~Shimizu,
Phys.\ Rev.\ D {\bf 66}, 115013 (2002)
[arXiv:hep-ph/0206110].

\bibitem{Brignole1}
A.~Brignole, A.~Rossi, 
\newblock  Phys.~Lett.~{\bf B566}, 217 (2003), hep-ph/0304081.

\bibitem{Arganda:2004bz}
  E.~Arganda, A.~M.~Curiel, M.~J.~Herrero and D.~Temes,
  Phys.\ Rev.\ D {\bf 71}, 035011 (2005)
  [arXiv:hep-ph/0407302].

\bibitem{Brignole2}
A.~Brignole and A.~Rossi,
  Nucl.\ Phys.\ B {\bf 701}, 3 (2004)
  [arXiv:hep-ph/0404211].

\bibitem{Paradisi:2005tk}
  P.~Paradisi,
  arXiv:hep-ph/0508054.

\bibitem{Parry:2005fp}
  J.~K.~Parry,
  arXiv:hep-ph/0510305.

\bibitem{pdg2004}
  S.~Eidelman {\it et al.}  [Particle Data Group],
  Phys.\ Lett.\ B {\bf 592}, 1 (2004).

\bibitem{Haber:1985rc}
H.~E. Haber and G.~L. Kane, \textsl{ The search for supersymmetry: probing
  physics beyond the standard model},
\newblock
\newblock Phys. Rept. \textbf{ 117}, 75 (1985).

\bibitem{Gunion:1986yn}
J.~F. Gunion and H.~E. Haber, \textsl{ Higgs bosons in supersymmetric models.
  1},
\newblock
\newblock Nucl. Phys. \textbf{ B272}, 1 (1986). [E: {\bf B402}, 569 (1993)]

\bibitem{Grossman:1997is}
Y.~Grossman and H.~E.~Haber,
Phys.\ Rev.\ Lett.\  {\bf 78} (1997) 3438
[arXiv:hep-ph/9702421].

\bibitem{Chankowski:2004jc}
P.~H.~Chankowski, J.~R.~Ellis, S.~Pokorski, M.~Raidal and K.~Turzynski,
Nucl.\ Phys.\ B {\bf 690} (2004) 279
[arXiv:hep-ph/0403180].

\bibitem{Bi:2003ea}
  X.~J.~Bi, B.~Feng and X.~m.~Zhang,
  arXiv:hep-ph/0309195.

\bibitem{review}
M.~C.~Gonzalez-Garcia and C.~Pena-Garay,
Phys.\ Rev.\ D {\bf 68} (2003) 093003
[arXiv:hep-ph/0306001].

\bibitem{Porod:2003um}
  W.~Porod,
  Comput.\ Phys.\ Commun.\  {\bf 153}, 275 (2003)
  [arXiv:hep-ph/0301101].


\bibitem{9604387}
  F.~Gabbiani, E.~Gabrielli, A.~Masiero, L.~Silvestrini,
Nucl. Phys. \textbf{ B477}, 321 (1996).

\bibitem{Chankowski:2005jh}
  P.~H.~Chankowski, O.~Lebedev and S.~Pokorski,
  arXiv:hep-ph/0502076.

\bibitem{Paradisi:2005fk}
  P.~Paradisi,
  arXiv:hep-ph/0505046.

\bibitem{Hollik}
W.~Hollik, in {\it Precision Tests of the Standard Electroweak
Model}, edited by P.~Langacker (World Scientific, Singapore,
1995), pp.~37--116;

\end{thebibliography}
\end{document}